\documentclass[runningheads,a4paper]{llncs}

\usepackage{amssymb}
\usepackage{graphicx}
\usepackage{amsmath, bm, mathtools,array}
\usepackage{textcomp}
\usepackage{hyperref}
\usepackage{verbatim}
\usepackage{lipsum} 
\usepackage{multirow}
\usepackage{siunitx}
\usepackage{diagbox}
\usepackage{dsfont}
\usepackage{colortbl}
\usepackage{booktabs}
\usepackage{subcaption}
\usepackage{booktabs}
\usepackage{tcolorbox}
\usepackage{lscape}

\DeclareMathOperator*{\argmin}{arg\,min}

\RequirePackage{color}\definecolor{BLUE}{rgb}{0,0,0}
\providecommand{\BLUE}[1]{{\color{BLUE}{#1}}}

\newcommand{\systemI}{\texttt{LCNN-trim-pad}}
\newcommand{\systemII}{\texttt{LCNN-attention}}
\newcommand{\systemIII}{\texttt{LCNN-LSTM-sum}}
\newcommand{\systemIV}{\texttt{RawNet2}}

\usepackage{url}
\urldef{\mailsa}\path|{wangxin, jyamagis}@nii.ac.jp|    
\newcommand{\keywords}[1]{\par\addvspace\baselineskip
\noindent\keywordname\enspace\ignorespaces#1}

\begin{document}

\mainmatter  

\title{A Practical Guide to Logical Access Voice Presentation Attack Detection}

\titlerunning{A Practical Guide to Logical Access Voice Presentation Attack Detection}

%
%
\author{Xin Wang%
\thanks{This work will appear as one chapter of the book Fake Media Generation and Detection, edited by Mahdi Khosravy, Isao Echizen, Noboru Babaguchi.}%
\and Junichi Yamagishi}
\authorrunning{Xin Wang, et al.}

\institute{National Institute of Informatics, Japan\\
\mailsa}

%
%

\maketitle

\begin{abstract}
Voice-based human-machine interfaces with an automatic speaker verification (ASV) component are commonly used in the market. However, the threat from presentation attacks is also growing since attackers can use recent speech synthesis technology to produce a natural-sounding voice of a victim. Presentation attack detection (PAD) for ASV, or speech anti-spoofing, is therefore indispensable. 
Research on voice PAD has seen significant progress since the early 2010s, including the advancement in PAD models, benchmark datasets, and evaluation campaigns. This chapter presents a practical guide to the field of voice PAD, with a focus on logical access attacks using text-to-speech and voice conversion algorithms and spoofing countermeasures based on artifact detection. It introduces the basic concept of voice PAD, explains the common techniques, and provides an experimental study using recent methods on a benchmark dataset. Code for the experiments is open-sourced.
\end{abstract}

\section{Introduction}
\label{sec:intro}
Speech- or voice-based human-machine interfaces have been used in commercial products and services. In many applications such as online banking services, the user's identity must be verified before the user can interact with the service further. This verification task can be addressed using automatic speaker verification (ASV) technology \cite{hansen2015speaker,kinnunen2010overview,wu2014study}. Given a speech segment uttered by the user, an ASV system estimates a score that indicates how likely the input speech is from the claimed speaker ID. The system then either accepts or rejects access from that user after comparing the score with a threshold. The ASV system in this case can be formulated as a binary classifier.

An ASV system can be deployed in online applications such as telephony- or Internet-based banking services. In another scenario, the user speaks in person directly to the ASV system deployed in a domestic environment. 
In both cases, ASV serves as a convenient alternative to the password-based access-control mechanism. 
However, ASV is vulnerable to fake or fraudulent voices \cite{evans2013spoofing}. A fraudster or imposter could create a fake voice that sounds like the voice of a victim using text-to-speech (TTS) or voice conversion (VC) technology. The fraudster then injects the fake voice into the ASV system and passes the verification. \BLUE{In 2012, it has been demonstrated that fraudulent voices from a speaker-adaptive TTS can pass the verification with an acceptance rate higher than 80\% \cite{de2012evaluation}. Both the TTS and ASV systems in that study were the start-of-the-art at that time. More recent studies showed that the current state-of-the-art ASV systems can be easily fooled by fraudulent voices produced by the latest TTS and VC systems \cite{WANG2020101114,yamagishi21_asvspoof}. }

\begin{figure}[t]
\centering
\includegraphics[trim=0 520 0 85, clip, width=0.8\textwidth]{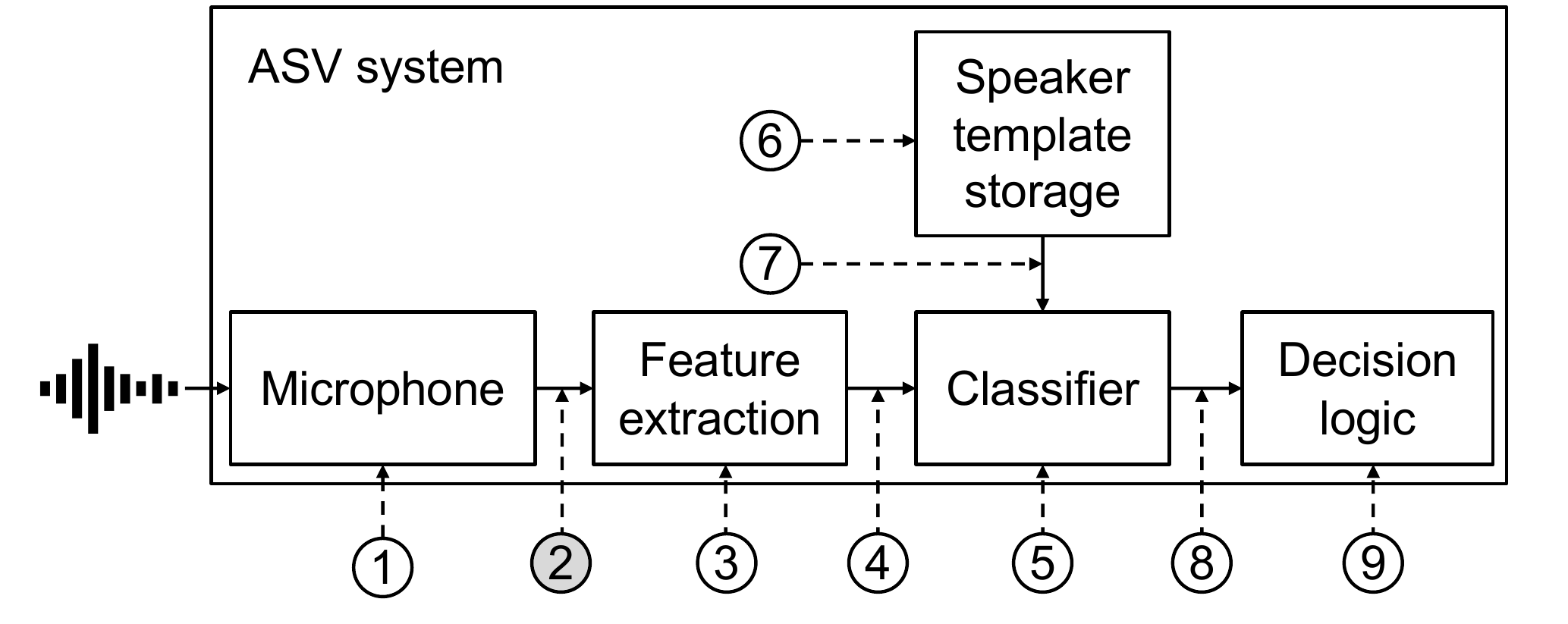}
\caption{Possible PAs to an ASV system. Type 2 on the transmission point corresponds to a logical access attack. Other details are discussed in \cite{sahidullah2019introduction}}
\label{fig:PAs}
\end{figure}

Using TTS and VC synthetic voices to attack ASV is one type of presentation attack (PA). 
According to the ISO/IEC standard \cite{ISOIEC30107}, PAs can be grouped as shown in Fig.\ref{fig:PAs}. PAs based on TTS and VC can be injected at the transmission point after the microphone  (i.e., No.2 in Fig.\ref{fig:PAs}). For example,  attackers may connect to an online banking service and send a synthetic voice signal to the ASV system while bypassing the microphone.
This type of PA is referred to as \emph{logical access} (LA) attack, and it is the main focus of this chapter\footnote{Another common type of attack is mounted in front of the microphone and is known as physical access. An attack in this case can be conducted by replying a pre-recorded voice of the victim in front of the ASV system. Note that the acronym PA refers to presentation attack rather than physical access in this chapter.}. 

Protecting ASV systems from the PAs is indispensable. This is referred to as presentation attack detection (PAD) in the ISO/IEC standard \cite{ISO-IEC-30107-1-PAD-Framework-160115}, where the PAD methods are categorized as either liveness detection, artifact detection, or other types. Liveness detection measures and analyzes `the anatomical information of involuntary or voluntary reactions to determine if a biometric sample is being captured from a living subject present' \cite[Sec. 3.3]{ISO-IEC-30107-1-PAD-Framework-160115}. Liveness information can be, for example, the pop noise by human breath \cite{shiota2015voice}, the movement of the articulators, and the associated time-difference-of-arrival of speech signals \cite{zhang2016voicelive}. Although it is possible to forge liveness information, a spoofed voice usually does not contain such information and thus can be discriminated from a real human voice. 
However, liveness detection usually requires specific microphones (e.g., micro electro mechanical systems microphones) or other hardware \cite{sahidullah2017robust,Zhang2017,zhang2020viblive}, which is not available in many applications. 

Most PAD methods in the speech processing research community are based on artifact detection, which is the topic of this chapter. It covers many works under an umbrella term called \emph{speech anti-spoofing}, with the goal to spot artifacts in a presented voice (or its statistics) that are untypical in real human voices. The task is straightforward: given an input speech waveform, PAD determines whether the waveform is uttered by a real human or a \emph{spoofed} waveform. The waveform uttered by a real human speaker is also known as being \emph{genuine} or \emph{bona fide}. The task of a common PAD system therefore can be formulated as a binary classification task:

\begin{tcolorbox}
Given an input waveform $\boldsymbol{o}_{1:T} = ({o}_1, {o}_2, \cdots, {o}_T)\in\mathbb{R}^{T}$ with $T$ sampling points, a PAD model is a function $f$ with a parameter set $\boldsymbol{\theta}$ that conducts
\begin{equation}
f_{\boldsymbol{\theta}}:\mathbb{R}^{T}\rightarrow \{\text{Bona fide}, \text{spoofed}\}
\end{equation}
\end{tcolorbox}

\begin{figure}[t]
\centering
\includegraphics[trim=0 440 0 85, clip, width=0.8\textwidth]{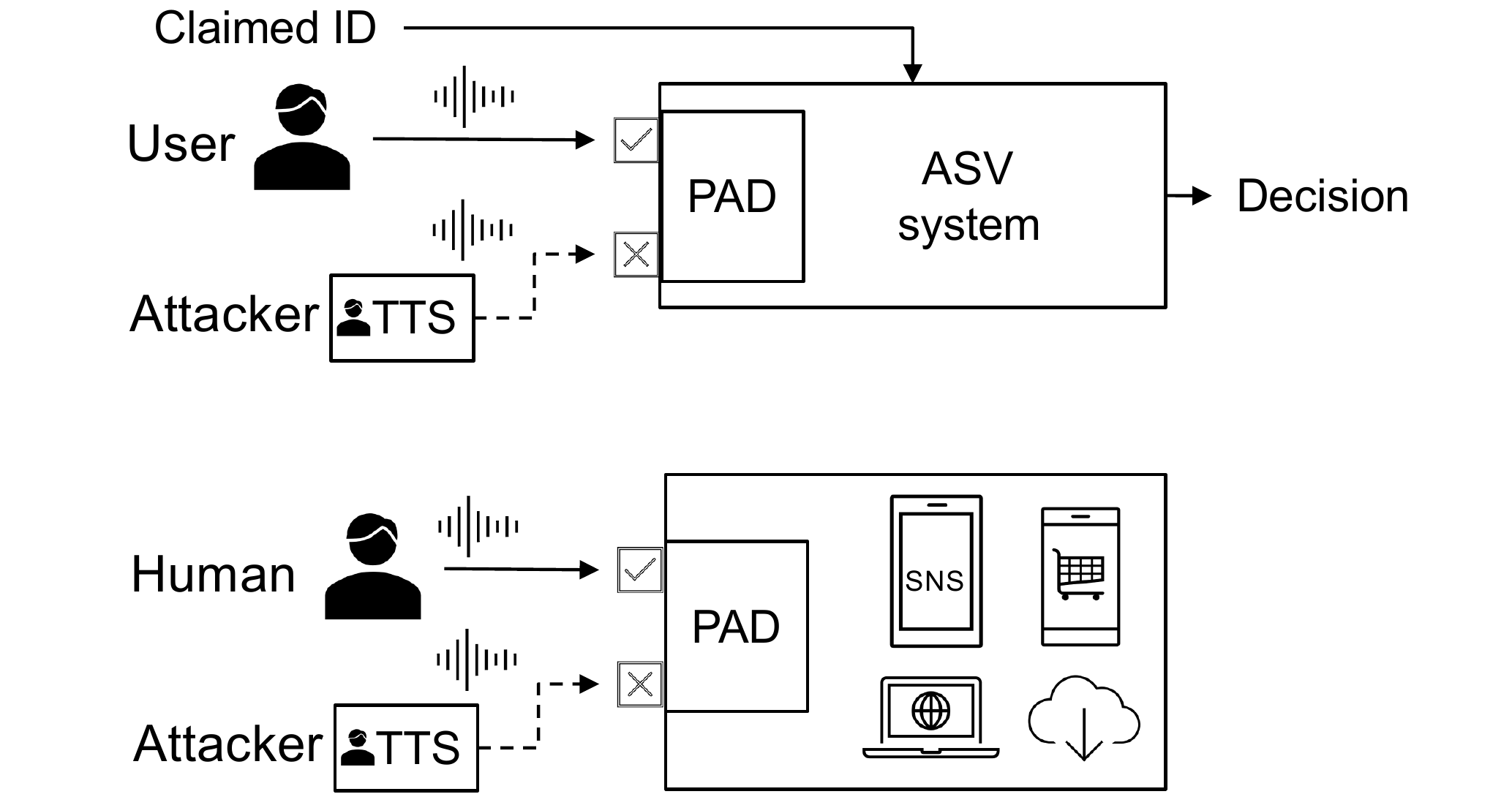}
\caption{Example PADs for ASV and other applications. Attackers can also use VC or other approaches to forge the voice of the victim.}
\label{fig:PAs2}
\end{figure}

The top panel in Fig.\ref{fig:PAs2} illustrates PAD that protects ASV systems from TTS-based attacks. Note that PAD is not only used to protect the ASV systems\footnote{Voice PAD can also be integrated with ASV \cite{khoury2014introducing,sizov2015joint}.}, it can also be applied to detect forged voices in other applications. One typical example is to detect so-called \emph{deepfake voices} without an ASV system in the context. In this case, the PAD could also be referred to as \emph{deepfake voice detection}. 
The fake voice does not need to be perceptually similar to the bona fide voice of a victim in order to fool the human users. 
Nevertheless, a good PAD should detect the fake voice no matter what the attack's goal is.  

Voice PAD has become an established research topic that gathers researchers from machine learning, speech processing, biometrics, and other related fields. While there have been other pilot studies, the first special session on ASV PAD was organized in the flagship conference of the International Speech Communication Association (ISCA) called INTERSPEECH in 2013\footnote{\url{https://www.isca-speech.org/archive/interspeech_2013/\#Sess025}}. Since then, there have been dedicated sessions not only at INTERSPEECH but also other major conferences in the speech research community, including the Speaker and Language Recognition Workshop of ISCA and ICASSP of IEEE. There have also been campaigns providing datasets and baselines, including the biennial ASVspoof challenges \cite{Kinnunen2017,Nautsch2021,wu2015asvspoof} and the BTAS 2016 Speaker Anti-spoofing Competition\cite{korshunov2016overview}.

While there have been excellent literature reviews on PAD \cite{kamble_sailor_patil_li_2020,Wu2015}, the rapid progress on both the attacker's and defender's sides requires an updated introduction. Therefore, this chapter focuses on recent techniques and describes the basic concepts and knowledge required to understand them. This chapter is also a practical guide to building high-performance PAD systems using open-sourced code and recipes. We highly recommend other literature reviews to grasp a broader view on this topic.

Sections~\ref{sec:attacker} and \ref{sec:defender} describe LA attackers and defenders, respectively. Section~\ref{sec:metric} describes evaluation metrics. Section~\ref{sec:camp} briefly explains the ASVspoof challenges, and Section~\ref{sec:exp} details the experiment using the challenges' data and a number of PAD systems\footnote{Experiments were extended from the authors' previous work \cite{wang2021comparative}, but detailed analysis, a new PAD model, and experiments on model fusion are newly added to this chapter. This chapter also includes detailed review on recent PAD models.}. This chapter ends with a discussion on the unsolved issues in Section~\ref{sec:discussion} and conclusion in Section~\ref{sec:conclusion}.

\section{Logical Access Presentation Attack}
\label{sec:attacker}

Presentation attack mounted at the transmission point can use TTS or VC systems to produce the speech waveform of a target speaker (i.e., victim). While the algorithms vary in detail, both TTS and VC can be written as a mapping function that converts one data sequence into another.

\begin{figure}[t]
\centering
\includegraphics[trim=10 500 10 85, clip, width=0.9\textwidth]{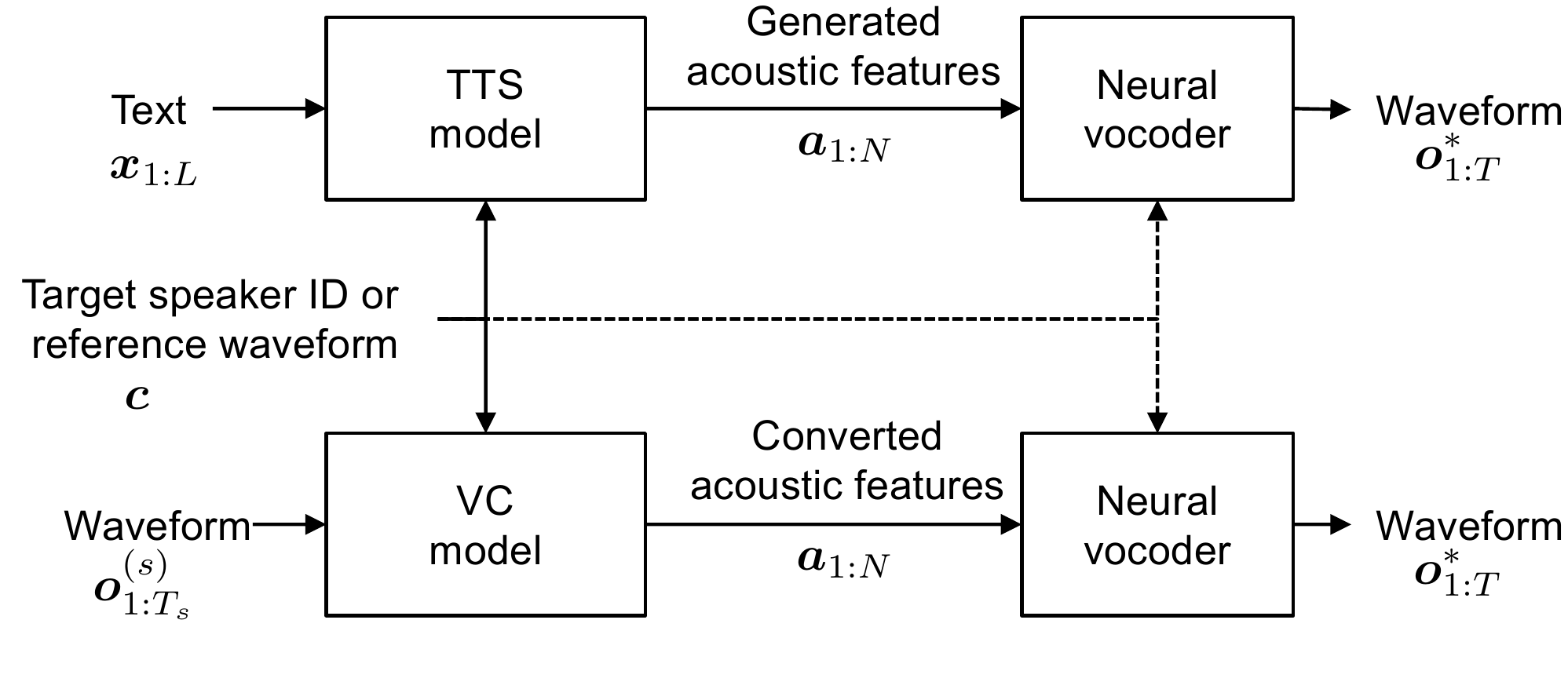}
\caption{General form of recent TTS and VC systems. Superscript $^{(s)}$ indicates the data from a source speaker $s$. Target speaker ID or reference waveform is an optional feature to the neural vocoder.}
\label{fig:attackers}
\end{figure}

\begin{align}
\text{TTS}: \boldsymbol{o}_{1:T}^{*} &= \arg\max_{\boldsymbol{o}_{1:T}} p(\boldsymbol{o}_{1:T} | \boldsymbol{x}_{1:L}, \boldsymbol{c}; \boldsymbol{\theta}) ,\label{eq:TTS}\\
\text{VC}: \boldsymbol{o}^{*}_{1:T}&= \arg\max_{\boldsymbol{o}_{1:T}} p(\boldsymbol{o}_{1:T} | \boldsymbol{o}^{(s)}_{1:T_s}, \boldsymbol{c}; \boldsymbol{\theta}) .\label{eq:VC}
\end{align}
For TTS, $\boldsymbol{x}_{1:L}=(\boldsymbol{x}_1, \boldsymbol{x}_2, \cdots, \boldsymbol{x}_L)$ denotes a text string with $L$ characters or other text units, and $\boldsymbol{c}$ is a general symbol that refers to either the discrete target speaker ID or a reference waveform uttered by the target speaker. The model parameter set is written as $\boldsymbol{\theta}$. \BLUE{The equation for VC has a similar form, but the input data, i.e., condition feature, is a waveform $\boldsymbol{o}^{(s)}_{1:T_s}$ from a source speaker $s$}.  For both TTS and VC, 
$\boldsymbol{o}^{*}_{1:T}$ denotes the generated waveform with $T$ sampling points. Note that the length of the input data sequence ($L$ or $T_s$) is not required to be equal to that of the output waveform.

Many TTS and VC systems can be divided into two parts. As Fig.\ref{fig:attackers} shows, they can use specific models to convert the input into an acoustic feature sequence $\boldsymbol{a}_{1:N}=(\boldsymbol{a}_1, \boldsymbol{a}_2, \cdots, \boldsymbol{a}_N)$. Given $\boldsymbol{a}_{1:N}$, a neural vocoder can be applied to producing the waveform. For example, one state-of-the-art TTS system uses an attention-based sequence-to-sequence model called Tacotron2 to convert the input text into a Mel-spectrogram \cite{shen2018natural}. It then uses a WaveNet vocoder \cite{wavenet,Tamamori2017} to produce the waveform. While the original Tacotron2 is built for a specific speaker, it can be extended to a multi-speaker TTS system by incorporating a speaker encoder to convert the reference waveform of the target speaker into an embedding vector \cite{jia2018transfer}. This embedding vector drives the TTS model to produce acoustic features that carry the identity of the target speaker. A VC system can be built in a similar way. In fact, the top VC system in the Voice Conversion challenge 2018 \cite{Lorenzo-Trueba2018} concatenates a speech recognition system with a TTS model. The waveform of the source speaker $\boldsymbol{o}^{(s)}_{1:T_s}$ is converted into a text string $\boldsymbol{x}_{1:L}$, after which $\boldsymbol{x}_{1:L}$ is transformed into the speech waveform $\boldsymbol{o}_{1:T}$ using the target speaker's TTS model and a WaveNet-based vocoder.

The two state-of-the-art TTS and VC systems have been included in the ASVspoof2019 LA database \cite{WANG2020101114}, i.e., A10 and A15, respectively.  As mentioned in Section~\ref{sec:intro}, the experimental study in \cite{WANG2020101114} has demonstrated the vulnerability of a state-of-the-art X-vector-based ASV system \cite{snyder2018x} to the spoofed waveforms from these TTS and VC systems. According to the scores assigned by the ASV system, the spoofed waveforms were more similar to the voice of the victim speaker than the bona fide speech. 
By using those TTS and VC systems, the attacker can easily pass the verification of the ASV system.

The ASVspoof 2019 LA database includes other TTS and VC systems, ranging from conventional statistical parametric speech synthesis to hybrid systems of TTS and VC. Since the release of the database, new TTS and VC methods have been published in recent years. Readers are recommended to read the latest literature review on TTS \cite{tan2021survey} and the summary paper of the latest Voice Conversion challenge \cite{Lorenzo-Trueba2018,Yi2020}.

\section{Logical Access Presentation Attack Detection}
\label{sec:defender}

PAD is a classification task, and a PAD model is usually trained in a supervised manner\footnote{It is also possible to train one-class classifiers that only require bona fide data \cite{alegre2013one}}. While many methodologies and tools can be applied to the classification task, this chapter focuses on the recent frameworks using front and back ends listed in Table \ref{tab:pad_types}. We treat the so-called end-to-end model as a PAD model with a dummy front end. Table \ref{tab:pad_examples} lists a few voice PAD models.

\begin{table}[t]
    \setlength{\tabcolsep}{5pt}
    \centering
    \caption{Typical PAD front and back ends. $\boldsymbol{o}_{1:T}\in\mathbb{R}^{T}$ and $y\in\{1,\cdots, C\}$ denote the input waveform and target label, respectively. Subscript is dropped to avoid notation clutter. $\mathbf{1}_{\boldsymbol{A}}(k)$ is an indicator function, and $\mathbf{1}_{\boldsymbol{A}}(k)=1$ if $k\in\boldsymbol{A}$, otherwise 0. ${x}_k$ denotes the $k$-th dimension of the vector $\boldsymbol{x}$. $v\in\{1,\cdots, C\}$ denotes the index of the bona fide label.}
    \resizebox{\textwidth}{!}{
\begin{tabular}{l l l}
\toprule
Front end & & \\
& Loss function for training & Feature extraction \\
\midrule
DSP $\mathcal{F}_{\boldsymbol{\phi}}$ & - & $\boldsymbol{a} = \mathcal{F}_{\boldsymbol{\phi}}(\boldsymbol{o})$\\
DNN $\mathcal{F}_{\boldsymbol{\theta_1, \theta_2}}$ & $\mathcal{L}(\boldsymbol{\theta}_1, \boldsymbol{\theta}_2) = -\sum_{k=1}^{C} \mathbf{1}_{\{y\}}(k)\log P(k|\mathcal{F}_{\boldsymbol{\theta}_1,\boldsymbol{\theta}_2}(\boldsymbol{o}))$ & $\boldsymbol{a} = \mathcal{F}_{\boldsymbol{\theta_1}}(\boldsymbol{o})$\\
Hybrid $\mathcal{F}_{\boldsymbol{\theta_1,\theta_2}}\circ\mathcal{F}_{\boldsymbol{\phi}}$ & $\mathcal{L}(\boldsymbol{\theta}_1, \boldsymbol{\theta}_2) = -\sum_{k=1}^{C} \mathbf{1}_{\{y\}}(k)\log P(k|\mathcal{F}_{\boldsymbol{\theta}_1, \boldsymbol{\theta}_2}(\mathcal{F}_{\boldsymbol{\phi}}(\boldsymbol{o})))$ & $\boldsymbol{a} = \mathcal{F}_{\boldsymbol{\theta_1}}(\mathcal{F}_{\boldsymbol{\phi}}(\boldsymbol{o}))$\\
Raw waveform & - & $\boldsymbol{a} = \boldsymbol{o}$\\
\bottomrule\\
\toprule
Back end & & \\
& Loss function for training & Scoring \\
\midrule
GMM $\boldsymbol{\lambda}^{(y)}$ & $\mathcal{L}(\boldsymbol{\lambda}^{(y)}) = -\log p(\boldsymbol{a}|\boldsymbol{\lambda}^{(y)})$ & $\log{\Big({p(\boldsymbol{a} | \boldsymbol{\lambda}^{(y=1)})}/{p(\boldsymbol{a} | \boldsymbol{\lambda}^{(y=0)})}\Big)}$ \\
Distance & - & $-(\boldsymbol{a} - \boldsymbol{\mu}^{(v)})^\top{\boldsymbol{\Sigma}^{(v)}}^{-1}(\boldsymbol{a} - \boldsymbol{\mu}^{(v)})$ \\
DNN $\mathcal{H}_{\boldsymbol{\theta_3}}$ & $\mathcal{L}(\boldsymbol{\theta}_3) = -\sum_{k=1}^{C} \mathbf{1}_{\{y\}}(k)\log P(k|\mathcal{H}_{\boldsymbol{\theta}_3}(\boldsymbol{a}))$ & $P(v|\mathcal{H}_{\boldsymbol{\theta}_3}(\boldsymbol{a}))$ \\
DNN $\mathcal{H}_{\boldsymbol{\theta_3}}$ & $\mathcal{L}(\boldsymbol{\theta}_3) = \sum_{k=1}^{C}\big(\mathcal{H}_{\boldsymbol{\theta}_3}(\boldsymbol{a})_k - \mathbf{1}_{\{y\}}(k)\big)^2$ & $\mathcal{H}_{\boldsymbol{\theta}_3}(\boldsymbol{a})_v$ \\
\bottomrule
\end{tabular}
    }
    \label{tab:pad_types}
\end{table}

\begin{table}
    \setlength{\tabcolsep}{5pt}
    \centering
    \caption{Example PAD models with different front and back ends. This table does not intend to list all the existing PAD models. LA2015, PA2017, LA2019, and PA2019 denote the ASVspoof 2015, ASVspoof 2017, ASVspoof 2019 LA, and ASVspoof 2019 PA databases, respectively. Note that equal error rates (EERs) on different databases are not comparable. The way to compute EER in be explained in Section~\ref{sec:metric}.}
    \resizebox{\textwidth}{!}{
    \begin{tabular}{l l l l l l}
    \toprule
    \multirow{2}{*}{Ref.} & \multirow{2}{*}{Front end} & \multicolumn{2}{l}{Back end} & \multirow{2}{*}{Database} & \multirow{2}{*}{EER (\%)} \\
    \cline{3-4}
    & & Model & Type & & \\
    \midrule
    \cite{zhang2017investigation} & Spec. & CNN & Fixed input size & LA2015 & 3.07 \\
                           & Spec. & RNN & Length-agnostic & LA2015 & 2.46 \\
                           & Spec. & CNN+RNN & Length-agnostic & LA2015 & 1.86\\
    \cite{tian2016spoofing} & GD & CNN & Seg. level & LA2015 & 1.40 \\ 
    \cite{muckenhirn2017end} & Wave & CNN SLP & Seg. level & LA2015 & 3.05 \\
                             & Wave & CNN MLP & Seg. level & LA2015 & 2.89 \\
    \cite{Sailor2017} & Wave + ConvRBM & GMM & - & LA2015 & 3.46\\
    \cite{yu2017dnn} & Spec + DNN(IGFCC) & GMM & - & LA2015 & 0.56 \\ 
    \cite{chen2015robust} & Fbank + DNN & M-distance & Frame level & LA2015 & 2.52 \\
    \cite{gomez2019gated} & STFT,MGD + GRCNN & PLAD & Utt. level & LA2015 & 0.02 \\
    \midrule
    \cite{Tom2018} & GD & ResNet+GAM att. & Fixed input size & PA2017 & 0.00 \\
    \cite{li2017study} & IMFCC & GMM & - &  PA2017 & 7.50 \\
                & LFCC & TDNN & Frame level & PA2017 & 8.16 \\
    \cite{Cai2017} & CQCC & GMM & -  & PA2017 & 19.18 \\ 
     & Spec. & ResNet & Frame level & PA2017 & 23.14 \\ 
    \cite{Lavrentyeva2017} & Spec. + LCNN & GMM &-   & PA2017 & 7.37 \\
                    & Spec. + LCNN-RNN &  GMM & -  & PA2017 & 10.69 \\
    \cite{Chen2017} & CQCC & ResNet & Frame level & PA2017 & 18.79 \\
    \cite{Alluri2017} & SFFCC & BLSTM & Length-agnostic & PA2017 & 17.82 \\
    \cite{lai2019attentive} & Spec. & Dilated ResNet & Fixed input size & PA2017 & 8.54 \\
    \midrule
    \cite{cheng2019replay} & CQT, MGD & ResNeWt & Fixed input size   & PA2019 & 0.39 \\
    \cite{Chettri2019} & Spec. & CNN & Fixed input size & PA2019 & 5.75 \\ 
    \cite{Yang2019} & CQT & CGCNN &  Length-agnostic & PA2019 & - \\
    \cite{Cai2019} & GD & ResNet &  Length-agnostic & PA2019 & 1.08 \\
    \cite{Lai2019} & Spec. & SENet34 & Fixed input size  & PA2019 & 1.29 \\
    \midrule
    \cite{Chettri2019} & Spec. & CNN & Fixed input size & LA2019 & 7.66 \\
    \cite{Yang2019} & CQT + VAE & CGCRNN & Length-agnostic & LA2019 &  - \\
    \cite{Lai2019} & Spec. & SENet34 & Fixed input size  & LA2019 & 11.75 \\
    \cite{Chen2020Odyssey} & LFB + ResNet & MLP & Utt. level & LA2019  & 1.81 \\
    \cite{Zeinali2019} & Spec., CQT & VGG & Seg. level & LA2019 & 10.52 \\
                & Wave  & SincNet + MLP & Seg. level & LA2019 & 20.11 \\
    \cite{tak2020end} & Wave & RawNet2 & Fixed input size & LA2019 & 5.64 \\
    \midrule
    \cite{dinkel2017end} & Wave  & LSTM & Length-agnostic & PA BTAS & - \\
    \bottomrule
    \end{tabular}
    }
    \label{tab:pad_examples}
\end{table}

\subsection{Front End: DSP-based Deterministic Approaches}
\label{sec:pad-frontend-dsp}
Many front ends are based on digital signal processing (DSP) algorithms that extract a sequence of acoustic features $\boldsymbol{a}_{1:N}$ from the input waveform $\boldsymbol{o}_{1:T}$. Accordingly, a front end is a function $\mathcal{F}_{\boldsymbol{\phi}}: \mathbb{R}^{T\times{1}}\rightarrow\mathbb{R}^{N\times{D}}$, where $N$ and $D$ denote the number of frames and feature dimensions in each frame, respectively. The letter $\boldsymbol{\phi}$ denotes the configurations or hyper-parameters of the front end, e.g., fast Fourier transform (FFT) bins, frame shift, and so on.

Many speech processing tasks have used Mel-frequency cepstrum coefficients (MFCC) \cite{davis1980comparison}. Linear frequency cepstrum coefficients (LFCC) \cite{davis1980comparison} and inverse MFCC (IMFCC) \cite{chakroborty2007improved} are closed related to the MFCC, but they use a different frequency scale and emphasize different frequency bands of the input waveform. Another popular front end is the constant Q cepstral coefficient (CQCC) \cite{todisco2016new,Todisco2017} that uses constant Q transform (CQT) rather than short time Fourier transform (STFT). Both CQCC and LFCC have been used in the ASVspoof challenges and are open-sourced. Rather than using cepstrum coefficients, many front ends only extract the spectrogram. Different from the cepstrum coefficients, linear filter bank feature (LFB) compresses the spectrogram without cepstral analysis; a similar approach is also used in extracting Mel-spectrogram.

There are also front ends based on phase information such as the group delay (GD) and modified group delay (MGD). These features are complements to the magnitude-based front ends introduced in the previous paragraph. There is a survey paper explaining the phase-based front ends in more detail \cite{kamble_sailor_patil_li_2020}.

The discussion so far suggests that most DSP-based front ends are based on short-time analysis. Each acoustic feature vector $\boldsymbol{a}_n, \forall{n}\in\{1,\cdots,N\}$ is extracted from a short segment of the input waveform, which is the so-called \emph{frame}. Accordingly, we can write $\boldsymbol{a}_{n} = F_{\boldsymbol{\phi}}(\boldsymbol{o}_{t_n:t_n+L})$, where $\boldsymbol{o}_{t_n:t_n+L}$ is the $n$-th waveform frame that starts at time step $t_n$ and ends at $t_n+L$.  
For example, $F_{\boldsymbol{\phi}}(\boldsymbol{o}_{t_n:t_n+L})$ for MFCC can be written as $F_{\boldsymbol{\phi}}(\boldsymbol{o}_{t_n:t_n+L}) :=  \text{DCT}\big(\log(\boldsymbol{W}_{\text{fb}}\cdot|\text{FFT}(\boldsymbol{o}_{t_n:t_n+L}\odot\boldsymbol{w}_{1:L})|^2)\big)$, where $\boldsymbol{w}_{1:L}$ is an analyzing window (e.g., Hann window) and $\boldsymbol{W}_{\text{fb}}$ is a transformation matrix whose rows are coefficients of the filter banks on the Mel-frequency scale. It is also a common practice to augment each frame $\boldsymbol{a}_n$ with 1st- and 2nd-order differentials (i.e., $\Delta$ and $\Delta^2$ components).

The explanation on MFCC indicates two properties on the DSP-based front end. First, the designer has to manually specify the configuration or hyper-parameter $\boldsymbol{\phi}$: frame length $L$, starting point $t_n$, number of FFT points, and so on. Second, most of the short-time analysis front ends produce $\boldsymbol{a}_n$ using only information from the current frame $\boldsymbol{o}_{t_n:t_n+L}$. The $\Delta$ and $\Delta^2$ components take adjacent frames into consideration, but they cannot reach frames outside of the neighborhood. To extract useful features from a longer waveform segment, the trainable front ends in the next section are more suitable.

\subsection{Front End: DNN-based Supervised Training Approach}
\label{sec:pad-dnn-supervised}
Deep neural networks (DNNs) can be used as a front end $\mathcal{F}_{\boldsymbol{\theta}_1}$. Different from a DSP-based one, the parameter set $\boldsymbol{\theta}_1$ of a DNN-based front end has to be learned from the data.
Such a trainable front end is preferred for three reasons. First, the DNN could be configured to extract acoustic features from a longer waveform segment. Second, the data-driven DNN with non-linear transformations is more powerful than linear operations in DSP-based front ends. Third, a DNN-based front end can easily be integrated with a DSP-based front end.

In most cases, $\mathcal{F}_{\boldsymbol{\theta}_1}$ is a sub-part of a DNN. For example, the speaker representation called X-vector \cite{snyder2018x} is extracted from the third last layer of a feedforward DNN. In this case, $\mathcal{F}_{\boldsymbol{\theta}_1}$ denotes the sub-part before the penultimate layer. The rest of the DNN parameter set is denoted as ${\boldsymbol{\theta}_2}$ in this chapter. 

The full parameter set $\{\boldsymbol{\theta}_1, \boldsymbol{\theta}_2\}$ can be learned from the data in a supervised way. 
Let $y\in\{1\cdots, C\}$ be the label of an input waveform $\boldsymbol{o}_{1:T}$, and the front end can be trained by minimizing the cross-entropy between the target label and the probability $p(k|\mathcal{F}_{\boldsymbol{\theta}_1, \boldsymbol{\theta}_2}(\boldsymbol{o}))$ computed by the DNN given the input waveform. This criterion can be written as
\begin{equation}
\mathcal{L}(\boldsymbol{\theta}_1, \boldsymbol{\theta}_2) = -\sum_{k=1}^{C} \mathbf{1}_{\{y\}}(k)\log P(k|\mathcal{F}_{\boldsymbol{\theta}_1,\boldsymbol{\theta}_2}(\boldsymbol{o})),
\label{eq:ce-frontend}
\end{equation}
where $\mathbf{1}_{\{y\}}(k)$ is an indicator function and $\mathbf{1}_{\{y\}}(k)=1$ when $k=y$ or $\mathbf{1}_{\{y\}}(k)=0$ in other cases. 
The target label $y$ can be binary (i.e., bona fide or spoof). When the spoofing algorithms in the training set are known, the target labels could also be the spoofing algorithm type. In other words, the training target of the DNN-based front end can be different from that of the PAD back end. After training, the sub-part of the DNN can be used to extract the acoustic features. 

How $\mathcal{F}_{\boldsymbol{\theta}_1}$ extracts the acoustic features from the input waveform $\boldsymbol{o}_{1:T}$ is the designer's choice. There are mainly two ways:
\begin{itemize}
    \item $\mathcal{F}_{\boldsymbol{\theta_1}}: \mathbb{R}^{T\times{1}}\rightarrow\mathbb{R}^{N\times{D}}$: a sequence of acoustic features $\boldsymbol{a}_{1:N}\in\mathbb{R}^{N\times{D}}$ is extracted, where $N$ and $D$ denote the number of frames and the dimensions of the vector in each frame, respectively.
    \item $\mathcal{F}_{\boldsymbol{\theta_1}}: \mathbb{R}^{T\times{1}}\rightarrow\mathbb{R}^{1\times{D}}$: a single utterance-level vector $\boldsymbol{a}\in\mathbb{R}^{1\times{D}}$ is extracted.
\end{itemize}  
The two choices require different DNN architectures. We introduce a few examples in the following paragraphs even though not all of them have been used for voice PAD.

\subsubsection{SincNet} SincNet is a convolutional neural network (CNN) with multiple trainable windowed-sinc filters in the input layer \cite{ravanelli2018speaker}. Each sinc filter is a time-invariant band-pass filter with trainable cut-off frequencies. 
The trained input layer extracts different frequency components from the input signal. The trainable sinc filter requires fewer parameters and is more interpretable than conventional convolution layers \cite{ravanelli2018speaker}. The sinc filter does not change the temporal length and dimension of the input waveform, i.e., $\mathbb{R}^{T\times{1}}\rightarrow\mathbb{R}^{T\times{1}}$. 
To derive acoustic features at the frame level, it is possible to define $\mathcal{F}_{\boldsymbol{\theta}_1}=\text{CNN}(\text{Sinc}(\boldsymbol{o}_{1:T}))$, where the CNN can compress the features through local temporal pooling. To derive a single vector, a global temporal pooling layer can be added to pool the features across all frames. \BLUE{The trainable sinc filter has been used in a few so-called end-to-end PAD models, including RawNet2 \cite{tak2020end} and PAD using raw differentiable architecture search \cite{ge2021raw}. }

\subsubsection{LEAF} LEAF is a trainable DNN-based front end with Gabor filters, a Gaussian low-pass pooling layer, and a compression function \cite{zeghidour2021leaf}. The pipeline of LEAF uses the trainable Gabor filters to perform filtering, the low-pass pooling layer to reduce the temporal length, and the compression function to reduce the dynamic range. Thus, LEAF defines $\mathcal{F}_{\boldsymbol{\theta}_1}: \mathbb{R}^{T\times{1}}\rightarrow\mathbb{R}^{N\times{D}}$, similarly to the DSP-based front ends with a short-time analysis. This front end has been testified in various applications such as speaker identification \cite{zeghidour2021leaf}, \BLUE{but only one work investigated LEAF for physical access PAD \cite{tomilov21_asvspoof}}.

\subsubsection{Wav2Spk} Wav2Spk is another discriminative-training-based front end \cite{lin2020wav2spk}. Compared with LEAF, Wave2Spk only uses basic neural network components (e.g., instance normalization and convolution) to convert the waveform into an embedding vector, $\mathcal{F}_{\boldsymbol{\theta}_1}: \mathbb{R}^{T\times{1}}\rightarrow\mathbb{R}^{1\times{D}}$. This new type of front end has demonstrated competitive results in a speaker verification task but not well investigated for PAD.

\subsubsection{DSP Plus Trainable Front Ends} 
While the aforementioned front ends directly take a waveform as input, most of the trainable front ends proposed for voice PAD tasks combine DSP- and DNN-based front ends as $\mathcal{F}_{\boldsymbol{\theta}_1}\circ\mathcal{F}_{\boldsymbol{\phi}}$. For example, in \cite{zhang2017investigation}, the front end is defined as $\mathcal{F}_{\boldsymbol{\theta}_1}\circ\mathcal{F}_{\boldsymbol{\phi}} := \text{DNN}_{\boldsymbol{\theta}_1}(\text{STFT}_{\boldsymbol{\phi}}(\boldsymbol{o}_{1:T}))$. Let $\tilde{\boldsymbol{a}}_{1:N}=\text{STFT}_{\boldsymbol{\phi}}(\boldsymbol{o}_{1:T})$ denote the spectrogram extracted through STFT. Then, a multi-layer DNN is used to generate a sequence of embedding vectors $\boldsymbol{a}_{1:N}$, where each $\boldsymbol{a}_{n}$ is produced given the spectra of the current frame and the adjacent five frames on both sides, i.e., $\boldsymbol{a}_{n}=\mathcal{F}_{\boldsymbol{\theta}_1}(\tilde{\boldsymbol{a}}_{n-5:n+5})$. 

Another advanced front end $\mathcal{F}_{\boldsymbol{\theta}_1}$ is the light CNN (LCNN) \cite{Lavrentyeva2017}. By combining the  LCNN-based front end and a simple Gaussian mixture model (GMM)-based back end, the resulting voice PAD model achieved the top performance in the ASVspoof2017 challenge. A similar approach was proposed in \cite{Chen2020Odyssey}, but the LCNN and STFT were replaced with ResNet and linear filter bank features, respectively. This model is one of top single systems in the ASVspoof2019 LA task.  Another related work focuses on learning trainable filter banks to compress the dimensions of a spectrogram \cite{yu2017dnn}, and the proposed model demonstrated state-of-the-art EER on the ASVspoof 2015 database.

\subsection{Front End: DNN-based Self-supervised Training Approach}
\label{sec:pad-dnn-unsupervised}
It is also possible to apply self-supervised training to the DNN-based front end. One advantage is that a large amount of data can be used even if they do not have target labels. This is a popular topic in speech processing community, and many models have been proposed: wav2vec \cite{Schneider2019a}, wav2vec2 \cite{NEURIPS2020_92d1e1eb}, VQ-wav2vec \cite{baevski2019vq}, contrastive predictive coding \cite{chung2021similarity}, auto-regressive predictive coding \cite{chung2021similarity}, and HuBERT \cite{hsu2021hubert}. Readers are encouraged to check the related references and other papers in major conferences (e.g., ICASSP and Interspeech special sessions, and NISP workshop\footnote{\url{https://icml-sas.gitlab.io/}}). 

\BLUE{For voice PAD, one recent work proposes using wav2vec and metric learning for PAD \cite{xie21_interspeech}. It achieved promising results on ASVspoof 2019 LA database and obtained an EER of 1.15\%. Another latest work focused on wav2vec 2.0 and investigated different pre-trained wav2vec 2.0 models, back end architectures, and training strategies to plug wav2vec 2.0 into PAD models \cite{wang2021investigating}. It was found that a good pre-trained wav2vec 2.0 model, either being used off-the-shelf or fine-tuned on the ASVspoof 2019 LA training set, outperformed a strong LFCC-based baseline on multiple test sets.  Since the pre-trained self-supervised models reply on external data, we cannot directly compare the results in that work with those reported in this chapter. However, using a self-supervised front end may be a potential solution to build better generalized PAD models. }

Before the recent boom of self-supervised models, there is a pioneering work using conditional Restricted Bolzman Machine (ConvRBM) as a trainable front end \cite{Sailor2017}. It showed competitive performance on the ASVspoof 2015 database.

\subsection{Back End: GMM Framework}
\label{sec:pad-gmm}
With the features from the front end, the back end conducts the classification.
The first type of back end is inspired by the GMM-based ASV \cite{reynolds2000speaker}. It assumes a null hypothesis $\mathcal{H}_0$ and an alternative hypothesis $\mathcal{H}_1$.
\begin{itemize}
    \item $\mathcal{H}_0$: input waveform $\boldsymbol{o}_{1:T}$ is bona fide;
    \item $\mathcal{H}_1$: input waveform $\boldsymbol{o}_{1:T}$ is spoofed.  
\end{itemize}
The training is conducted through maximizing the likelihood of two GMMs over the bona fide and spoofed data, respectively. Let $\boldsymbol{\lambda}^{(y=1)}$ be the parameter set of the GMM for bona fide data. $\boldsymbol{\lambda}^{(y=1)}$ is estimated by maximizing $p(\mathcal{F}_{\boldsymbol{\theta}_1}(\boldsymbol{o}_{1:T})|\boldsymbol{\lambda}^{(y=1)})$ over all the bona fide data $\boldsymbol{o}_{1:T}$ in the training set. $\boldsymbol{\lambda}^{(y=0)}$ can be estimated in a similar way using the spoofed data. The maximum-likelihood estimation of the GMM parameter set can be conducted using the expectation-maximization algorithm \cite{dempster1977maximum}. The front end $\mathcal{F}$ can be either a DSP or DNN-based one. We simply use $\mathcal{F}_{\boldsymbol{\theta}_1}$ for explanation in this section.

Given the trained GMMs and an input trial $\boldsymbol{o}$,
a log likelihood ratio (LLR) $r=\log{\frac{p(\mathcal{F}_{\boldsymbol{\theta}_1}(\boldsymbol{o}) | \boldsymbol{\lambda}^{(y=1)})}{p(\mathcal{F}_{\boldsymbol{\theta}_1}(\boldsymbol{o}) | \boldsymbol{\lambda}^{(y=0)})}}$ is computed and compared with a threshold $\tau$: 
\begin{align}
    \begin{cases}
    r \geq \tau, & \text{accept }\mathcal{H}_0, \\ 
    r < \tau, &  \text{reject }\mathcal{H}_0
    \end{cases}
\end{align}
The threshold $\tau$ is determined with task-specific prior knowledge. When comparing the model performance on benchmark data sets, we can also sweep the threshold (i.e., changing the operation point) or determine one that demonstrates the discriminating power of the model. The evaluation metric will be covered in more detail in Section~\ref{sec:metric}.

As mentioned in previous sections, the front end $\mathcal{F}_{\boldsymbol{\theta}_1}(\boldsymbol{o}_{1:T})$ could derive a single utterance-level vector or a sequence of features vectors. In the former case where $\boldsymbol{a}=\mathcal{F}_{\boldsymbol{\theta}_1}(\boldsymbol{o}_{1:T})$, the GMM can be directly applied to model $p(\boldsymbol{a}|\boldsymbol{\lambda}^{(y)})$. In the latter case where $\boldsymbol{a}_{1:N}=\mathcal{F}_{\boldsymbol{\theta}_1}(\boldsymbol{o}_{1:T})$, a practical choice is to decompose $p(\mathcal{F}_{\boldsymbol{\theta}_1}(\boldsymbol{o}_{1:T})|\boldsymbol{\lambda}^{(y)})$ as 
\begin{align}
 p(\mathcal{F}_{\boldsymbol{\theta}_1}(\boldsymbol{o}_{1:T})|\boldsymbol{\lambda}^{(y)}) = \prod_{n=1}^{N} p(\boldsymbol{a}_{n}|\boldsymbol{\lambda}^{(y)}) = \prod_{n=1}^{N} \sum_{m=1}^{M} \pi_m^{(y)} \mathcal{N}(\boldsymbol{a}_{n};\boldsymbol{\mu}_m^{(y)}, \boldsymbol{\Sigma}_m^{(y)}), 
 \label{eq:gmm}
\end{align}
where $\boldsymbol{\lambda}^{(y)} = \{\pi_m^{(y)}, \boldsymbol{\mu}_m^{(y)}, \boldsymbol{\Sigma}_m^{(y)}\}_{m=1}^{M}$ is the GMM parameter set for class $y$. As Eq.~(\ref{eq:gmm}) demonstrates, the GMM assumes statistical independence among $\{\boldsymbol{a}_{1}, \cdots, \boldsymbol{a}_{N}\}$. While this assumption makes it straightforward to compute scores for input waveforms with varied lengths, it ignores the temporal dependency that could be useful to the PAD task. Other back ends could provide potentially better solutions. Nevertheless, the GMM-based back end has been used in many voice PAD models, including the ASVspoof challenge baselines \cite{yamagishi21_asvspoof,Todisco2019,Kinnunen2017,wu2015asvspoof}.

\subsection{Back End: Distance-based Approach}
The distance-based scoring framework is closely related to the GMM back end.
In general, let us consider $C$ target classes, where $C=2$ if the target labels are $\{\text{bona fide}, \text{spoof}\}$. 
Suppose the GMM is reduced to a single Gaussian distribution, which has a parameter set $\boldsymbol{\lambda}^{(c)} = \{\boldsymbol{\mu}^{(c)}, \boldsymbol{\Sigma}^{(c)}\}, \forall{c}\in\{1, \cdots, C\}$. If the front end has computed an utterance-level embedding  $\boldsymbol{a} = \mathcal{F}_{\boldsymbol{\theta}_1}(\boldsymbol{o}_{1:T})$, 
we can easily compute the log likelihood of the $c$-th Gaussian on the input trial as
\begin{align}
\log \mathcal{N}(\mathcal{F}_{\boldsymbol{\theta}_1}(\boldsymbol{o}_{1:T});\boldsymbol{\mu}^{(c)}, \boldsymbol{\Sigma}^{(c)}) = -\frac{1}{2}(\boldsymbol{a} - \boldsymbol{\mu}^{(c)})^\top{\boldsymbol{\Sigma}^{(c)}}^{-1}(\boldsymbol{a} - \boldsymbol{\mu}^{(c)}) + \text{const}
 \label{eq:m-dis-1}
\end{align}
The first term on the right hand side is the negative Mahalanobis distance (M-distance). Since the constant term is irrelevant to the input trial, we compute the score of the input trial on the $c$-th class as
\begin{equation}
    s_c(\boldsymbol{o}_{1:T})=-\frac{1}{2}(\boldsymbol{a} - \boldsymbol{\mu}^{(c)})^\top{\boldsymbol{\Sigma}^{(c)}}^{-1}(\boldsymbol{a} - \boldsymbol{\mu}^{(c)})
\end{equation}
The score used in the decision logic can be set as $r = s_{\text{bona fide}}$. $s_c$ can also be used to compute other metrics such as out-of-distribution distances \cite{Lee2018mdistance}.

The M-distance originates from a hypothesis test called the Hotelling test \cite[ch 6.5.4.3.]{heckert2002handbook} and has been widely used in outlier detection \cite{bulusu2020anomalous}. It is potentially useful for voice PAD because unknown spoofing attacks are likely to be outliers in the training data's distributions. 
The M-distance has been applied to voice PAD \cite{chen2015robust} and achieved the 3rd best performance in the ASVspoof 2015 challenge. However, the M-distance assumes a multivariate Gaussian distribution for the embedding vector $\boldsymbol{a}$, which could be inconsistent with the actual data distribution. Furthermore, estimation of the Gaussian covariance matrix $\boldsymbol{\Sigma}$ is sensitive to outliers. More robust estimators such as the minimum
covariance determinant (MCD) \cite{hubert2010minimum,rousseeuw1999fast} can be used.

\subsection{Back End: DNNs with Discriminative Training}
\label{sec:backend_discriminative}
Many advanced PAD models use DNN-based back ends. A major sub-category is a DNN with a discriminative training strategy. 
Given the target label $y$ of input waveform $\boldsymbol{o}_{1:T}$, the DNN is directly trained to maximize the probability $P(y|\boldsymbol{o}_{1:T}; \boldsymbol{\theta}_3)$. We use $\boldsymbol{\theta}_3$ to denote the parameter set of the DNN-based back end so that it can be differentiated from ($\boldsymbol{\theta}_1, \boldsymbol{\theta}_2$) of the DNN-based front end. 

The topology of the DNN is a designer's choice\footnote{It is also possible to conduct a neural architecture search without manually specifying the network structure. This has been investigated on voice PAD \cite{ge2021raw}.}, but it also depends on the dimensions of the front end's output. Without loss of generality, we assume that there are $C$ unique target labels. We then use $\boldsymbol{p}=[p_1, \cdots, p_k, \cdots, p_C]^\top\in\mathbb{R}^{C}$ to denote the probability vector, where the $k$-th dimension is the probability of being class $k$, i.e., ${p}_k=p(k|\boldsymbol{o}_{1:T}, \boldsymbol{\theta}_3), \forall{k}\in\{1, \cdots, C\}$. We further use $\mathcal{H}_{\boldsymbol{\theta}_3}$ to denote the non-linear transformation by the DNN in the back end.

\begin{figure}[t]
\centering
\includegraphics[trim=00 160 00 80, clip, width=0.95\textwidth]{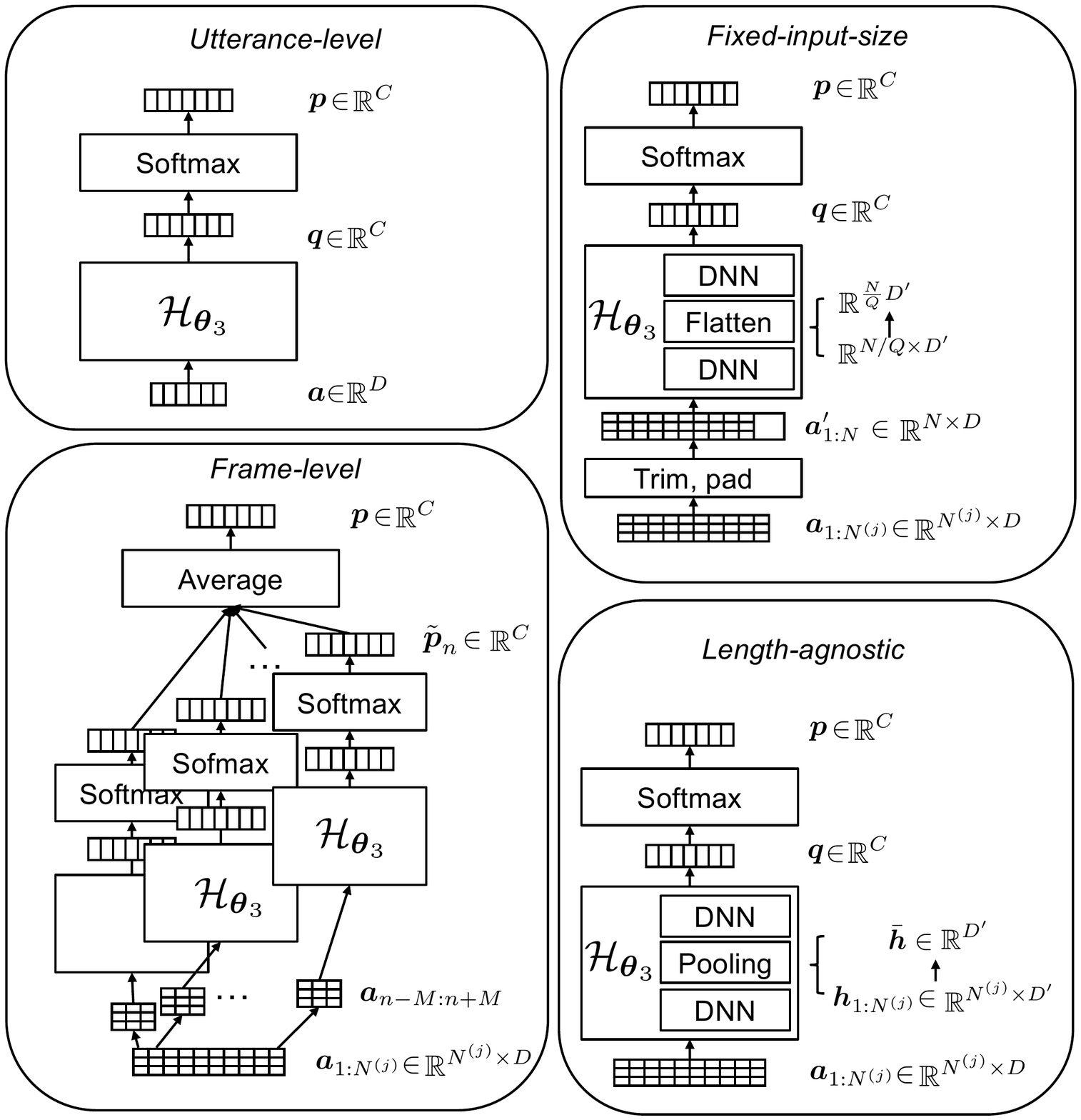}
\caption{Illustration of different types of back ends. Segmental-level back end is similar to the frame-level one and is excluded from the figure.}
\label{fig:back_ends}
\end{figure}

We now group the DNN-based back ends on the basis of how they map the front end's output to the probability vector $\boldsymbol{p}$. 

\subsubsection{Utterance-level Back End:} $\boldsymbol{a}\in\mathbb{R}^{D}\overset{\text{DNN}}{\rightarrow}\boldsymbol{p}\in\mathbb{R}^{C}$

\noindent Given an utterance-level feature $\boldsymbol{a}=\mathcal{F}_{\boldsymbol{\theta}_1}(\boldsymbol{o}_{1:T})\in\mathbb{R}^{D}$ from a trained front end $\mathcal{F}_{\boldsymbol{\theta}_1}$, the DNN $\mathcal{H}_{\boldsymbol{\theta}_3}$ in the back end first converts $\boldsymbol{a}$ into an activation vector $\boldsymbol{q} = \mathcal{H}_{\boldsymbol{\theta}_3}(\boldsymbol{a})\in\mathbb{R}^{C}$. After that, the vector $\boldsymbol{q}$ is converted into  $\boldsymbol{p}=\text{Softmax}(\mathcal{H}_{\boldsymbol{\theta}_3}(\boldsymbol{a}))$ through a softmax function \cite{bridle1990training}. The $k$-th dimension of $\boldsymbol{p}$ is computed as
\begin{equation}
    p_k = \frac{\exp(q_k)}{\sum_{c=1}^{C}\exp(q_c)} =  \frac{\exp(\mathcal{H}_{\boldsymbol{\theta}_3}(\boldsymbol{a})_k)}{\sum_{c=1}^{C}\exp(\mathcal{H}_{\boldsymbol{\theta}_3}(\boldsymbol{a})_c)},
\end{equation}
where $q_k := \mathcal{H}_{\boldsymbol{\theta}_3}(\boldsymbol{a})_k$ denotes the $k$-th dimension of $\boldsymbol{q}$.
Given $p_k$, the cross-entropy-based training loss over the input waveform can be computed as 
\begin{equation}
\mathcal{L}(\boldsymbol{\theta}_3) = -\sum_{k=1}^{C}\mathbf{1}_{\{y\}}(k)\log p_k.
\label{eq:ce-backend-1}
\end{equation}
The equation above explain the vanilla cross-entropy criterion. In Section~\ref{sec:training-criterion}, we will introduce more advanced softmax functions and training criteria.

During scoring, we can use ${p}_{y}$ or $q_y$ as the score of the input waveform, where $y$ is the index of the bona fide label. We can stretch or transform the score in various ways, but the score must be a scalar number in $\mathbb{R}$, and a higher score should indicate a higher probability of being bona fide. 

Many types of DNNs can be used as the utterance-level back end since the task is simply $\mathbb{R}^{D}\rightarrow\mathbb{R}^{C}$. In \cite{Chen2020Odyssey}, a network with a single fully-connected layer is used to convert the utterance-level embedding into the probability vector $\boldsymbol{p}$, and the model achieved good performance on the ASVspoof 2019 LA database (see Tabs. \ref{tab:pad_examples} and \ref{tab:results_all}).


\subsubsection{Frame-level Back End:} $\mathbb{R}^{N^{(j)}\times{D}}\overset{\text{DNN}}{\rightarrow}\mathbb{R}^{N^{(j)}\times{C}}\overset{\text{mean}}{\dashrightarrow}\mathbb{R}^{{C}}$

\noindent When the output of the front end for the $j$-th waveform is a sequence of acoustic features $\boldsymbol{a}^{(j)}_{1:N^{(j)}}=(\boldsymbol{a}^{(j)}_{1}, \cdots, \boldsymbol{a}^{(j)}_{N^{(j)}})\in\mathbb{R}^{N^{(j)}\times{D}}$, the back end has to convert the sequence $\boldsymbol{a}^{(j)}_{1:N^{(j)}}$ into the probability vector $\boldsymbol{p}^{(j)}\in\mathbb{R}^{C}$. We add a superscript $^{(j)}$ to emphasize the fact that the sequence length $N^{(j)}$ varies for different trials. 

The conversion process can be conducted using a frame-level back end with two parts. The first part is a DNN that converts $\boldsymbol{a}^{(j)}_{1:N^{(j)}}$ into a sequence of probability vectors $\tilde{\boldsymbol{p}}^{(j)}_{1:N^{(j)}}=(\tilde{\boldsymbol{p}}^{(j)}_1, \cdots, \tilde{\boldsymbol{p}}^{(j)}_{N^{(j)}})\in\mathbb{R}^{N^{(j)}\times{C}}$. The second part is simple averaging that produces $\boldsymbol{p}^{(j)} = \frac{1}{N^{(j)}}\sum_{n=1}^{N^{(j)}}\tilde{\boldsymbol{p}}^{(j)}_{n}$.  

Given the utterance-level score $\boldsymbol{p}^{(j)}$, we can use the training criterion in Eq.~(\ref{eq:ce-backend-1}). It is also possible to compute cross-entropy at the frame level. Given a single training trial $j$, this training criterion is written as
\begin{equation}
\mathcal{L}(\boldsymbol{\theta}_3) = -\sum_{n=1}^{N^{(j)}}\sum_{k=1}^{C}\mathbf{1}_{\{y^{(j)}_n\}}(k)\log \tilde{p}^{(j)}_{n,k},
\label{eq:ce-backend-2}
\end{equation}
where $\tilde{p}^{(j)}_{n,k}$ is the $k$-th dimension of $\tilde{\boldsymbol{p}}^{(j)}_{n}$ and denotes the probability of being class $k$ at the $n$-th frame.

A DNN $\mathcal{H}_{\boldsymbol{\theta}_3}$ can be used in the frame-level back end as long as it supports the conversion $\mathbb{R}^{N^{(j)}\times{D}}\overset{\text{DNN}}{\rightarrow}\mathbb{R}^{N^{(j)}\times{C}}$. The straightforward way is to independently conduct $\mathbb{R}^{{D}}\overset{\text{DNN}}{\rightarrow}\mathbb{R}^{{C}}$ for each frame $n\in\{1, \cdots, N^{(j)}\}$. For example, a DNN with linear transformations and non-linear activation functions can be used to do the following conversion for each $n$:
\begin{align}
\tilde{\boldsymbol{p}}_{n}^{(j)} &= \text{Softmax}(\mathcal{H}_{\boldsymbol{\theta}_3}(\boldsymbol{a}_n^{(j)})) \\
&=\text{Softmax}(\boldsymbol{W}_1\tanh(\boldsymbol{W}_2\tanh(\cdots\tanh(\boldsymbol{W}_L\boldsymbol{a}_n^{(j)})))).
\end{align}
Obviously, the DNN only uses $\boldsymbol{a}_n^{(j)}$ to compute $\tilde{\boldsymbol{p}}_n^{(j)}$, ignoring the possible dependency of $\tilde{\boldsymbol{p}}_n^{(j)}$ on the acoustic features from other frames. 

Therefore, when predicting $\tilde{\boldsymbol{p}}_n^{(j)}$, it is more practical to feed in not only $\boldsymbol{a}_n^{(j)}$ but also adjacent frames. 
The mapping can be written as $\mathbb{R}^{(2M+1)\times{D}}\overset{\text{DNN}}{\rightarrow}\mathbb{R}^{{C}}$, where $M$ is a fixed number of frames preceding (or following) the current frame step. This can be done by simply concatenating the input frames:
\begin{equation}
\tilde{\boldsymbol{p}}_{n}^{(j)} = \text{Softmax}(\mathcal{H}_{\boldsymbol{\theta}_3}([\boldsymbol{a}_{n-{M}}^{(j)},\cdots, \boldsymbol{a}_n^{(j)}, \cdots, \boldsymbol{a}_{n+M}^{(j)}]^\top)) 
\end{equation}
For example, a PAD model in \cite{zhang2017investigation} concatenates 11 frames $\boldsymbol{a}_{n-5:n+5}$ as the input to a multilayer perceptron. It is also possible to use a CNN or time-delay DNN (TDNN) \cite{waibel1989phoneme} to process the $2M+1$ input frames.

\subsubsection{Segment-level Back End:} $\mathbb{R}^{N^{(j)}\times{D}}\overset{\text{DNN}}{\rightarrow}\mathbb{R}^{N^{(j)/Q}\times{C}}\overset{\text{mean}}{\dashrightarrow}\mathbb{R}^{{C}}$

\noindent The segment-level back end is closely related to the frame-level back end. The difference is that the DNN can reduce the temporal resolution of the output sequence by a factor $Q$. The mapping can be written as $\mathbb{R}^{N^{(j)}\times{D}}\overset{\text{DNN}}{\rightarrow}\mathbb{R}^{N^{(j)/Q}\times{C}}$. The reduction of temporal resolution, or down-sampling, can be easily implemented using 1D convolution with a larger stride or local pooling operation. Furthermore, the down-sampling from $N^{(j)}$ to $N^{(j)}/Q$ can be done in multiple stages, e.g.,  $N^{(j)} \rightarrow N^{(j)}/2 \rightarrow N^{(j)}/4  \rightarrow \cdots \rightarrow N^{(j)}/Q$.

A potential advantage of the segment-level back end is that it `extracts useful information from a long temporal span without concatenating a large number of neighboring frames' \cite{tian2016spoofing}. As listed in Table \ref{tab:pad_examples}, many PAD models use segment-level back ends. In \cite{tian2016spoofing}, a 1D CNN with a max-pooling-based down-sampling layer was used, and the $Q$ was set to 20. It showed competitive performance in detecting the most difficult spoofing attack in the ASVspoof 2015 challenge.

\subsubsection{Fixed-input-size Back End:} $\mathbb{R}^{N\times{D}}\overset{\text{DNN}}{\rightarrow}\mathbb{R}^{\frac{N}{Q}{D}'}\overset{\text{DNN}}{\rightarrow}\mathbb{R}^{C}$.

\noindent Advanced DNNs such as ResNet \cite{he2016deep} and LCNN \cite{wu2018light} were originally proposed for image processing tasks. They are designed to process images with a fixed height and width. When ResNet or LCNN is used in a frame-level (or segment-level) PAD back end, its input data at each frame step has a fixed size $\mathbb{R}^{(2M+1)\times{D}}$ and can be directly processed as an image. However, an additional step is required to average the DNN's output $\tilde{\boldsymbol{p}}_n$ at every frame step. 

A disadvantage of the frame- or segment-level back end is that the output $\tilde{\boldsymbol{p}}_n$ only depends on a fixed number of input frames. It is presumably better if the input $\boldsymbol{a}_{1:N^{(j)}}\in\mathbb{R}^{N^{(j)}\times{D}}$ can be directly converted into a single probability vector $\boldsymbol{p}\in\mathbb{R}^{C}$ for the whole input trial. 

Unfortunately, most image-oriented DNNs cannot conduct $\boldsymbol{a}_{1:N^{(j)}}\in\mathbb{R}^{N^{(j)}\times{D}}\rightarrow\boldsymbol{p}\in\mathbb{R}^{C}$ because the length of $\boldsymbol{a}_{1:N^{(j)}}$ is not fixed.
One straightforward method is to use a `unified representation' that trims or pads $\boldsymbol{a}_{1:N^{(j)}}$ into a fixed-size tensor $\boldsymbol{a}'_{1:N}\in\mathbb{R}^{N\times{D}}$. This fixed-size tensor can be directly processed by ResNet or LCNN as a single-channel image.  

In many voice PAD models, the fixed-input-size back end includes three parts: a 2D CNN that convert the input `image' into a smaller feature map, i.e., $\mathbb{R}^{N\times{D}}\overset{\text{DNN}}{\rightarrow}\mathbb{R}^{\frac{N}{Q}\times{D}'}$, a flattening layer that flattens the feature map into a vector $\mathbb{R}^{\frac{N}{Q}\times{D}'} \rightarrow \mathbb{R}^{\frac{N}{Q}{D}'}$, and a feedforward part that converts the flattened vector into the probability vector $\boldsymbol{p}$, i.e., $\mathbb{R}^{\frac{N}{Q}{D}'} \rightarrow \mathbb{R}^{C}$. 
The factor $Q$ denotes the down-sampling factor. 
The back end can be trained using the cross-entropy criterion similar to Eq.~(\ref{eq:ce-backend-1}). During scoring, the input trial can be trimmed or padded again, after which a single score can be computed by the back end.

The fixed-input-size back end has been used in many advanced voice PAD models, for example, the top single system in the ASVspoof 2019 LA and PA tasks \cite{lavrentyeva2019stc,cheng2019replay}. The RawNet2 baseline in ASVspoof2021  \cite{tak2020end} also uses the same strategy and sets the input waveform length to $64,000$. 

The above trim-and-pad strategy either discards useful information or adds redundancy to $\boldsymbol{a}_{1:N^{(j)}}$ \cite{chettri2020dataset}. When $N^{(j)}<N$, many models simply take the first $N$ frames by setting $\boldsymbol{a}'_{1:N} = \boldsymbol{a}_{1:N}$ \cite{lavrentyeva2019stc}. Others randomly take a chunk by  $\boldsymbol{a}'_{1:N} = \boldsymbol{a}_{k:k+N}$, where $k$ is a random integer \cite{zhang2020one}. In both cases, only part of the information in $\boldsymbol{a}_{1:N^{(j)}}$ is consumed by the back end. When $N^{(j)}>N$, zero-padding or duplication can be used, but it causes artifacts \cite{chettri2020dataset}.

\subsubsection{Length-agnostic Back End:}
$\mathbb{R}^{N^{(j)}\times{D}}\overset{\text{DNN}}{\rightarrow}\mathbb{R}^{N^{(j)}\times{D}'}\overset{\text{pooling}}{\rightarrow}\mathbb{R}^{{D}'}\overset{\text{DNN}}{\rightarrow}\mathbb{R}^{C}$

\noindent A more flexible way to handle varied-length input is to augment the DNN with an operation that can reduce the varied-length hidden feature sequence into a single vector. 
Suppose the input $\boldsymbol{a}_{1:N^{(j)}}$ has passed through convolution layers and been converted to a hidden feature sequence $\boldsymbol{h}_{1:N^{(j)}}\in\mathbb{R}^{N^{(j)}\times{D}'}$. Global average pooling can be used to average $\boldsymbol{h}_{1:N^{(j)}}$ as  $\bar{\boldsymbol{h}}=\frac{1}{N^{(j)}}\sum_{n=1}^{N^{(j)}}\boldsymbol{h}_{n}$. After that, $\bar{\boldsymbol{h}}$ can be converted into the probability vector $\boldsymbol{p}\in\mathbb{R}^{C}$ through additional neural layers. It is also possible to reduce the temporal resolution of $\boldsymbol{h}_{1:N^{(j)}}$.

Note that the average pooling on $\boldsymbol{h}_{1:N^{(j)}}$ is different from that in the frame-level back end. First, the pooling operation here is on hidden features rather than the probability vectors $\tilde{\boldsymbol{p}}_n$ of each frame. Furthermore, the cross-entropy-based training criterion is defined on the whole utterance similar to Eq.~(\ref{eq:ce-backend-1}), and it is different from the frame-level cross-entropy defined in Eq.~(\ref{eq:ce-backend-2}). This difference is similar to that between the D-vector \cite{variani2014deep} and X-vector \cite{snyder2018x}. Several voice PAD models have used the pooling strategy to build a length-agnostic back end \cite{Cai2019,Lai2019,Yang2019}. 

The average pooling can be extended to attention-based pooling. Let us re-write the average pooling as $\bar{\boldsymbol{h}}=\sum_{n=1}^{N^{(j)}}w_n\boldsymbol{h}_{n}$, where $w_n = \frac{1}{N^{(j)}}, \forall{n}\in\{1,\cdots,N^{(j)}\}$. Accordingly, we can replace the fixed $w_n$ with a weight computed from an attention mechanism. For example, we can use $\boldsymbol{w} = \text{Softmax}(\boldsymbol{H}\boldsymbol{w}')$, where $\boldsymbol{w}=[w_1, w_2, \cdots, w_{N^{(j)}}]^{\top}\in\mathbb{R}^{N^{(j)}}, \boldsymbol{H}=[\boldsymbol{h}_1, \cdots, \boldsymbol{h}_{N^{(j)}}]^{\top}\in\mathbb{R}^{N^{(j)}\times{D'}}$, and where $\boldsymbol{w}'\in\mathbb{R}^{D'}$ is the trainable parameter of the attention mechanism. 
It has been shown that attention-based pooling can achieve reasonably good performance on a benchmark dataset \cite{wang2021comparative}. 
There are also studies on PAD using attention mechanism for feature selection \cite{lai2019attentive,Tom2018,ma2021improved,ling21_interspeech}.

Rather than pooling, we can also use the output vector at the last frame step if the hidden feature sequence $\boldsymbol{h}_{1:N^{(j)}}$ is produced from a uni-directional recurrent layer. In other words, we set $\bar{\boldsymbol{h}}=\boldsymbol{h}_{N^{(j)}}$. If the recurrent layer is bi-directional, we can concatenate the hidden vectors at the two ends by $\bar{\boldsymbol{h}}=[\boldsymbol{h}_{1}; \boldsymbol{h}_{N^{(j)}}]^\top$. This idea is similar to the LSTM-based D-vector \cite{rahman2018attention} for speaker verification and has been used in some PAD back ends \cite{Alluri2017,zhang2017investigation}.

\subsection{End-to-end Models}
\label{sec:end-to-end}
We can treat end-to-end models as a combination of a neural back end and a dummy front end. From another viewpoint, we can also split an end-to-end model into front and back ends. 
The term `end-to-end' is based on the fact that the front and back ends are jointly trained. Ideally, the model should conduct $\boldsymbol{o}_{1:T}\in\mathbb{R}^{T\times{1}}\rightarrow\boldsymbol{p}\in\mathbb{R}^{C}$. 

The raw waveform CLDNN is one end-to-end PAD model \cite{dinkel2017end}. The front end part uses convolutional and LSTM layers for $\boldsymbol{o}_{1:T}\in\mathbb{R}^{T\times{1}}\rightarrow\boldsymbol{h}\in\mathbb{R}^{D}$. Given the utterance-level representation $\boldsymbol{h}$, a DNN is used to produce the probability vector $\boldsymbol{p}$. A recently proposed model called RawNet2 followed a similar approach but combined SincNet with a CNN-RNN block. 
Although RawNet2 is configured to process a fixed length of input waveform, it is theoretically length-agnostic.
RawNet2 has demonstrated competitive performance on the ASVspoof 2019 database \cite{tak2020end}.

\subsection{Training Criterion for DNN}
\label{sec:training-criterion}
The previous sections briefly mentioned the discriminative training criterion based on the cross entropy. This section explains the training criterion in more detail and introduces variants and alternatives. Without loss of generality, we follow the description on the utterance-level back end in Section~\ref{sec:backend_discriminative} and expand it by describing the training loss over a set of input trials instead of a single trial.

\subsubsection{Cross-entropy with Vanilla Softmax}
\label{sec:loss_softmax}
\textcolor{white}{:}

\noindent Let $\mathcal{D}=\{\cdots, \{\boldsymbol{o}_{1:T_j}^{(j)}, y_j \}, \cdots\}$ be a training data set with $|\mathcal{D}|$ pairs of a waveform $\boldsymbol{o}_{1:T_j}^{(j)}$ and a label $y_j\in\{1, 2, \cdots, C\}$. Given the front end $\mathcal{F}_{\boldsymbol{\theta}_1}$, the $j$-th trial is converted by the front end into $\boldsymbol{a}^{(j)} = \mathcal{F}_{\boldsymbol{\theta}_1}(\boldsymbol{o}_{1:T_j}^{(j)})$. 
The DNN in the back end then computes a hidden feature vector $\boldsymbol{h}^{(j)} = \text{DNN}_{\boldsymbol{\theta}_3}(\boldsymbol{a}^{(j)})\in\mathbb{R}^{h}$, transforms it with a linear transformation as $\boldsymbol{q}^{(j)} = \boldsymbol{W}\boldsymbol{h}^{(j)}\in\mathbb{R}^{C}$, and passes $\boldsymbol{q}^{(j)}$ through a softmax layer. The output is a probability vector $\boldsymbol{p}^{(j)} = [{p}^{(j)}_1, {p}^{(j)}_2, \cdots, {p}^{(j)}_C]^\top$, where ${p}^{(j)}_k$ is the probability of the $j$-th trial being class $k$:
\begin{equation}
    {p}^{(j)}_k = \frac{\exp({q}^{(j)}_k)}{\sum_{i=1}^{C}\exp({q}^{(j)}_{i})} = \frac{\exp(\boldsymbol{w}_{k}^{\top}\boldsymbol{h}^{(j)})}{\sum_{i=1}^{C}\exp(\boldsymbol{w}_{i}^{\top}\boldsymbol{h}^{(j)})},
    \label{}
\end{equation}
and where $\boldsymbol{w}_k$ is the $w$-th row of $\boldsymbol{W}$. The cross-entropy over the dataset $\mathcal{D}$ can be written as
\begin{align}
\mathcal{L}^{(\text{ce})}(\boldsymbol{\theta}_3) &= -\frac{1}{|\mathcal{D}|}\sum_{j=1}^{|\mathcal{D}|}\sum_{k=1}^{C} \mathbf{1}_{\{y\}}(k) \log p_{k}^{(j)}.
\end{align}
Note that, different from the description in Section~\ref{sec:backend_discriminative}, we separate the linear transformation $\boldsymbol{W}\boldsymbol{h}^{(j)}\in\mathbb{R}^{C}$ from the rest of the DNN. While it is unnecessary for the vanilla softmax, the linear transformation is fundamental for more advanced training criteria explained in the next subsections.

\subsubsection{Cross-entropy with Margin-based Softmax}
\label{sec:loss_margin_softmax}
\textcolor{white}{:}

\noindent Cross-entropy with vanilla softmax is widely used in the discriminative training of DNNs. Although the network learns to discriminate the data of different classes, the training criterion does not separate them further. Consider a special case where $C=2$, the decision boundary between the two classes, i.e., the hyper-plane where $p_1^{(j)}=p_2^{(j)}$, is $(\boldsymbol{w}_1 - \boldsymbol{w}_2)^\top\boldsymbol{q}^{(j)}=0$. As long as the training data of classes 1 and 2 stay on different sides of the decision boundary, the network is judged to be well trained.

The advanced training criterion not only drives the data to stay on the correct side of the decision boundary but also encourages the data to be far away from it. In other words, it introduces extra margins between the data and the decision boundary. A general form for the margin-based softmax \cite{deng2019arcface} can be written as:
\begin{equation} 
p_{k}^{(j)} = \frac{\exp\big({\alpha[\cos (m_1 \theta_{j, k} + m_2) - m_3]}\big)}{ \exp\big({\alpha [\cos (m_1 \theta_{j, k} + m_2) - m_3]}\big) + \sum_{i=1, i\neq k}^{C} \exp\big({\alpha\cos\theta_{j, i}}\big)}, 
\label{eq:margin_softmax}
\end{equation}
where $\theta_{j, k}$ is the angle between the length-normalized weight vector $\widehat{\boldsymbol{w}}_k = \boldsymbol{w}_k/||\boldsymbol{w}_k||$ and hidden feature vector $\widehat{\boldsymbol{h}}^{(j)} = \boldsymbol{h}^{(j)}/||\boldsymbol{h}^{(j)}||$:
\begin{equation}
    \theta_{j, k} = \arccos(\widehat{\boldsymbol{w}}_k^\top\widehat{\boldsymbol{h}}^{(j)}),
\label{eq:margin_softmax_cos}    
\end{equation}

The hyper-parameter set $(\alpha, m_1, m_2, m_3)$ is the key to defining margins.
When $(\alpha=1, m_1=1=1, m_2=0, m_3=0)$, the general form is reduced to the vanilla softmax except that the vector product $\boldsymbol{w}_k^\top\boldsymbol{h}^{(j)}$ is replaced with $\widehat{\boldsymbol{w}}_k^\top\widehat{\boldsymbol{h}}^{(j)}$. Thus, no margin is introduced. Let us consider $C=2$, the decision boundary of the vanilla softmax is $\theta_{j,1}=\theta_{j,2}$. Now let us consider $(\alpha=1, m_1=1, m_2>0, m_3=0)$. When $y_j=1$, the data is correctly classified only if $p_{1}^{(j)} > 0.5$. With Eq.~(\ref{eq:margin_softmax}), this condition can be re-written as $\cos(\theta_{j,1} + m_2) > \cos(\theta_{j,2})$. In other words, the decision boundary is $\theta_{j,1} + m_2 = \theta_{j,2}$, and the $j$-th trial can be classified as class 1 only if the angle $\theta_{j,1}$ is smaller than $\theta_{j,2} - m_2$. The $m_2$ is the extra margin. Similarly, the decision boundary for data of class 2 becomes $\theta_{j,2} + m_2 = \theta_{j,1}$. This margin-based softmax therefore encourages the training data of different classes to be more separable.

For voice PAD, the margin-softmax \cite{liu2017sphereface} defined by $(m_1>1, m_2=0, m_3=0)$ has been used in an LCNN-based model \cite{lavrentyeva2019stc}, while an additive-margin (AM) softmax \cite{wang2018cosface,wang2018additive} defined by $(m_1=1, m_2=0, m_3>0)$ has been used in other works \cite{zhang2020one,Chen2020Odyssey}. 
The AM-softmax could be further extended, for example, by defining a $m_{3,k}$ for each class $k$. This so-called one-class (OC) softmax has demonstrated good performance on the ASVspoof2019 LA task \cite{zhang2020one}.

\subsubsection{Regression-based Training Criterion}
\label{seq:loss_func_p2s}
\textcolor{white}{:}

\noindent The margin-based softmax is sensitive to the hyper-parameter setting \cite{zhang2019p2sgrad}. An alternative is a mean square loss (MSE) between $\cos\theta_{j, k}$ and a scalar target value, which is introduced in \cite{wang2021comparative}:
\begin{equation}
\mathcal{L}^{(\text{p2s})}(\boldsymbol{\theta}_3) = \frac{1}{|\mathcal{D}|}\sum_{j=1}^{|\mathcal{D}|}\sum_{k=1}^{C} (\cos\theta_{j,k} - \mathbf{1}_{\{y^{(j)}_n\}}(k)) ^ 2.
\label{eq:p2sgrad}
\end{equation}
Although this loss is based on MSE, the network's output here is a cosine similarity $\cos\theta_{j, k} = \widehat{\boldsymbol{w}}_k^\top\widehat{\boldsymbol{h}}^{(j)} $ rather than an unconstrained value. Due to usage of the cosine, the gradient ${\partial\mathcal{L}^{(\text{p2s})}}/{\partial\boldsymbol{o}_j}$ is identical to the probability-to-similarity gradient (P2SGrad) \cite{zhang2019p2sgrad}, and it points to the same optimal direction as the margin-based softmax. We follow the convention in \cite{wang2021comparative} and refer to this loss function as \textit{MSE-for-P2SGrad}. 

\subsection{Regularization for DNN-based PAD}
The DNN is prone to over-fitting, and DNN-based voice PAD models are more vulnerable to over-fitting because the training set only covers part of the spoofing attacks in the evaluation set. From another viewpoint, the DNN-based voice PAD models should be generalizable to unseen attacks not in the training set. Regularization is therefore indispensable in PAD models. 

While the basic early stopping has been widely used, many PAD models also apply dropout \cite{srivastava2014dropout}, a technique that randomly sets a subset of a hidden layer's output to zero. Many effective PAD models use very a high dropout rate (i.e., the percentage of output nodes that are zeroed). For example, the top three single systems in the ASVspoof 2019 LA task, namely T45 \cite{lavrentyeva2019stc}, T24 \cite{Chen2020Odyssey} and T39 \cite{Chettri2019}, used a dropout rate larger than 50\% on some of the hidden layers. 
Similar trends can be found in the top system of the ASVspoof 2017 challenge with a high drop rate of 70\% \cite{Lavrentyeva2017} and another high-performance system in the ASVspoof 2015 challenge with a dropout rate of 60\% \cite{gomez2019gated}. While it is not always reported in the literature, for PAD models using a fixed-size-input CNN, the dropout can be applied on the fully-connected layer after the CNN \cite{wang2021comparative}. It is also practical to apply dropout on every hidden layer \cite{zhang2017investigation}.

Other potential regularization techniques include \emph{mixup} \cite{zhang2017mixup}. This technique randomly creates new pairs of input and output data by linear interpolation $x' = \lambda x_1 + (1-\lambda) x_2$, where $\lambda$ is a random scalar drawn from a Beta distribution. This augmentation-based regularization technique has been used in voice PAD \cite{zhang2017mixup}. Another useful augmentation-based regularization method is the mask in the frequency and time domains \cite{Chen2020Odyssey}.

\subsection{Model Fusion}
The techniques explained so far enable us to build a voice PAD model. 
However, a single classifier may not capture all the variations in the spoofed speech, especially the spoofed data created from various TTS and VC algorithms. Model fusion combines multiple `weak' classifiers into a potentially stronger one. This is a mature technique in the machine learning field and is also referred to as model ensemble or a meta-algorithm. 

While many fusion algorithms exist, the one widely used in the voice PAD community is score fusion. Suppose a set of PAD models $\{\mathcal{P}^{(1)}_{\boldsymbol{\theta}}, \cdots, \mathcal{P}^{(M)}_{\boldsymbol{\theta}}\}$ has been trained, where each model maps the input waveform $\boldsymbol{o}_{1:T}$ into a score $s^{(m)} = \mathcal{P}^{(m)}_{\boldsymbol{\theta}}(\boldsymbol{o}_{1:T})$. The score fusion algorithm computes a merged score $s$ as 
\begin{equation}
    s = \sum_{m=1}^{M} w_m \mathcal{P}^{(m)}_{\boldsymbol{\theta}}(\boldsymbol{o}_{1:T}).
\end{equation}
The weight $w_m$ can be uniform $w_m = \frac{1}{M}$ or tuned on a development set. 

In voice PAD, it is a common practice to fuse classifiers with varied front and back ends. For example, the top primary systems in the ASVspoof 2019 LA task combine single models using FFT, DCT, raw waveform, ResNet, MobileNet, LCNN, and so on. As the results of the ASVspoof2019 challenge demonstrated, the PAD models merged from multiple single models achieved better performance than the single models \cite{Nautsch2021}. 

\section{Evaluation Metrics}
\label{sec:metric}
\subsection{EER and DET Curve}
Voice PAD is currently treated as a binary classification task. 
As Section~\ref{sec:pad-gmm} describes, the classifier computes a score $r$ for the input trial and compares $r$ with a threshold $\tau$. The trial is classified as being \emph{bona fide} if $r\geq \tau$, otherwise \emph{spoofed}. Accordingly, two types of classification error can occur:
\begin{itemize}
    \item \emph{False rejection}: falsely classifying a bona fide trial as being spoofed;
    \item \emph{False acceptance}: falsely classifying a spoofed trial as being bona fide.
\end{itemize}

\begin{figure}[t]
\centering
\includegraphics[trim=0 480 0 85, clip, width=0.9\textwidth]{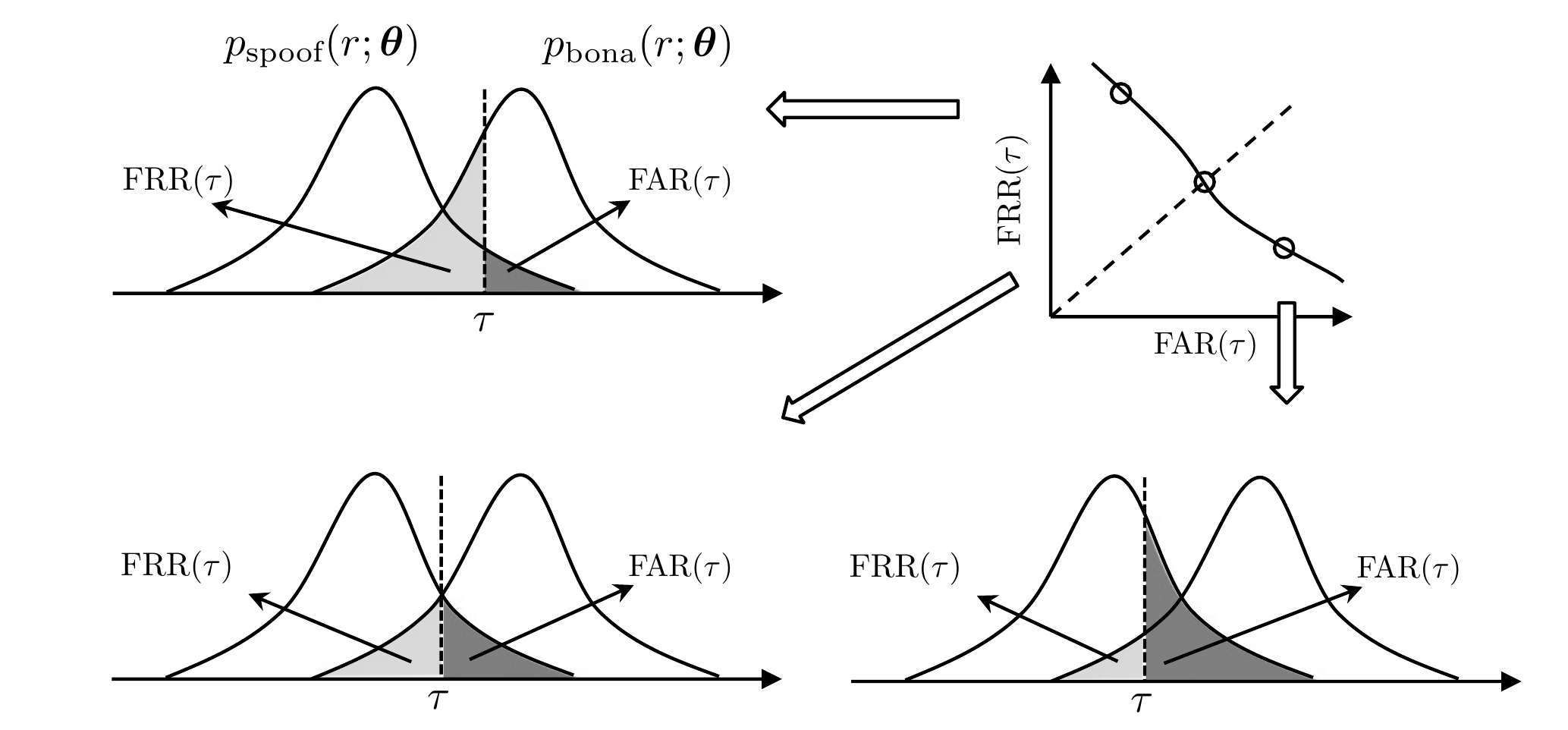}
\caption{Illustration of $\text{FRR}(\tau)$, $\text{FAR}(\tau)$, and DET curve.}
\label{fig:far_frr_det}
\end{figure}

Without loss of generality, let us use $p_{\text{bona}}(r; \boldsymbol{\theta})$ and $p_{\text{spoof}}(r; \boldsymbol{\theta})$ to denote the score distributions of the bona fide and spoofed trials in a dataset, respectively. The $\boldsymbol{\theta}$ denotes the trained parameter set of the PAD model. Fig.\ref{fig:far_frr_det} illustrates examples of the score distributions. 
The classification threshold $\tau$ marks the region to claim \emph{bona fide} and \emph{spoof}. The shaded area of $p_{\text{bona}}(r; \boldsymbol{\theta})$ on the left side of $\tau$ denotes false rejection. In contrast, the area of $p_{\text{spoofed}}(r; \boldsymbol{\theta})$ on the right side of $\tau$ denotes false acceptance. Accordingly, the false rejection rate (FRR) and false acceptance rate (FAR) can be computed as
\begin{align}
    \text{FRR}(\tau) &= \int_{-\infty}^{\tau} p_{\text{bona}}(r; \boldsymbol{\theta}) dr, \\
    \text{FAR}(\tau) &= \int_{\tau}^{\infty} p_{\text{spoofed}}(r; \boldsymbol{\theta}) dr.
\end{align}
Since the exact score distributions are unknown, the integration is approximated by counting:
\begin{align}
    \text{FRR}(\tau) &\approx \frac{1}{|\mathcal{D}_{\text{bona}}|}\sum_{j\in\mathcal{D}_{\text{bona}}}\mathbf{1}(r_j < \tau), \\
    \text{FAR}(\tau) &\approx \frac{1}{|\mathcal{D}_{\text{spoof}}|}\sum_{j\in\mathcal{D}_{\text{spoof}}}\mathbf{1}(r_j \geq \tau), 
\end{align}
where $j$ is the index of a trial, and $\mathcal{D}_{\text{bona}}$ and $\mathcal{D}_{\text{spoof}}$ denote the sets of bona fide and spoofed trials, respectively. 
Note that, while FRR and FAR are widely used in the ASV and voice PAD community,  many binary classification tasks may use other metrics such as true rejection rate. 

Both $\text{FRR}(\tau)$ and $\text{FAR}(\tau)$ are functions of $\tau$, and there is a trade-off between the two metrics; regardless if we move $\tau$ to the right or left, one of the metrics decreases while the other increases. This is plotted in Fig.\ref{fig:far_frr_det}. Determining a proper $\tau$ requires application-specific knowledge. The reason is that the loss caused by a false rejection is usually different from a false acceptance. Nevertheless, both $\text{FRR}(\tau)$ and $\text{FAR}(\tau)$ reflect the model performance. A verbose option to display the model performance is to sweep $\tau$ and plot the values of $\text{FRR}(\tau)$ and $\text{FAR}(\tau)$ on a 2D space. This is known as the detection error trade-off (DET) curve \cite{martin1997det}. One example is plotted in the top-right corner of Fig. \ref{fig:far_frr_det}. Note that both axes of the DET curve are warped with a probit function $f(x)=\sqrt{2}\text{erf}^{-1}(2x-1)$, where $\text{erf}^{-1}$ denotes the inverse Gauss error function\footnote{The inverse Gauss error function is implemented in scipy.special.erfinv of the SciPy library \url{https://docs.scipy.org/doc/}.}. More concrete DET curves in the experiments are plotted in Fig. \ref{fig:det_variance}. 

While being informative, the DET curve is not handy when a single metric is desired. One metric to summarize the discriminative power of the PAD model is the EER. In theory, EER refers to the joint between the DET curve and the anti-diagonal line (see the middle circle on the DET curve in Fig.\ref{fig:far_frr_det}). In practice\footnote{This method is used in the ASVspoof challenge EER evaluation package. The link is available in Appendix.}, the EER can be computed by
\begin{equation}
\text{EER} = \frac{\text{FRR}(\tau') + \text{FAR}(\tau')}{2},
\end{equation}
where 
\begin{equation}
    \tau' = \argmin_{\tau} |\text{FRR}(\tau) - \text{FAR}(\tau)|.
\end{equation} 
Although the EER has been deprecated in the recent  ISO/IEC standard, it is still used as a convenient metric in the research community.

\subsection{Min t-DCF}
Both the EER and the DET curve focus on the PAD model. However, falsely classified trials by the voice PAD model have different impacts on the ASV model. Particularly, some falsely accepted spoofed trials could present no threat to the ASV model. Therefore, a new metric is necessary when evaluating the PAD and ASV models jointly. 

Candidate metrics include the tandem detection cost functions (t-DCF) \cite{kinnunen2018t,kinnunen2020tandem}. The latest one used in the ASVspoof challenge \cite{Nautsch2021} is the ASV-constrained minimum normalized t-DCF (min t-DCF), which is from the perspective of a PAD developer `who cannot interact with the ASV system' \cite[Sec.III.D]{kinnunen2020tandem}. In other words, it assumes that the ASV system is a fixed black box. 
The min t-DCF is then defined as 
\begin{equation}
    \text{min t-DCF} = \min_\tau \frac{C_0 + C_1 \text{FRR}(\tau) + C_2 \text{FAR}(\tau)}{\text{t-DCF}_{\text{default}}},
\end{equation}
where $(C_0, C_1, C_2)$ is a hyper-parameter set determined by the performance of the ASV system, the prior probability of non-target, target, and spoofed trials, and the cost of incorrect decisions in ASV and CM systems. The denominator $\text{t-DCF}_{\text{default}} = C_0 + \min(\{C_1, C_2\})$ is the t-DCF of a default PAD model that either accepts or rejects all the input trials. The denominator normalizes the min t-DCF value so that it will be a value between the ASV floor $\frac{C_0}{\text{t-DCF}_{\text{default}}}$ and 1. A lower min t-DCF indicates a better performance.

In the ASVspoof challenges, the value of $(C_0, C_1, C_2)$ is computed on a state-of-the-art ASV system and distributed by the organizers. The toolkit to compute the min t-DCF value is open-sourced (see Appendix), and the ASV scores used to compute the $(C_0, C_1, C_2)$ are available in the publicly available database \cite{WANG2020101114}.

Note that in the ASVspoof 2019 challenge, a legacy version of the t-DCF metric was used. The legacy version has a few more hyper-parameters than the version above \cite{kinnunen2020tandem}. Both versions of the t-DCT values will be reported in the experiment of this chapter.

\subsection{Multiple Runs and Statistical Analysis}
The evaluation metrics above measure the discriminative power of a PAD model. However, a PAD model's performance is affected by the optimization process. For DNNs-based PAD models, the optimization process is based on stochastic gradient descent and is known to be sensitive to the initial state \cite{madhyastha2019model}. Furthermore, even with the same initial state, the optimization is affected by hardware and software. 
Only a single round of training and evaluation would not reliably reflect the model's performance. Therefore, an experiment with multiple independent training and evaluation rounds is preferred.
We may change the initial state for different rounds by using different random seeds for the random initialization of the DNN's weights.

However, as \cite{wang2021comparative} points out,  not all the difference is statistically significant after multiple rounds of training and evaluation. To measure the statistical significance, we could conduct pair-wise statistical analysis using the methodology in \cite{bengio2004statistical}. For a pair of models $(A, B)$ in the comparison, we can compute the value of a $z$-statistic using 
\begin{equation}
z=\frac{2|\text{EER}_A - \text{EER}_B|}{\sqrt{\Big[\text{EER}_A (1-\text{EER}_A)+\text{EER}_B (1-\text{EER}_B)\Big]\frac{|\mathcal{D}_{\text{bona}|} + |\mathcal{D}_{\text{spoof}}|}{|\mathcal{D}_{\text{bona}}||\mathcal{D}_{\text{spoof}}|}}},
\end{equation}
where $|\mathcal{D}_{\text{bona}}|$ and $|\mathcal{D}_{\text{spoof}}|$ denote the number of bonafide and spoof trials, respectively.  
We then compare $z$ with a value $Z_{\alpha/2}$ determined by a significance level $\alpha=0.05$. 
$z\geq Z_{\alpha/2}$ suggests a statistically significant difference between $A$ and $B$ at the significance level of $\alpha=0.05$. 

The more tests we conduct, the more likely that we will at least observe one significant difference\footnote{Suppose the probability to observe a significant difference is $P$, then the probability of observing at least one significant difference among $N$ tests is $1-(1-P)^{N}$.}. Therefore, we have to adjust the value $Z_{\alpha/2}$ to make a more conservative claim. A conservative strategy known as Bonferroni correction is to use $Z_{\alpha/2}/N$, where $N$ is the number of statistical tests. A less conservative strategy is the Holm-Bonferroni correction \cite{holm1979simple}. Suppose we have conducted $N$ tests and collect the scores $\{z_1, z_2, \cdots, z_N\}$. We can sort the scores from the lowest to highest and compare each $z_n$ with $Z_{\alpha/2}/i$ where $i$ is the ranking of $z_n$ in the sorted list.

\section{Datasets and Campaigns}
\label{sec:camp}
A major benchmark campaign on voice PAD is the biennial event called the ASVspoof challenge \cite{Kinnunen2017,Nautsch2021,wu2015asvspoof}.  The task of the challenge evolves with time. In the first challenge in 2015 \cite{wu2015asvspoof}, the task was to detect spoofing attacks from the by then popular speech synthesis and voice conversion algorithms. In 2017 \cite{Kinnunen2017}, the task was to detect real replayed speech for physical access attack. In 2019 \cite{Nautsch2021}, more advanced TTS and VC algorithms were included in the LA task. The tasks in 2021 were more challenging \cite{yamagishi21_asvspoof}: LA introduced the impact of codecs, and a new task called deepfake detection was included with data from other domains. The databases used in the challenges are publicly available, and the links can be found in Appendix. 

There are other publicly available databases. One is the AVspoof database \cite{korshunov2016overview} and its extension called VoicePA \cite{korshunov2018use}. Another one is the Spoofing and Anti-spoofing (SAS) \cite{wu2015sas} database. \BLUE{More recently, a database called WaveFake was released \cite{frank2021wavefake}. It covers a few advanced neural vocoders and includes not only English but also Japanese speech data. Another database for Mandarin speech also covers advanced TTS and VC systems \cite{zhang2021fmfcc}.}

\section{Experiments on the ASVspoof2019 LA task}
\label{sec:exp}
This section explains the experiment on the ASVspoof2019 LA database. This database was selected because it includes spoofing speech from the advanced TTS and VC algorithms but does not involve codec and channel variation. Such a database is ideal for experiments in this practical guide.  Experiments on the LCNN-based single models have been published in \cite{wang2021comparative}. We re-use the results from that work but added more detailed analysis, one additional system using RawNet2, and experiments using model fusion.

\subsection{Experimental Models}
We selected a few representative techniques from those introduced in Section~\ref{sec:defender}. The front ends used in the experiment included LFB, LFCC, and spectrogram. The training criteria covered AM-softmax, OC-softmax, vanilla softmax, and the MSE-for-P2SGrad. The back end was based on LCNN but with either a trim-and-pad strategy, average pooling, or attentive pooling layer. The first back end uses a fixed-size input through the trim-and-pad strategy, while the latter two are length-agnostic. Additionally, the end-to-end RawNet2 was included. 

We recap the configurations used in \cite{wang2021comparative} in the following paragraphs. Among the front ends, LFCC followed the ASVspoof 2019 baseline recipe \cite{Todisco2019}: 60 dimensions, 20ms for frame length, 10ms for frame shift, 512 FFT points, a linearly spaced triangle filter bank of 20 channels. However, the first dimension was replaced with log spectral energy. The LFB was similarly configured but the number of linear filter bank channels was 60. The spectrogram was similar and had 257 dimensions. 
Only LFCC includes delta and delta-delta components.

We configured the LCNN-based back ends in the following ways:
\begin{itemize}
\item \systemI: The same LCNN as \cite{lavrentyeva2019stc} but the input size was fixed to $N=750$ frames \cite{zhang2020one}. Shorter inputs were padded with zeros. Longer inputs were trimmed by $\tilde{\boldsymbol{a}}_{1:N} = \boldsymbol{a}_{n:n+N}$, where $n$ is a random integer.
\item \systemII: The same LCNN as \systemI, but the layers after the CNN part (i.e., those after `MaxPool\_28' in \cite{lavrentyeva2019stc}) were replaced with a single-head-attention-based pooling layer \cite{Zhu2018} and an FC layer.
\item \systemIII: The same LCNN as \systemI, but the layers after the CNN part were replaced with two Bi-LSTM layers, an average pooling layer, and an FC layer. A skip connection was added over the two Bi-LSTM layers. 
\end{itemize}
When the back end is combined with the spectrogram, a trainable FC layer initialized with coefficients of the linear filter bank for LFB was added before the LCNN. This layer compressed the spectrogram to 60-dimensional hidden features, and it stabilized the learning curve of the CM. 

Among the training criteria, AM-softmax used a hyper-parameter set $(\alpha=20, m_1=0, m_2=0, m_3=0.9)$ \cite{zhang2020one}, while OC-softmax used $(\alpha=20, m_{3,1}=0.9, m_{3,2}=0.2)$ \cite{zhang2020one}.
For AM-softmax, OC-softmax, and MSE-for-P2SGrad, the last FC layer had a layer size of 64. In other words, the dimension of $\widehat{\boldsymbol{h}}$ in Eq.(\ref{eq:margin_softmax_cos}) was 64.

We built the experimental PAD models by combining the aforementioned components, but the combination is non-exhaustive due to the limited computation resources for experiments. The RawNet2 followed the original recipe \cite{tak2020end} and was only trained using the vanilla softmax\footnote{RawNet code was based on the ASVspoof2021 official implementation but with a minor fix. See more details in the code released with this chapter.}.

\subsection{Training Recipe}
\label{sec:exp-recipe}
The training recipe was borrowed from our previous work \cite{wang2021comparative}: the Adam optimizer  \cite{kingma2014adam} with hyper-parameters setting to $\beta_1=0.9, \beta_2=0.999, \epsilon=10^{-8}$,  a learning rate initialized with $3\times10^{-4}$ and halved for every ten epochs \cite{zhang2020one}, and a mini-batch size of 64. The mini-batch size was reduced to 8 if the required GPU memory of a large mini-batch was beyond the GPU card's capacity. There was no constraint on the ratio of bona fide versus spoof trials in one mini-batch. Neither voice activity detection or feature normalization was used. 

Since the length of the speech trials is not uniform, a mini-batch cannot be directly created by stacking trials of varied length. 
One strategy is to pad the trials to the length of the longest trial in the mini-batch. However, trials randomly selected for a mini-batch could have diverse lengths, and the short trials in the mini-batch have to be padded with long segments of zeros, which affects the back ends with recurrent layers. 
A customized method used in this chapter creates the mini-batch using randomly selected trials with similar lengths\footnote{The mini-batch sampler and collating functions are customized, not the default functions provided by Pytorch. See more details in the released code.}. Since the trials in a mini-batch have a similar duration, the padded length is relatively short. 
During the evaluation, the batch size was set to one, and padding was not necessary. This customized method was used for all the experimental models.

To examine the variance of model performance, we trained and evaluated each model independently for six times, and the random seed for the $k$-th run was $10^{k-1}$. All the experiments were done using an Nvidia Tesla P100 card. The results are reproducible with the same random seed and GPU environment.

\subsection{Results of Single Systems}
Table \ref{tab:eer} lists the EERs and min t-DCF values on the ASVspoof 2019 LA evaluation set. The results of the statistical analysis are plotted in Fig.~\ref{fig:sig_test}. White and dark grey indicate insignificant and significant statistical differences, respectively.

\subsubsection{Intra-model Variance} The first observation is the intra-model variance. The results in Table \ref{tab:eer} demonstrate that the EER of the same model can be quite different when the initial state was changed with a different random seed. For example, LFCC-\systemI-AM-softmax was similar to the top-2 single system (T45) in the ASVspoof2019 LA task, and its best EER (3.04\%) among the six rounds was better than that reported in the challenge (5.06\% for T45). However, its EER was as low as 7.06\% in the worst round. Furthermore, Fig.\ref{fig:sig_test} shows that the block of LFCC-\systemI-AM-softmax on the anti-diagonal line is almost in dark grey. This indicates that, if we select two from the six rounds, in most cases, the difference between the two is statistically significant.

A similar intra-model variance can be observed for other models, as shown from other non-white blocks on the anti-diagonal line of Fig.\ref{fig:sig_test}. 
Although it is expected to see intra-model differences in DNN-based classifiers, it should be noted that the intra-model difference between the best and worst runs of one model could be larger than the differences across different models. For example, in the condition with LFCC and AM-softmax, the six runs of \systemIII\ achieved EERs (2.46\%, 3.02\%, 3.23\%, 3.34\%, 3.41\%, 3.86\%), while \systemI\ achieved (3.04\%, 3.52\%, 4.73\%, 4.89\%, 5.85\%, 7.06\%). The difference between the best and worst EERs of \systemI\ is 4.02\%, while that between the best runs of \systemI\ and \systemIII\ is merely 0.58\%.  

The aforementioned comparison suggests that it is risky to compare the DNN-based voice PAD with only a single round of training and evaluation. The result of a single round may not fully reveal the model capability if the model started from an `unlucky' initial state. Furthermore, a single round of training and evaluation provides no evidence about the model's performance when it is trained using a different starting point or on a different machine. 

\BLUE{Intra-model variance is a general issue of DNNs, and it has been reported in natural language processing \cite{madhyastha2019model} and computer vision communities \cite{prabhu2021uncertainty}. How to reduce the intra-model variance is an ongoing research topic.}

\subsubsection{On the Front End} 
Despite the intra-model variance, the results suggest a few potential techniques for PAD on the ASVspoof2019 LA database. 

\begin{figure}[!t]
\centering
\includegraphics[trim=0 100 0 180, clip, width=0.9\textwidth]{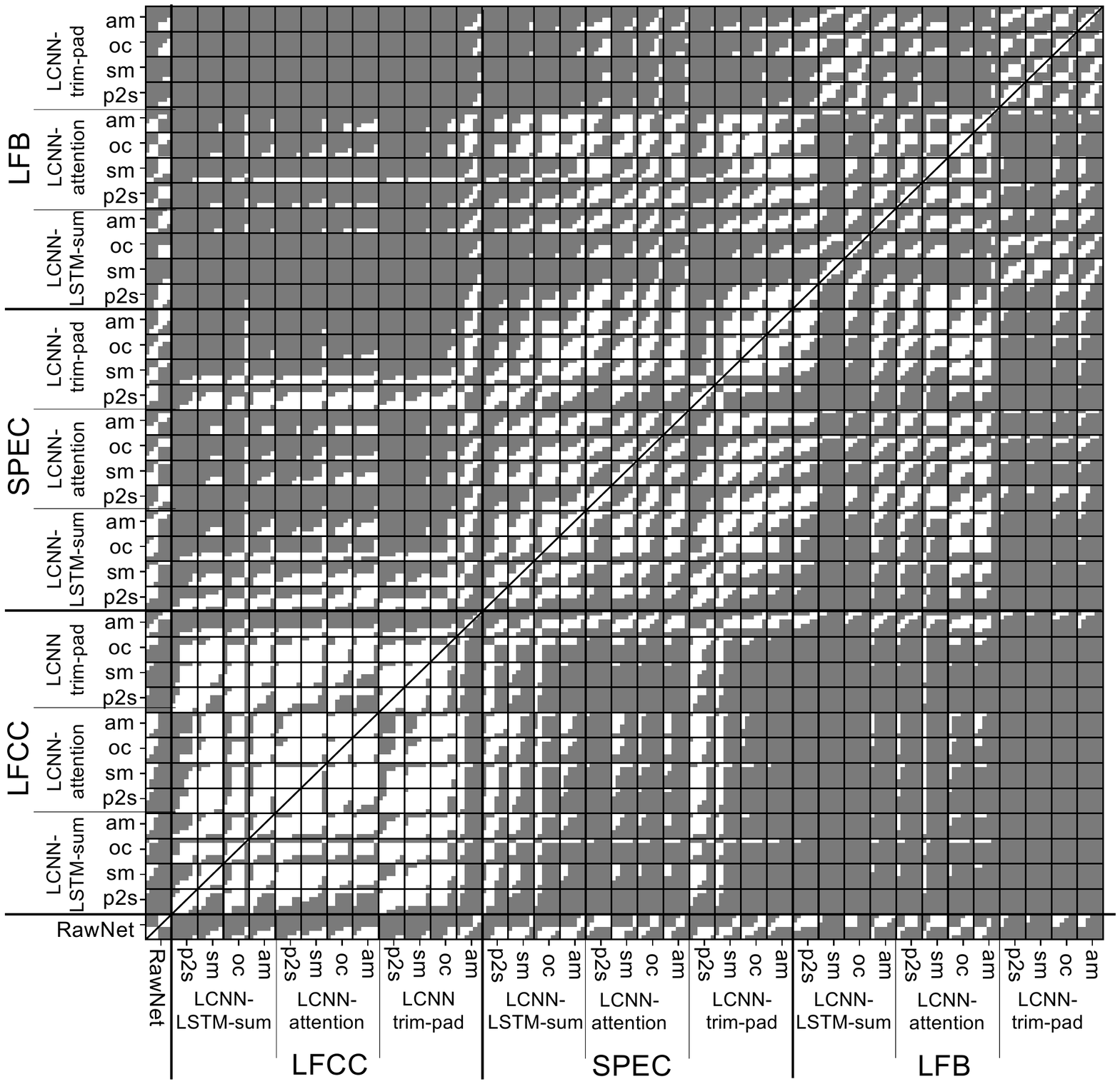}
\caption{Results of statistical significance test on EERs using a significance level of $\alpha=0.05$ with Holm-Bonferroni correction. Each square in the black frames contains $6\times6$ entries, denoting pair-wise tests between six training-evaluation rounds of two models. 
A significant difference is indicated by dark grey, otherwise by white. The square on the anti-diagonal line shows the intra-model differences. Because a new RawNet model is added, results in this table were re-computed and not exactly the same as that in \cite{wang2021comparative}.}
\label{fig:sig_test}
\end{figure}

On the front end, the LFCC had better performance than the other two. Furthermore, as Fig.\ref{fig:sig_test} demonstrates, models using LFCC have smaller intra-model variances. This can be observed also from the color codes used in Table \ref{tab:eer}. Compared with RawNet, using the LFCC front end seems to be a good choice.

\begin{landscape}
\begin{table*}[t!]
\caption{EER and min t-DCF on the ASVspoof2019 LA evaluation set. The first and second sub-tables except the results of RawNet2 are first reported in \cite[Table 1]{wang2021comparative}. For visualization, the results of six training-evaluation rounds were sorted in accordance with the EER from low (I) to high (VI). A darker cell indicates a higher EER or min t-DCF. The ASV floor, i.e., the v2.0 min t-DCF with a perfect PAD model, is 0.0627 \cite{Nautsch2021}.}
\begin{center}
\resizebox{1.4\textwidth}{!}{
\setlength{\tabcolsep}{2pt}
\begin{tabular}{cc|cccccc|cccccc|cccccc|cccccc}
\multicolumn{26}{c}{EERs}   \\
\toprule
  \multicolumn{2}{c}{} &  \multicolumn{6}{c}{AM-softmax} &  \multicolumn{6}{c}{OC-softmax} &  \multicolumn{6}{c}{Softmax} &  \multicolumn{6}{c}{MSE-for-P2SGrad}   \\
  \cmidrule(r){3-26}
\multicolumn{2}{c}{}   &  I  &  II  &  III   &  IV   &  V   &  \multicolumn{1}{c}{VI}   & I  &  II  &  III   &  IV   &  V   &   \multicolumn{1}{c}{VI}      & I  &  II  &  III   &  IV   &  V   &   \multicolumn{1}{c}{VI}       &  I  &  II  &  III   &  IV   &  V   &   \multicolumn{1}{c}{VI}   \\ 
\midrule

 \multirow{3}{*}{\rotatebox[origin=c]{0}{LFB}}  &  \systemI    & \cellcolor[rgb]{0.71, 0.71, 0.71} 5.59 & \cellcolor[rgb]{0.62, 0.62, 0.62} 6.39 & \cellcolor[rgb]{0.56, 0.56, 0.56} 7.12 & \cellcolor[rgb]{0.50, 0.50, 0.50} 7.79 & \cellcolor[rgb]{0.38, 0.38, 0.38} 9.33 & \cellcolor[rgb]{0.27, 0.27, 0.27} 10.72 & \cellcolor[rgb]{0.66, 0.66, 0.66} 5.98 & \cellcolor[rgb]{0.59, 0.59, 0.59} 6.76 & \cellcolor[rgb]{0.56, 0.56, 0.56} 7.02 & \cellcolor[rgb]{0.55, 0.55, 0.55} 7.14 & \cellcolor[rgb]{0.36, 0.36, 0.36} 9.56 & \cellcolor[rgb]{0.22, 0.22, 0.22} 11.34 & \cellcolor[rgb]{0.56, 0.56, 0.56} 7.00 & \cellcolor[rgb]{0.53, 0.53, 0.53} 7.40 & \cellcolor[rgb]{0.43, 0.43, 0.43} 8.69 & \cellcolor[rgb]{0.41, 0.41, 0.41} 8.89 & \cellcolor[rgb]{0.34, 0.34, 0.34} 9.86 & \cellcolor[rgb]{0.33, 0.33, 0.33} 10.13 & \cellcolor[rgb]{0.58, 0.58, 0.58} 6.81 & \cellcolor[rgb]{0.54, 0.54, 0.54} 7.25 & \cellcolor[rgb]{0.47, 0.47, 0.47} 8.10 & \cellcolor[rgb]{0.46, 0.46, 0.46} 8.28 & \cellcolor[rgb]{0.45, 0.45, 0.45} 8.41 & \cellcolor[rgb]{0.32, 0.32, 0.32} 10.21\\ 
                                                 &  \systemII   & \cellcolor[rgb]{0.82, 0.82, 0.82} 4.26 & \cellcolor[rgb]{0.82, 0.82, 0.82} 4.32 & \cellcolor[rgb]{0.79, 0.79, 0.79} 4.69 & \cellcolor[rgb]{0.74, 0.74, 0.74} 5.23 & \cellcolor[rgb]{0.44, 0.44, 0.44} 8.55 & \cellcolor[rgb]{0.01, 0.01, 0.01} \textcolor[rgb]{0.6,0.6,0.6}{14.86} & \cellcolor[rgb]{0.84, 0.84, 0.84} 4.01 & \cellcolor[rgb]{0.80, 0.80, 0.80} 4.54 & \cellcolor[rgb]{0.80, 0.80, 0.80} 4.57 & \cellcolor[rgb]{0.74, 0.74, 0.74} 5.27 & \cellcolor[rgb]{0.73, 0.73, 0.73} 5.33 & \cellcolor[rgb]{0.69, 0.69, 0.69} 5.76 & \cellcolor[rgb]{0.90, 0.90, 0.90} 3.34 & \cellcolor[rgb]{0.76, 0.76, 0.76} 5.07 & \cellcolor[rgb]{0.70, 0.70, 0.70} 5.70 & \cellcolor[rgb]{0.67, 0.67, 0.67} 5.91 & \cellcolor[rgb]{0.67, 0.67, 0.67} 5.93 & \cellcolor[rgb]{0.67, 0.67, 0.67} 5.93 & \cellcolor[rgb]{0.85, 0.85, 0.85} 3.99 & \cellcolor[rgb]{0.77, 0.77, 0.77} 4.87 & \cellcolor[rgb]{0.69, 0.69, 0.69} 5.75 & \cellcolor[rgb]{0.68, 0.68, 0.68} 5.85 & \cellcolor[rgb]{0.64, 0.64, 0.64} 6.20 & \cellcolor[rgb]{0.53, 0.53, 0.53} 7.36\\ 
                                                 &  \systemIII  & \cellcolor[rgb]{0.83, 0.83, 0.83} 4.24 & \cellcolor[rgb]{0.74, 0.74, 0.74} 5.27 & \cellcolor[rgb]{0.69, 0.69, 0.69} 5.71 & \cellcolor[rgb]{0.64, 0.64, 0.64} 6.23 & \cellcolor[rgb]{0.56, 0.56, 0.56} 7.03 & \cellcolor[rgb]{0.56, 0.56, 0.56} 7.10 & \cellcolor[rgb]{0.68, 0.68, 0.68} 5.81 & \cellcolor[rgb]{0.61, 0.61, 0.61} 6.51 & \cellcolor[rgb]{0.57, 0.57, 0.57} 6.89 & \cellcolor[rgb]{0.51, 0.51, 0.51} 7.64 & \cellcolor[rgb]{0.39, 0.39, 0.39} 9.15 & \cellcolor[rgb]{0.32, 0.32, 0.32} 10.24 & \cellcolor[rgb]{0.56, 0.56, 0.56} 7.04 & \cellcolor[rgb]{0.49, 0.49, 0.49} 7.93 & \cellcolor[rgb]{0.44, 0.44, 0.44} 8.52 & \cellcolor[rgb]{0.39, 0.39, 0.39} 9.16 & \cellcolor[rgb]{0.37, 0.37, 0.37} 9.53 & \cellcolor[rgb]{0.34, 0.34, 0.34} 9.84 & \cellcolor[rgb]{0.76, 0.76, 0.76} 5.06 & \cellcolor[rgb]{0.75, 0.75, 0.75} 5.17 & \cellcolor[rgb]{0.70, 0.70, 0.70} 5.69 & \cellcolor[rgb]{0.69, 0.69, 0.69} 5.75 & \cellcolor[rgb]{0.64, 0.64, 0.64} 6.17 & \cellcolor[rgb]{0.61, 0.61, 0.61} 6.50\\ 
\midrule
 \multirow{3}{*}{\rotatebox[origin=c]{0}{SPEC}} &  \systemI    & \cellcolor[rgb]{0.77, 0.77, 0.77} 4.84 & \cellcolor[rgb]{0.76, 0.76, 0.76} 5.02 & \cellcolor[rgb]{0.74, 0.74, 0.74} 5.23 & \cellcolor[rgb]{0.68, 0.68, 0.68} 5.87 & \cellcolor[rgb]{0.67, 0.67, 0.67} 5.90 & \cellcolor[rgb]{0.64, 0.64, 0.64} 6.25 & \cellcolor[rgb]{0.81, 0.81, 0.81} 4.41 & \cellcolor[rgb]{0.80, 0.80, 0.80} 4.60 & \cellcolor[rgb]{0.77, 0.77, 0.77} 4.84 & \cellcolor[rgb]{0.76, 0.76, 0.76} 5.04 & \cellcolor[rgb]{0.73, 0.73, 0.73} 5.37 & \cellcolor[rgb]{0.63, 0.63, 0.63} 6.35 & \cellcolor[rgb]{0.92, 0.92, 0.92} 3.09 & \cellcolor[rgb]{0.88, 0.88, 0.88} 3.58 & \cellcolor[rgb]{0.78, 0.78, 0.78} 4.72 & \cellcolor[rgb]{0.77, 0.77, 0.77} 4.82 & \cellcolor[rgb]{0.74, 0.74, 0.74} 5.22 & \cellcolor[rgb]{0.71, 0.71, 0.71} 5.56 & \cellcolor[rgb]{0.93, 0.93, 0.93} 2.94 & \cellcolor[rgb]{0.90, 0.90, 0.90} 3.22 & \cellcolor[rgb]{0.88, 0.88, 0.88} 3.59 & \cellcolor[rgb]{0.87, 0.87, 0.87} 3.73 & \cellcolor[rgb]{0.81, 0.81, 0.81} 4.49 & \cellcolor[rgb]{0.78, 0.78, 0.78} 4.74\\ 
                                                 &  \systemII   & \cellcolor[rgb]{0.84, 0.84, 0.84} 4.02 & \cellcolor[rgb]{0.84, 0.84, 0.84} 4.08 & \cellcolor[rgb]{0.76, 0.76, 0.76} 4.99 & \cellcolor[rgb]{0.74, 0.74, 0.74} 5.22 & \cellcolor[rgb]{0.71, 0.71, 0.71} 5.57 & \cellcolor[rgb]{0.47, 0.47, 0.47} 8.20 & \cellcolor[rgb]{0.84, 0.84, 0.84} 4.05 & \cellcolor[rgb]{0.80, 0.80, 0.80} 4.55 & \cellcolor[rgb]{0.77, 0.77, 0.77} 4.94 & \cellcolor[rgb]{0.70, 0.70, 0.70} 5.67 & \cellcolor[rgb]{0.63, 0.63, 0.63} 6.31 & \cellcolor[rgb]{0.44, 0.44, 0.44} 8.47 & \cellcolor[rgb]{0.85, 0.85, 0.85} 3.92 & \cellcolor[rgb]{0.84, 0.84, 0.84} 4.04 & \cellcolor[rgb]{0.81, 0.81, 0.81} 4.42 & \cellcolor[rgb]{0.77, 0.77, 0.77} 4.91 & \cellcolor[rgb]{0.77, 0.77, 0.77} 4.95 & \cellcolor[rgb]{0.60, 0.60, 0.60} 6.62 & \cellcolor[rgb]{0.78, 0.78, 0.78} 4.70 & \cellcolor[rgb]{0.76, 0.76, 0.76} 5.03 & \cellcolor[rgb]{0.71, 0.71, 0.71} 5.60 & \cellcolor[rgb]{0.68, 0.68, 0.68} 5.88 & \cellcolor[rgb]{0.63, 0.63, 0.63} 6.35 & \cellcolor[rgb]{0.61, 0.61, 0.61} 6.50\\ 
                                                 &  \systemIII  & \cellcolor[rgb]{0.85, 0.85, 0.85} 3.96 & \cellcolor[rgb]{0.84, 0.84, 0.84} 4.04 & \cellcolor[rgb]{0.81, 0.81, 0.81} 4.38 & \cellcolor[rgb]{0.80, 0.80, 0.80} 4.52 & \cellcolor[rgb]{0.75, 0.75, 0.75} 5.13 & \cellcolor[rgb]{0.67, 0.67, 0.67} 5.97 & \cellcolor[rgb]{0.94, 0.94, 0.94} 2.81 & \cellcolor[rgb]{0.89, 0.89, 0.89} 3.41 & \cellcolor[rgb]{0.80, 0.80, 0.80} 4.49 & \cellcolor[rgb]{0.80, 0.80, 0.80} 4.50 & \cellcolor[rgb]{0.79, 0.79, 0.79} 4.65 & \cellcolor[rgb]{0.77, 0.77, 0.77} 4.91 & \cellcolor[rgb]{0.90, 0.90, 0.90} 3.29 & \cellcolor[rgb]{0.88, 0.88, 0.88} 3.56 & \cellcolor[rgb]{0.86, 0.86, 0.86} 3.82 & \cellcolor[rgb]{0.81, 0.81, 0.81} 4.45 & \cellcolor[rgb]{0.80, 0.80, 0.80} 4.61 & \cellcolor[rgb]{0.72, 0.72, 0.72} 5.44 & \cellcolor[rgb]{0.97, 0.97, 0.97} 2.37 & \cellcolor[rgb]{0.93, 0.93, 0.93} 2.91 & \cellcolor[rgb]{0.92, 0.92, 0.92} 3.00 & \cellcolor[rgb]{0.85, 0.85, 0.85} 3.94 & \cellcolor[rgb]{0.82, 0.82, 0.82} 4.26 & \cellcolor[rgb]{0.81, 0.81, 0.81} 4.37\\ 
\midrule 
 \multirow{3}{*}{\rotatebox[origin=c]{0}{LFCC}} &  \systemI    & \cellcolor[rgb]{0.92, 0.92, 0.92} 3.04 & \cellcolor[rgb]{0.88, 0.88, 0.88} 3.52 & \cellcolor[rgb]{0.78, 0.78, 0.78} 4.73 & \cellcolor[rgb]{0.77, 0.77, 0.77} 4.89 & \cellcolor[rgb]{0.68, 0.68, 0.68} 5.85 & \cellcolor[rgb]{0.56, 0.56, 0.56} 7.06 & \cellcolor[rgb]{0.93, 0.93, 0.93} 2.93 & \cellcolor[rgb]{0.93, 0.93, 0.93} 2.93 & \cellcolor[rgb]{0.93, 0.93, 0.93} 2.95 & \cellcolor[rgb]{0.92, 0.92, 0.92} 2.99 & \cellcolor[rgb]{0.86, 0.86, 0.86} 3.84 & \cellcolor[rgb]{0.85, 0.85, 0.85} 4.00 & \cellcolor[rgb]{0.96, 0.96, 0.96} 2.54 & \cellcolor[rgb]{0.94, 0.94, 0.94} 2.73 & \cellcolor[rgb]{0.94, 0.94, 0.94} 2.77 & \cellcolor[rgb]{0.93, 0.93, 0.93} 2.91 & \cellcolor[rgb]{0.92, 0.92, 0.92} 3.08 & \cellcolor[rgb]{0.89, 0.89, 0.89} 3.47 & \cellcolor[rgb]{0.97, 0.97, 0.97} 2.31 & \cellcolor[rgb]{0.96, 0.96, 0.96} 2.46 & \cellcolor[rgb]{0.95, 0.95, 0.95} 2.64 & \cellcolor[rgb]{0.95, 0.95, 0.95} 2.65 & \cellcolor[rgb]{0.92, 0.92, 0.92} 3.09 & \cellcolor[rgb]{0.91, 0.91, 0.91} 3.11\\ 
                                                 &  \systemII   & \cellcolor[rgb]{0.92, 0.92, 0.92} 2.99 & \cellcolor[rgb]{0.88, 0.88, 0.88} 3.48 & \cellcolor[rgb]{0.88, 0.88, 0.88} 3.60 & \cellcolor[rgb]{0.88, 0.88, 0.88} 3.61 & \cellcolor[rgb]{0.88, 0.88, 0.88} 3.62 & \cellcolor[rgb]{0.85, 0.85, 0.85} 3.92 & \cellcolor[rgb]{0.93, 0.93, 0.93} 2.91 & \cellcolor[rgb]{0.93, 0.93, 0.93} 2.96 & \cellcolor[rgb]{0.89, 0.89, 0.89} 3.36 & \cellcolor[rgb]{0.89, 0.89, 0.89} 3.40 & \cellcolor[rgb]{0.88, 0.88, 0.88} 3.61 & \cellcolor[rgb]{0.87, 0.87, 0.87} 3.71 & \cellcolor[rgb]{0.91, 0.91, 0.91} 3.18 & \cellcolor[rgb]{0.91, 0.91, 0.91} 3.19 & \cellcolor[rgb]{0.90, 0.90, 0.90} 3.24 & \cellcolor[rgb]{0.90, 0.90, 0.90} 3.29 & \cellcolor[rgb]{0.89, 0.89, 0.89} 3.39 & \cellcolor[rgb]{0.84, 0.84, 0.84} 4.12 & \cellcolor[rgb]{0.94, 0.94, 0.94} 2.72 & \cellcolor[rgb]{0.94, 0.94, 0.94} 2.81 & \cellcolor[rgb]{0.93, 0.93, 0.93} 2.91 & \cellcolor[rgb]{0.92, 0.92, 0.92} 3.03 & \cellcolor[rgb]{0.90, 0.90, 0.90} 3.22 & \cellcolor[rgb]{0.90, 0.90, 0.90} 3.30\\ 
                                                 &  \systemIII  & \cellcolor[rgb]{0.96, 0.96, 0.96} 2.46 & \cellcolor[rgb]{0.92, 0.92, 0.92} 3.02 & \cellcolor[rgb]{0.90, 0.90, 0.90} 3.23 & \cellcolor[rgb]{0.90, 0.90, 0.90} 3.34 & \cellcolor[rgb]{0.89, 0.89, 0.89} 3.41 & \cellcolor[rgb]{0.86, 0.86, 0.86} 3.86 & \cellcolor[rgb]{0.97, 0.97, 0.97} 2.23 & \cellcolor[rgb]{0.96, 0.96, 0.96} 2.42 & \cellcolor[rgb]{0.93, 0.93, 0.93} 2.96 & \cellcolor[rgb]{0.93, 0.93, 0.93} 2.96 & \cellcolor[rgb]{0.93, 0.93, 0.93} 2.96 & \cellcolor[rgb]{0.79, 0.79, 0.79} 4.64 & \cellcolor[rgb]{0.95, 0.95, 0.95} 2.67 & \cellcolor[rgb]{0.93, 0.93, 0.93} 2.94 & \cellcolor[rgb]{0.91, 0.91, 0.91} 3.20 & \cellcolor[rgb]{0.89, 0.89, 0.89} 3.40 & \cellcolor[rgb]{0.89, 0.89, 0.89} 3.45 & \cellcolor[rgb]{0.88, 0.88, 0.88} 3.53 & \cellcolor[rgb]{1.00, 1.00, 1.00} 1.92 & \cellcolor[rgb]{0.99, 0.99, 0.99} 2.09 & \cellcolor[rgb]{0.96, 0.96, 0.96} 2.43 & \cellcolor[rgb]{0.96, 0.96, 0.96} 2.50 & \cellcolor[rgb]{0.95, 0.95, 0.95} 2.62 & \cellcolor[rgb]{0.92, 0.92, 0.92} 3.10\\ 
\midrule 
&  \systemIV   &       &       &       &       &       &       &       &       &       &       &       &       & \cellcolor[rgb]{0.92, 0.92, 0.92} 3.70 & \cellcolor[rgb]{0.90, 0.90, 0.90} 4.00 & \cellcolor[rgb]{0.89, 0.89, 0.89} 4.30 & \cellcolor[rgb]{0.80, 0.80, 0.80} 5.95 & \cellcolor[rgb]{0.80, 0.80, 0.80} 6.04 & \cellcolor[rgb]{0.78, 0.78, 0.78} 6.26 &       &       &       &       &       &      \\ 
 \bottomrule
 \\
\multicolumn{26}{c}{min t-DCF (legacy version used in the ASVspoof 2019 challenge)}   \\
\toprule
 \multirow{3}{*}{\rotatebox[origin=c]{0}{LFB}}  &  \systemI         & \cellcolor[rgb]{0.72, 0.72, 0.72} 0.129 & \cellcolor[rgb]{0.54, 0.54, 0.54} 0.168 & \cellcolor[rgb]{0.41, 0.41, 0.41} 0.204 & \cellcolor[rgb]{0.38, 0.38, 0.38} 0.212 & \cellcolor[rgb]{0.30, 0.30, 0.30} 0.235 & \cellcolor[rgb]{0.30, 0.30, 0.30} 0.236 & \cellcolor[rgb]{0.59, 0.59, 0.59} 0.156 & \cellcolor[rgb]{0.50, 0.50, 0.50} 0.179 & \cellcolor[rgb]{0.52, 0.52, 0.52} 0.173 & \cellcolor[rgb]{0.53, 0.53, 0.53} 0.171 & \cellcolor[rgb]{0.42, 0.42, 0.42} 0.200 & \cellcolor[rgb]{0.26, 0.26, 0.26} 0.246 & \cellcolor[rgb]{0.55, 0.55, 0.55} 0.166 & \cellcolor[rgb]{0.51, 0.51, 0.51} 0.175 & \cellcolor[rgb]{0.44, 0.44, 0.44} 0.195 & \cellcolor[rgb]{0.35, 0.35, 0.35} 0.222 & \cellcolor[rgb]{0.32, 0.32, 0.32} 0.229 & \cellcolor[rgb]{0.33, 0.33, 0.33} 0.227 & \cellcolor[rgb]{0.53, 0.53, 0.53} 0.171 & \cellcolor[rgb]{0.38, 0.38, 0.38} 0.212 & \cellcolor[rgb]{0.42, 0.42, 0.42} 0.200 & \cellcolor[rgb]{0.46, 0.46, 0.46} 0.190 & \cellcolor[rgb]{0.36, 0.36, 0.36} 0.217 & \cellcolor[rgb]{0.37, 0.37, 0.37} 0.215\\ 
                                                &  \systemII        & \cellcolor[rgb]{0.76, 0.76, 0.76} 0.118 & \cellcolor[rgb]{0.77, 0.77, 0.77} 0.117 & \cellcolor[rgb]{0.86, 0.86, 0.86} 0.093 & \cellcolor[rgb]{0.83, 0.83, 0.83} 0.101 & \cellcolor[rgb]{0.41, 0.41, 0.41} 0.203 & \cellcolor[rgb]{0.01, 0.01, 0.01} \textcolor[rgb]{0.6,0.6,0.6}{0.331} & \cellcolor[rgb]{0.79, 0.79, 0.79} 0.111 & \cellcolor[rgb]{0.75, 0.75, 0.75} 0.121 & \cellcolor[rgb]{0.75, 0.75, 0.75} 0.122 & \cellcolor[rgb]{0.62, 0.62, 0.62} 0.148 & \cellcolor[rgb]{0.59, 0.59, 0.59} 0.155 & \cellcolor[rgb]{0.57, 0.57, 0.57} 0.160 & \cellcolor[rgb]{0.96, 0.96, 0.96} 0.064 & \cellcolor[rgb]{0.88, 0.88, 0.88} 0.087 & \cellcolor[rgb]{0.73, 0.73, 0.73} 0.126 & \cellcolor[rgb]{0.88, 0.88, 0.88} 0.088 & \cellcolor[rgb]{0.80, 0.80, 0.80} 0.108 & \cellcolor[rgb]{0.86, 0.86, 0.86} 0.092 & \cellcolor[rgb]{0.87, 0.87, 0.87} 0.089 & \cellcolor[rgb]{0.80, 0.80, 0.80} 0.110 & \cellcolor[rgb]{0.74, 0.74, 0.74} 0.124 & \cellcolor[rgb]{0.76, 0.76, 0.76} 0.119 & \cellcolor[rgb]{0.74, 0.74, 0.74} 0.123 & \cellcolor[rgb]{0.71, 0.71, 0.71} 0.131\\ 
                                                &  \systemIII       & \cellcolor[rgb]{0.87, 0.87, 0.87} 0.091 & \cellcolor[rgb]{0.66, 0.66, 0.66} 0.141 & \cellcolor[rgb]{0.63, 0.63, 0.63} 0.146 & \cellcolor[rgb]{0.43, 0.43, 0.43} 0.197 & \cellcolor[rgb]{0.44, 0.44, 0.44} 0.195 & \cellcolor[rgb]{0.43, 0.43, 0.43} 0.197 & \cellcolor[rgb]{0.57, 0.57, 0.57} 0.160 & \cellcolor[rgb]{0.48, 0.48, 0.48} 0.184 & \cellcolor[rgb]{0.55, 0.55, 0.55} 0.165 & \cellcolor[rgb]{0.43, 0.43, 0.43} 0.198 & \cellcolor[rgb]{0.39, 0.39, 0.39} 0.210 & \cellcolor[rgb]{0.11, 0.11, 0.11} \textcolor[rgb]{0.6,0.6,0.6}{0.285} & \cellcolor[rgb]{0.49, 0.49, 0.49} 0.182 & \cellcolor[rgb]{0.42, 0.42, 0.42} 0.201 & \cellcolor[rgb]{0.43, 0.43, 0.43} 0.197 & \cellcolor[rgb]{0.43, 0.43, 0.43} 0.197 & \cellcolor[rgb]{0.32, 0.32, 0.32} 0.231 & \cellcolor[rgb]{0.34, 0.34, 0.34} 0.224 & \cellcolor[rgb]{0.63, 0.63, 0.63} 0.148 & \cellcolor[rgb]{0.63, 0.63, 0.63} 0.147 & \cellcolor[rgb]{0.59, 0.59, 0.59} 0.156 & \cellcolor[rgb]{0.56, 0.56, 0.56} 0.163 & \cellcolor[rgb]{0.53, 0.53, 0.53} 0.172 & \cellcolor[rgb]{0.48, 0.48, 0.48} 0.183\\ 
\midrule
 \multirow{3}{*}{\rotatebox[origin=c]{0}{SPEC}} &  \systemI         & \cellcolor[rgb]{0.72, 0.72, 0.72} 0.129 & \cellcolor[rgb]{0.71, 0.71, 0.71} 0.132 & \cellcolor[rgb]{0.71, 0.71, 0.71} 0.130 & \cellcolor[rgb]{0.70, 0.70, 0.70} 0.133 & \cellcolor[rgb]{0.66, 0.66, 0.66} 0.141 & \cellcolor[rgb]{0.58, 0.58, 0.58} 0.158 & \cellcolor[rgb]{0.75, 0.75, 0.75} 0.123 & \cellcolor[rgb]{0.75, 0.75, 0.75} 0.122 & \cellcolor[rgb]{0.75, 0.75, 0.75} 0.122 & \cellcolor[rgb]{0.71, 0.71, 0.71} 0.130 & \cellcolor[rgb]{0.70, 0.70, 0.70} 0.134 & \cellcolor[rgb]{0.58, 0.58, 0.58} 0.158 & \cellcolor[rgb]{0.89, 0.89, 0.89} 0.085 & \cellcolor[rgb]{0.84, 0.84, 0.84} 0.099 & \cellcolor[rgb]{0.75, 0.75, 0.75} 0.123 & \cellcolor[rgb]{0.71, 0.71, 0.71} 0.132 & \cellcolor[rgb]{0.68, 0.68, 0.68} 0.137 & \cellcolor[rgb]{0.62, 0.62, 0.62} 0.148 & \cellcolor[rgb]{0.89, 0.89, 0.89} 0.085 & \cellcolor[rgb]{0.88, 0.88, 0.88} 0.088 & \cellcolor[rgb]{0.85, 0.85, 0.85} 0.095 & \cellcolor[rgb]{0.81, 0.81, 0.81} 0.106 & \cellcolor[rgb]{0.73, 0.73, 0.73} 0.126 & \cellcolor[rgb]{0.73, 0.73, 0.73} 0.127\\ 
                                                &  \systemII        & \cellcolor[rgb]{0.76, 0.76, 0.76} 0.119 & \cellcolor[rgb]{0.77, 0.77, 0.77} 0.117 & \cellcolor[rgb]{0.71, 0.71, 0.71} 0.131 & \cellcolor[rgb]{0.60, 0.60, 0.60} 0.153 & \cellcolor[rgb]{0.67, 0.67, 0.67} 0.139 & \cellcolor[rgb]{0.59, 0.59, 0.59} 0.156 & \cellcolor[rgb]{0.82, 0.82, 0.82} 0.105 & \cellcolor[rgb]{0.73, 0.73, 0.73} 0.126 & \cellcolor[rgb]{0.66, 0.66, 0.66} 0.141 & \cellcolor[rgb]{0.60, 0.60, 0.60} 0.153 & \cellcolor[rgb]{0.56, 0.56, 0.56} 0.162 & \cellcolor[rgb]{0.44, 0.44, 0.44} 0.194 & \cellcolor[rgb]{0.80, 0.80, 0.80} 0.109 & \cellcolor[rgb]{0.80, 0.80, 0.80} 0.109 & \cellcolor[rgb]{0.77, 0.77, 0.77} 0.115 & \cellcolor[rgb]{0.67, 0.67, 0.67} 0.140 & \cellcolor[rgb]{0.66, 0.66, 0.66} 0.141 & \cellcolor[rgb]{0.52, 0.52, 0.52} 0.173 & \cellcolor[rgb]{0.69, 0.69, 0.69} 0.135 & \cellcolor[rgb]{0.74, 0.74, 0.74} 0.125 & \cellcolor[rgb]{0.65, 0.65, 0.65} 0.143 & \cellcolor[rgb]{0.72, 0.72, 0.72} 0.129 & \cellcolor[rgb]{0.61, 0.61, 0.61} 0.153 & \cellcolor[rgb]{0.48, 0.48, 0.48} 0.183\\ 
                                                &  \systemIII       & \cellcolor[rgb]{0.81, 0.81, 0.81} 0.106 & \cellcolor[rgb]{0.83, 0.83, 0.83} 0.102 & \cellcolor[rgb]{0.77, 0.77, 0.77} 0.116 & \cellcolor[rgb]{0.76, 0.76, 0.76} 0.118 & \cellcolor[rgb]{0.69, 0.69, 0.69} 0.136 & \cellcolor[rgb]{0.70, 0.70, 0.70} 0.134 & \cellcolor[rgb]{0.92, 0.92, 0.92} 0.077 & \cellcolor[rgb]{0.88, 0.88, 0.88} 0.088 & \cellcolor[rgb]{0.76, 0.76, 0.76} 0.118 & \cellcolor[rgb]{0.72, 0.72, 0.72} 0.130 & \cellcolor[rgb]{0.75, 0.75, 0.75} 0.121 & \cellcolor[rgb]{0.73, 0.73, 0.73} 0.126 & \cellcolor[rgb]{0.88, 0.88, 0.88} 0.087 & \cellcolor[rgb]{0.81, 0.81, 0.81} 0.105 & \cellcolor[rgb]{0.83, 0.83, 0.83} 0.102 & \cellcolor[rgb]{0.75, 0.75, 0.75} 0.122 & \cellcolor[rgb]{0.75, 0.75, 0.75} 0.122 & \cellcolor[rgb]{0.63, 0.63, 0.63} 0.146 & \cellcolor[rgb]{0.97, 0.97, 0.97} 0.060 & \cellcolor[rgb]{0.92, 0.92, 0.92} 0.077 & \cellcolor[rgb]{0.91, 0.91, 0.91} 0.079 & \cellcolor[rgb]{0.85, 0.85, 0.85} 0.097 & \cellcolor[rgb]{0.84, 0.84, 0.84} 0.097 & \cellcolor[rgb]{0.77, 0.77, 0.77} 0.117\\ 
\midrule
 \multirow{3}{*}{\rotatebox[origin=c]{0}{LFCC}} &  \systemI         & \cellcolor[rgb]{0.95, 0.95, 0.95} 0.068 & \cellcolor[rgb]{0.90, 0.90, 0.90} 0.081 & \cellcolor[rgb]{0.90, 0.90, 0.90} 0.083 & \cellcolor[rgb]{0.86, 0.86, 0.86} 0.092 & \cellcolor[rgb]{0.87, 0.87, 0.87} 0.090 & \cellcolor[rgb]{0.76, 0.76, 0.76} 0.118 & \cellcolor[rgb]{0.95, 0.95, 0.95} 0.068 & \cellcolor[rgb]{0.95, 0.95, 0.95} 0.068 & \cellcolor[rgb]{0.94, 0.94, 0.94} 0.071 & \cellcolor[rgb]{0.88, 0.88, 0.88} 0.087 & \cellcolor[rgb]{0.82, 0.82, 0.82} 0.103 & \cellcolor[rgb]{0.84, 0.84, 0.84} 0.099 & \cellcolor[rgb]{0.95, 0.95, 0.95} 0.069 & \cellcolor[rgb]{0.95, 0.95, 0.95} 0.069 & \cellcolor[rgb]{0.94, 0.94, 0.94} 0.071 & \cellcolor[rgb]{0.93, 0.93, 0.93} 0.074 & \cellcolor[rgb]{0.93, 0.93, 0.93} 0.074 & \cellcolor[rgb]{0.84, 0.84, 0.84} 0.099 & \cellcolor[rgb]{0.98, 0.98, 0.98} 0.056 & \cellcolor[rgb]{0.97, 0.97, 0.97} 0.061 & \cellcolor[rgb]{0.95, 0.95, 0.95} 0.068 & \cellcolor[rgb]{0.96, 0.96, 0.96} 0.065 & \cellcolor[rgb]{0.93, 0.93, 0.93} 0.074 & \cellcolor[rgb]{0.93, 0.93, 0.93} 0.075\\ 
                                                &  \systemII        & \cellcolor[rgb]{0.93, 0.93, 0.93} 0.074 & \cellcolor[rgb]{0.93, 0.93, 0.93} 0.073 & \cellcolor[rgb]{0.94, 0.94, 0.94} 0.071 & \cellcolor[rgb]{0.92, 0.92, 0.92} 0.077 & \cellcolor[rgb]{0.94, 0.94, 0.94} 0.072 & \cellcolor[rgb]{0.93, 0.93, 0.93} 0.074 & \cellcolor[rgb]{0.96, 0.96, 0.96} 0.066 & \cellcolor[rgb]{0.93, 0.93, 0.93} 0.074 & \cellcolor[rgb]{0.93, 0.93, 0.93} 0.074 & \cellcolor[rgb]{0.92, 0.92, 0.92} 0.076 & \cellcolor[rgb]{0.89, 0.89, 0.89} 0.084 & \cellcolor[rgb]{0.89, 0.89, 0.89} 0.084 & \cellcolor[rgb]{0.95, 0.95, 0.95} 0.068 & \cellcolor[rgb]{0.90, 0.90, 0.90} 0.081 & \cellcolor[rgb]{0.92, 0.92, 0.92} 0.076 & \cellcolor[rgb]{0.92, 0.92, 0.92} 0.076 & \cellcolor[rgb]{0.88, 0.88, 0.88} 0.086 & \cellcolor[rgb]{0.83, 0.83, 0.83} 0.101 & \cellcolor[rgb]{0.91, 0.91, 0.91} 0.079 & \cellcolor[rgb]{0.91, 0.91, 0.91} 0.080 & \cellcolor[rgb]{0.91, 0.91, 0.91} 0.080 & \cellcolor[rgb]{0.91, 0.91, 0.91} 0.079 & \cellcolor[rgb]{0.90, 0.90, 0.90} 0.082 & \cellcolor[rgb]{0.87, 0.87, 0.87} 0.091\\ 
                                                &  \systemIII       & \cellcolor[rgb]{0.98, 0.98, 0.98} 0.057 & \cellcolor[rgb]{0.95, 0.95, 0.95} 0.068 & \cellcolor[rgb]{0.91, 0.91, 0.91} 0.078 & \cellcolor[rgb]{0.92, 0.92, 0.92} 0.077 & \cellcolor[rgb]{0.91, 0.91, 0.91} 0.079 & \cellcolor[rgb]{0.95, 0.95, 0.95} 0.068 & \cellcolor[rgb]{0.96, 0.96, 0.96} 0.064 & \cellcolor[rgb]{0.96, 0.96, 0.96} 0.064 & \cellcolor[rgb]{0.94, 0.94, 0.94} 0.073 & \cellcolor[rgb]{0.92, 0.92, 0.92} 0.076 & \cellcolor[rgb]{0.92, 0.92, 0.92} 0.076 & \cellcolor[rgb]{0.73, 0.73, 0.73} 0.127 & \cellcolor[rgb]{0.96, 0.96, 0.96} 0.064 & \cellcolor[rgb]{0.96, 0.96, 0.96} 0.063 & \cellcolor[rgb]{0.95, 0.95, 0.95} 0.066 & \cellcolor[rgb]{0.90, 0.90, 0.90} 0.083 & \cellcolor[rgb]{0.94, 0.94, 0.94} 0.073 & \cellcolor[rgb]{0.92, 0.92, 0.92} 0.075 & \cellcolor[rgb]{1.00, 1.00, 1.00} 0.052 & \cellcolor[rgb]{0.98, 0.98, 0.98} 0.058 & \cellcolor[rgb]{0.95, 0.95, 0.95} 0.067 & \cellcolor[rgb]{0.99, 0.99, 0.99} 0.055 & \cellcolor[rgb]{0.95, 0.95, 0.95} 0.067 & \cellcolor[rgb]{0.93, 0.93, 0.93} 0.075\\
\midrule
&  \systemIV        & &       &       &       &       &       &       &       &       &       &       &       & \cellcolor[rgb]{0.89, 0.89, 0.89} 0.103 & \cellcolor[rgb]{0.86, 0.86, 0.86} 0.114 & \cellcolor[rgb]{0.84, 0.84, 0.84} 0.123 & \cellcolor[rgb]{0.71, 0.71, 0.71} 0.176 & \cellcolor[rgb]{0.80, 0.80, 0.80} 0.142 & \cellcolor[rgb]{0.73, 0.73, 0.73} 0.167 &       &       &       &       &       &  \\
  \bottomrule \\
  \multicolumn{26}{c}{min t-DCF (version 2.0)}   \\
\toprule
\multirow{3}{*}{\rotatebox[origin=c]{0}{LFB}}  &  \systemI         & \cellcolor[rgb]{0.83, 0.83, 0.83} 0.183 & \cellcolor[rgb]{0.73, 0.73, 0.73} 0.220 & \cellcolor[rgb]{0.62, 0.62, 0.62} 0.254 & \cellcolor[rgb]{0.60, 0.60, 0.60} 0.261 & \cellcolor[rgb]{0.54, 0.54, 0.54} 0.283 & \cellcolor[rgb]{0.54, 0.54, 0.54} 0.284 & \cellcolor[rgb]{0.76, 0.76, 0.76} 0.209 & \cellcolor[rgb]{0.70, 0.70, 0.70} 0.230 & \cellcolor[rgb]{0.72, 0.72, 0.72} 0.225 & \cellcolor[rgb]{0.72, 0.72, 0.72} 0.223 & \cellcolor[rgb]{0.63, 0.63, 0.63} 0.250 & \cellcolor[rgb]{0.52, 0.52, 0.52} 0.294 & \cellcolor[rgb]{0.73, 0.73, 0.73} 0.219 & \cellcolor[rgb]{0.71, 0.71, 0.71} 0.227 & \cellcolor[rgb]{0.65, 0.65, 0.65} 0.246 & \cellcolor[rgb]{0.58, 0.58, 0.58} 0.271 & \cellcolor[rgb]{0.56, 0.56, 0.56} 0.278 & \cellcolor[rgb]{0.56, 0.56, 0.56} 0.275 & \cellcolor[rgb]{0.72, 0.72, 0.72} 0.223 & \cellcolor[rgb]{0.60, 0.60, 0.60} 0.262 & \cellcolor[rgb]{0.63, 0.63, 0.63} 0.250 & \cellcolor[rgb]{0.66, 0.66, 0.66} 0.241 & \cellcolor[rgb]{0.59, 0.59, 0.59} 0.267 & \cellcolor[rgb]{0.60, 0.60, 0.60} 0.264\\ 
                                               &  \systemII        & \cellcolor[rgb]{0.85, 0.85, 0.85} 0.173 & \cellcolor[rgb]{0.86, 0.86, 0.86} 0.172 & \cellcolor[rgb]{0.91, 0.91, 0.91} 0.149 & \cellcolor[rgb]{0.89, 0.89, 0.89} 0.157 & \cellcolor[rgb]{0.62, 0.62, 0.62} 0.253 & \cellcolor[rgb]{0.37, 0.37, 0.37} 0.373 & \cellcolor[rgb]{0.87, 0.87, 0.87} 0.167 & \cellcolor[rgb]{0.85, 0.85, 0.85} 0.176 & \cellcolor[rgb]{0.84, 0.84, 0.84} 0.177 & \cellcolor[rgb]{0.78, 0.78, 0.78} 0.201 & \cellcolor[rgb]{0.76, 0.76, 0.76} 0.208 & \cellcolor[rgb]{0.75, 0.75, 0.75} 0.213 & \cellcolor[rgb]{0.97, 0.97, 0.97} 0.123 & \cellcolor[rgb]{0.93, 0.93, 0.93} 0.144 & \cellcolor[rgb]{0.83, 0.83, 0.83} 0.181 & \cellcolor[rgb]{0.92, 0.92, 0.92} 0.145 & \cellcolor[rgb]{0.88, 0.88, 0.88} 0.164 & \cellcolor[rgb]{0.91, 0.91, 0.91} 0.149 & \cellcolor[rgb]{0.92, 0.92, 0.92} 0.146 & \cellcolor[rgb]{0.87, 0.87, 0.87} 0.166 & \cellcolor[rgb]{0.84, 0.84, 0.84} 0.179 & \cellcolor[rgb]{0.85, 0.85, 0.85} 0.175 & \cellcolor[rgb]{0.84, 0.84, 0.84} 0.178 & \cellcolor[rgb]{0.82, 0.82, 0.82} 0.186\\ 
                                               &  \systemIII       & \cellcolor[rgb]{0.92, 0.92, 0.92} 0.148 & \cellcolor[rgb]{0.80, 0.80, 0.80} 0.195 & \cellcolor[rgb]{0.78, 0.78, 0.78} 0.200 & \cellcolor[rgb]{0.64, 0.64, 0.64} 0.248 & \cellcolor[rgb]{0.65, 0.65, 0.65} 0.246 & \cellcolor[rgb]{0.64, 0.64, 0.64} 0.247 & \cellcolor[rgb]{0.75, 0.75, 0.75} 0.213 & \cellcolor[rgb]{0.68, 0.68, 0.68} 0.235 & \cellcolor[rgb]{0.74, 0.74, 0.74} 0.217 & \cellcolor[rgb]{0.64, 0.64, 0.64} 0.248 & \cellcolor[rgb]{0.61, 0.61, 0.61} 0.260 & \cellcolor[rgb]{0.44, 0.44, 0.44} 0.330 & \cellcolor[rgb]{0.69, 0.69, 0.69} 0.233 & \cellcolor[rgb]{0.63, 0.63, 0.63} 0.251 & \cellcolor[rgb]{0.64, 0.64, 0.64} 0.247 & \cellcolor[rgb]{0.64, 0.64, 0.64} 0.248 & \cellcolor[rgb]{0.56, 0.56, 0.56} 0.279 & \cellcolor[rgb]{0.57, 0.57, 0.57} 0.273 & \cellcolor[rgb]{0.78, 0.78, 0.78} 0.201 & \cellcolor[rgb]{0.78, 0.78, 0.78} 0.200 & \cellcolor[rgb]{0.76, 0.76, 0.76} 0.209 & \cellcolor[rgb]{0.74, 0.74, 0.74} 0.216 & \cellcolor[rgb]{0.72, 0.72, 0.72} 0.224 & \cellcolor[rgb]{0.68, 0.68, 0.68} 0.234\\ 
\midrule
\multirow{3}{*}{\rotatebox[origin=c]{0}{SPEC}} &  \systemI         & \cellcolor[rgb]{0.83, 0.83, 0.83} 0.183 & \cellcolor[rgb]{0.82, 0.82, 0.82} 0.186 & \cellcolor[rgb]{0.82, 0.82, 0.82} 0.185 & \cellcolor[rgb]{0.82, 0.82, 0.82} 0.187 & \cellcolor[rgb]{0.80, 0.80, 0.80} 0.194 & \cellcolor[rgb]{0.76, 0.76, 0.76} 0.211 & \cellcolor[rgb]{0.84, 0.84, 0.84} 0.178 & \cellcolor[rgb]{0.84, 0.84, 0.84} 0.177 & \cellcolor[rgb]{0.84, 0.84, 0.84} 0.177 & \cellcolor[rgb]{0.82, 0.82, 0.82} 0.185 & \cellcolor[rgb]{0.82, 0.82, 0.82} 0.188 & \cellcolor[rgb]{0.76, 0.76, 0.76} 0.211 & \cellcolor[rgb]{0.93, 0.93, 0.93} 0.142 & \cellcolor[rgb]{0.90, 0.90, 0.90} 0.155 & \cellcolor[rgb]{0.84, 0.84, 0.84} 0.178 & \cellcolor[rgb]{0.82, 0.82, 0.82} 0.186 & \cellcolor[rgb]{0.81, 0.81, 0.81} 0.192 & \cellcolor[rgb]{0.78, 0.78, 0.78} 0.202 & \cellcolor[rgb]{0.93, 0.93, 0.93} 0.142 & \cellcolor[rgb]{0.92, 0.92, 0.92} 0.145 & \cellcolor[rgb]{0.91, 0.91, 0.91} 0.152 & \cellcolor[rgb]{0.88, 0.88, 0.88} 0.162 & \cellcolor[rgb]{0.83, 0.83, 0.83} 0.181 & \cellcolor[rgb]{0.83, 0.83, 0.83} 0.182\\ 
                                               &  \systemII        & \cellcolor[rgb]{0.85, 0.85, 0.85} 0.175 & \cellcolor[rgb]{0.86, 0.86, 0.86} 0.173 & \cellcolor[rgb]{0.82, 0.82, 0.82} 0.185 & \cellcolor[rgb]{0.77, 0.77, 0.77} 0.206 & \cellcolor[rgb]{0.80, 0.80, 0.80} 0.193 & \cellcolor[rgb]{0.76, 0.76, 0.76} 0.209 & \cellcolor[rgb]{0.88, 0.88, 0.88} 0.161 & \cellcolor[rgb]{0.84, 0.84, 0.84} 0.180 & \cellcolor[rgb]{0.80, 0.80, 0.80} 0.195 & \cellcolor[rgb]{0.77, 0.77, 0.77} 0.207 & \cellcolor[rgb]{0.75, 0.75, 0.75} 0.215 & \cellcolor[rgb]{0.65, 0.65, 0.65} 0.245 & \cellcolor[rgb]{0.87, 0.87, 0.87} 0.165 & \cellcolor[rgb]{0.87, 0.87, 0.87} 0.165 & \cellcolor[rgb]{0.86, 0.86, 0.86} 0.171 & \cellcolor[rgb]{0.80, 0.80, 0.80} 0.194 & \cellcolor[rgb]{0.80, 0.80, 0.80} 0.195 & \cellcolor[rgb]{0.72, 0.72, 0.72} 0.225 & \cellcolor[rgb]{0.81, 0.81, 0.81} 0.190 & \cellcolor[rgb]{0.84, 0.84, 0.84} 0.179 & \cellcolor[rgb]{0.79, 0.79, 0.79} 0.197 & \cellcolor[rgb]{0.83, 0.83, 0.83} 0.183 & \cellcolor[rgb]{0.77, 0.77, 0.77} 0.206 & \cellcolor[rgb]{0.69, 0.69, 0.69} 0.234\\ 
                                               &  \systemIII       & \cellcolor[rgb]{0.88, 0.88, 0.88} 0.162 & \cellcolor[rgb]{0.89, 0.89, 0.89} 0.158 & \cellcolor[rgb]{0.86, 0.86, 0.86} 0.172 & \cellcolor[rgb]{0.85, 0.85, 0.85} 0.174 & \cellcolor[rgb]{0.81, 0.81, 0.81} 0.190 & \cellcolor[rgb]{0.82, 0.82, 0.82} 0.188 & \cellcolor[rgb]{0.95, 0.95, 0.95} 0.135 & \cellcolor[rgb]{0.92, 0.92, 0.92} 0.146 & \cellcolor[rgb]{0.85, 0.85, 0.85} 0.174 & \cellcolor[rgb]{0.83, 0.83, 0.83} 0.184 & \cellcolor[rgb]{0.85, 0.85, 0.85} 0.177 & \cellcolor[rgb]{0.84, 0.84, 0.84} 0.180 & \cellcolor[rgb]{0.93, 0.93, 0.93} 0.144 & \cellcolor[rgb]{0.88, 0.88, 0.88} 0.161 & \cellcolor[rgb]{0.89, 0.89, 0.89} 0.158 & \cellcolor[rgb]{0.84, 0.84, 0.84} 0.177 & \cellcolor[rgb]{0.84, 0.84, 0.84} 0.177 & \cellcolor[rgb]{0.78, 0.78, 0.78} 0.200 & \cellcolor[rgb]{0.98, 0.98, 0.98} 0.119 & \cellcolor[rgb]{0.95, 0.95, 0.95} 0.135 & \cellcolor[rgb]{0.94, 0.94, 0.94} 0.137 & \cellcolor[rgb]{0.90, 0.90, 0.90} 0.154 & \cellcolor[rgb]{0.90, 0.90, 0.90} 0.154 & \cellcolor[rgb]{0.86, 0.86, 0.86} 0.172\\ 
\midrule
\multirow{3}{*}{\rotatebox[origin=c]{0}{LFCC}} &  \systemI         & \cellcolor[rgb]{0.96, 0.96, 0.96} 0.126 & \cellcolor[rgb]{0.94, 0.94, 0.94} 0.138 & \cellcolor[rgb]{0.94, 0.94, 0.94} 0.141 & \cellcolor[rgb]{0.92, 0.92, 0.92} 0.149 & \cellcolor[rgb]{0.92, 0.92, 0.92} 0.147 & \cellcolor[rgb]{0.85, 0.85, 0.85} 0.174 & \cellcolor[rgb]{0.96, 0.96, 0.96} 0.127 & \cellcolor[rgb]{0.96, 0.96, 0.96} 0.127 & \cellcolor[rgb]{0.96, 0.96, 0.96} 0.130 & \cellcolor[rgb]{0.93, 0.93, 0.93} 0.144 & \cellcolor[rgb]{0.89, 0.89, 0.89} 0.159 & \cellcolor[rgb]{0.90, 0.90, 0.90} 0.156 & \cellcolor[rgb]{0.96, 0.96, 0.96} 0.127 & \cellcolor[rgb]{0.96, 0.96, 0.96} 0.127 & \cellcolor[rgb]{0.96, 0.96, 0.96} 0.129 & \cellcolor[rgb]{0.95, 0.95, 0.95} 0.132 & \cellcolor[rgb]{0.95, 0.95, 0.95} 0.132 & \cellcolor[rgb]{0.90, 0.90, 0.90} 0.156 & \cellcolor[rgb]{0.99, 0.99, 0.99} 0.116 & \cellcolor[rgb]{0.98, 0.98, 0.98} 0.120 & \cellcolor[rgb]{0.96, 0.96, 0.96} 0.126 & \cellcolor[rgb]{0.97, 0.97, 0.97} 0.124 & \cellcolor[rgb]{0.95, 0.95, 0.95} 0.132 & \cellcolor[rgb]{0.95, 0.95, 0.95} 0.133\\ 
                                               &  \systemII        & \cellcolor[rgb]{0.95, 0.95, 0.95} 0.132 & \cellcolor[rgb]{0.96, 0.96, 0.96} 0.131 & \cellcolor[rgb]{0.96, 0.96, 0.96} 0.129 & \cellcolor[rgb]{0.95, 0.95, 0.95} 0.135 & \cellcolor[rgb]{0.96, 0.96, 0.96} 0.130 & \cellcolor[rgb]{0.95, 0.95, 0.95} 0.132 & \cellcolor[rgb]{0.97, 0.97, 0.97} 0.124 & \cellcolor[rgb]{0.95, 0.95, 0.95} 0.132 & \cellcolor[rgb]{0.95, 0.95, 0.95} 0.132 & \cellcolor[rgb]{0.95, 0.95, 0.95} 0.134 & \cellcolor[rgb]{0.93, 0.93, 0.93} 0.142 & \cellcolor[rgb]{0.93, 0.93, 0.93} 0.141 & \cellcolor[rgb]{0.96, 0.96, 0.96} 0.127 & \cellcolor[rgb]{0.94, 0.94, 0.94} 0.139 & \cellcolor[rgb]{0.95, 0.95, 0.95} 0.134 & \cellcolor[rgb]{0.95, 0.95, 0.95} 0.134 & \cellcolor[rgb]{0.93, 0.93, 0.93} 0.143 & \cellcolor[rgb]{0.89, 0.89, 0.89} 0.157 & \cellcolor[rgb]{0.94, 0.94, 0.94} 0.137 & \cellcolor[rgb]{0.94, 0.94, 0.94} 0.137 & \cellcolor[rgb]{0.94, 0.94, 0.94} 0.137 & \cellcolor[rgb]{0.94, 0.94, 0.94} 0.137 & \cellcolor[rgb]{0.94, 0.94, 0.94} 0.139 & \cellcolor[rgb]{0.92, 0.92, 0.92} 0.148\\ 
                                               &  \systemIII       & \cellcolor[rgb]{0.99, 0.99, 0.99} 0.116 & \cellcolor[rgb]{0.96, 0.96, 0.96} 0.127 & \cellcolor[rgb]{0.95, 0.95, 0.95} 0.136 & \cellcolor[rgb]{0.95, 0.95, 0.95} 0.135 & \cellcolor[rgb]{0.94, 0.94, 0.94} 0.137 & \cellcolor[rgb]{0.96, 0.96, 0.96} 0.126 & \cellcolor[rgb]{0.97, 0.97, 0.97} 0.122 & \cellcolor[rgb]{0.97, 0.97, 0.97} 0.123 & \cellcolor[rgb]{0.96, 0.96, 0.96} 0.131 & \cellcolor[rgb]{0.95, 0.95, 0.95} 0.134 & \cellcolor[rgb]{0.95, 0.95, 0.95} 0.134 & \cellcolor[rgb]{0.83, 0.83, 0.83} 0.182 & \cellcolor[rgb]{0.97, 0.97, 0.97} 0.123 & \cellcolor[rgb]{0.97, 0.97, 0.97} 0.122 & \cellcolor[rgb]{0.97, 0.97, 0.97} 0.125 & \cellcolor[rgb]{0.94, 0.94, 0.94} 0.141 & \cellcolor[rgb]{0.96, 0.96, 0.96} 0.131 & \cellcolor[rgb]{0.95, 0.95, 0.95} 0.133 & \cellcolor[rgb]{1.00, 1.00, 1.00} 0.112 & \cellcolor[rgb]{0.99, 0.99, 0.99} 0.117 & \cellcolor[rgb]{0.96, 0.96, 0.96} 0.126 & \cellcolor[rgb]{0.99, 0.99, 0.99} 0.115 & \cellcolor[rgb]{0.97, 0.97, 0.97} 0.125 & \cellcolor[rgb]{0.95, 0.95, 0.95} 0.133\\ 
\midrule
&  \systemIV        & &       &       &       &       &       &       &       &       &       &       &       & \cellcolor[rgb]{0.89, 0.89, 0.89} 0.159 & \cellcolor[rgb]{0.86, 0.86, 0.86} 0.170 & \cellcolor[rgb]{0.84, 0.84, 0.84} 0.178 & \cellcolor[rgb]{0.71, 0.71, 0.71} 0.228 & \cellcolor[rgb]{0.80, 0.80, 0.80} 0.196 & \cellcolor[rgb]{0.73, 0.73, 0.73} 0.219 &       &       &       &       &       &      \\ 
\bottomrule
\end{tabular}
}
\end{center}
\vspace{-6mm}
\label{tab:eer}
\end{table*}%
\end{landscape}

\subsubsection{On the Training Criteria} 
AM-softmax achieved good performance in the best run, but their performance severely degraded in the worst run. For example, the best run of LFB-\systemII\ with AM-softmax achieved an EER of 4.26\%, but the worst run was the highest among all the results (14.86\%). Similar trends can be observed from OC-softmax. In contrast, results from vanilla softmax seem to be less varied. The LFB-\systemII\ with vanilla softmax achieved EERs between 3.34\% and 5.93\%. 

Experiments here used the hyper-parameter settings from another work \cite{zhang2020one} for AM-softmax and OC-softmax. These hyper-parameter settings were carefully selected by the original authors. It is possible that both criteria could achieve better performance if they are further tuned for the back end used in this experiment. However, searching for better hyper-parameter settings is laborious. The hyper-parameter-free vanilla softmax and P2SGrad may be preferred in practice. 

\subsubsection{On the Back End} 
The results on LCNNs with the trim-and-pad strategy (\systemI) are comparable with those using average pooling (\systemIII) and attention (\systemII). Note that all three types of back ends are based on LCNN. They differ mainly in the manner of merging the hidden features. Therefore, the results suggest that even the LCNN designed for 2D image processing can be easily used to process 1D speech data. It is not necessary to force the input to have a unified length. 

Many models use the trim-and-pad strategy mainly for practical reasons. As the results suggested, the average pooling is simple and effective in merging the varied-length input into a single vector. Furthermore, the customized mini-batch creation method in Section~\ref{sec:exp-recipe} is also easy to implement. It is straightforward to switch to the  length-agnostic back end. 

Another disadvantage of the trim-and-pad strategy is the large model size. Models using \systemI\ had more than $860$k parameters, and the single fully-connected layer after the flattening operation took $710$k parameters. This single layer is dense and large because it is required to convert the flattened hidden feature vector into a small vector, i.e., 
$\mathbb{R}^{\frac{N}{Q}{D}'}\overset{\text{DNN}}{\rightarrow}\mathbb{R}^{C}$ in Fig.~\ref{fig:back_ends}. Its parameter size was therefore proportional to the fixed length $N$ of the input acoustic features. Since this $N$ has to be large enough to cover sufficient frames of input acoustic features, the fully-connected layer becomes large. In contrast, the fully-connected layers in \systemII\ and \systemIII\ only conduct $\mathbb{R}^{D'}\rightarrow\mathbb{R}^{C}$. The hidden feature is agnostic to the input feature length because of the average or attention-based pooling layer. 
\systemII\ and \systemIII\ had roughly $190 \pm 30$k and $290 \pm 30$k parameters, respectively. They are much smaller than those based on the trim-and-pad strategy.

\subsubsection{Combinations of Components} 
Among the possible combinations of the components used in the experiment, the models using LFCC and MSE-for-P2SGrad achieved EERs in the range between 1.92\% and 3.30\%. Combining LFCC and MSE-for-P2SGrad worked well regardless of which back end was used. This could be reasonable since the three types of back ends are all based on LCNN but with different pooling strategies. Among the three, the LFCC-\systemIII-p2s achieved the lowest EERs in all  six rounds.
In general, the LCNN with average pooling, the LFCC front end, and the MSE-for-P2SGrad are potential techniques for voice PAD on the ASVspoof2019 LA database.

\begin{figure}[t]
\centering
\includegraphics[width=\textwidth]{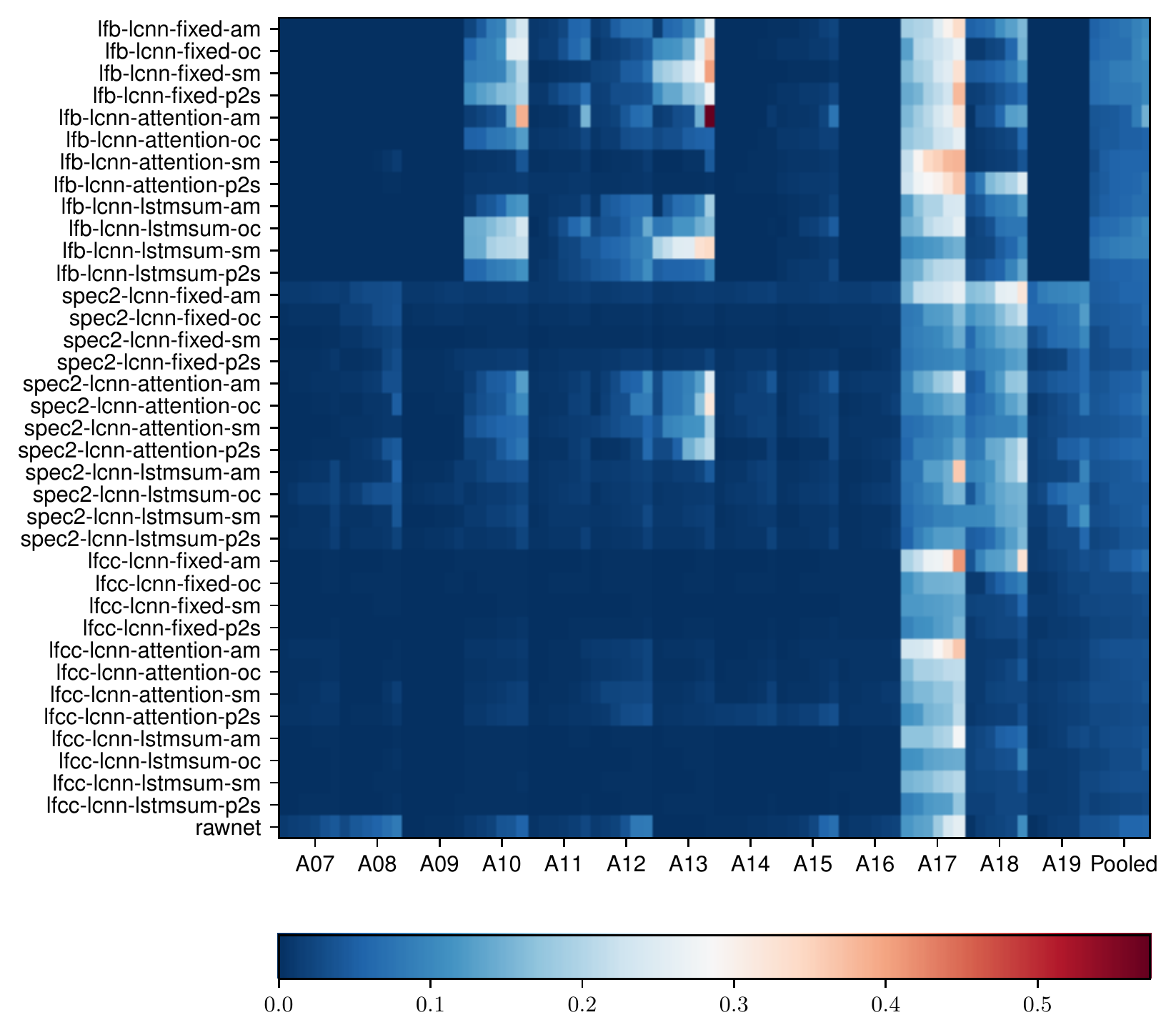}
\caption{Decomposed and pooled EER and mint t-DCF values. Each row has $14\times 6$ columns, where $14$ denotes $13$ attacks plus one pooled result, and where $6$ denotes the six training-evaluation rounds.}
\label{fig:decomposed_eer}
\end{figure}

\subsubsection{EER over Individual Attack} 
Fig.\ref{fig:decomposed_eer} plots the EERs of each model on each attacker in the evaluation set. Note that A16 and A19 are known attacks that also appear in the training and development sets, even though the trials are disjointed. Other details on the attackers can be found in \cite{WANG2020101114}.

A few messages can be observed from the decomposed EERs. First, A17 is difficult to detect for all the experimental models. This difficult attack has been known in the ASVspoof 2019 LA task \cite{Nautsch2021} and is expected. Second, the impact of the front end is observable. LFB-based models achieved worse performance in detecting A10 and A13. In contrast, the spectrogram-based models achieved higher EERs on A18. The LFCC-based models were generally better at detecting all attacks except A17.  
Third, among the LFCC-based models, those based on attentive pooling (\systemII) were worse than the others particularly on detecting A10 and A10. Similar trends can be observed in spectrogram-based models but not in LFB-based ones. 

Overall, the front end leaves a `finger print' in the decomposed EERs. In other words, models using different front ends have varied responses to the attackers in the database.

\begin{figure}[t]
\centering
\includegraphics[width=1.0\textwidth]{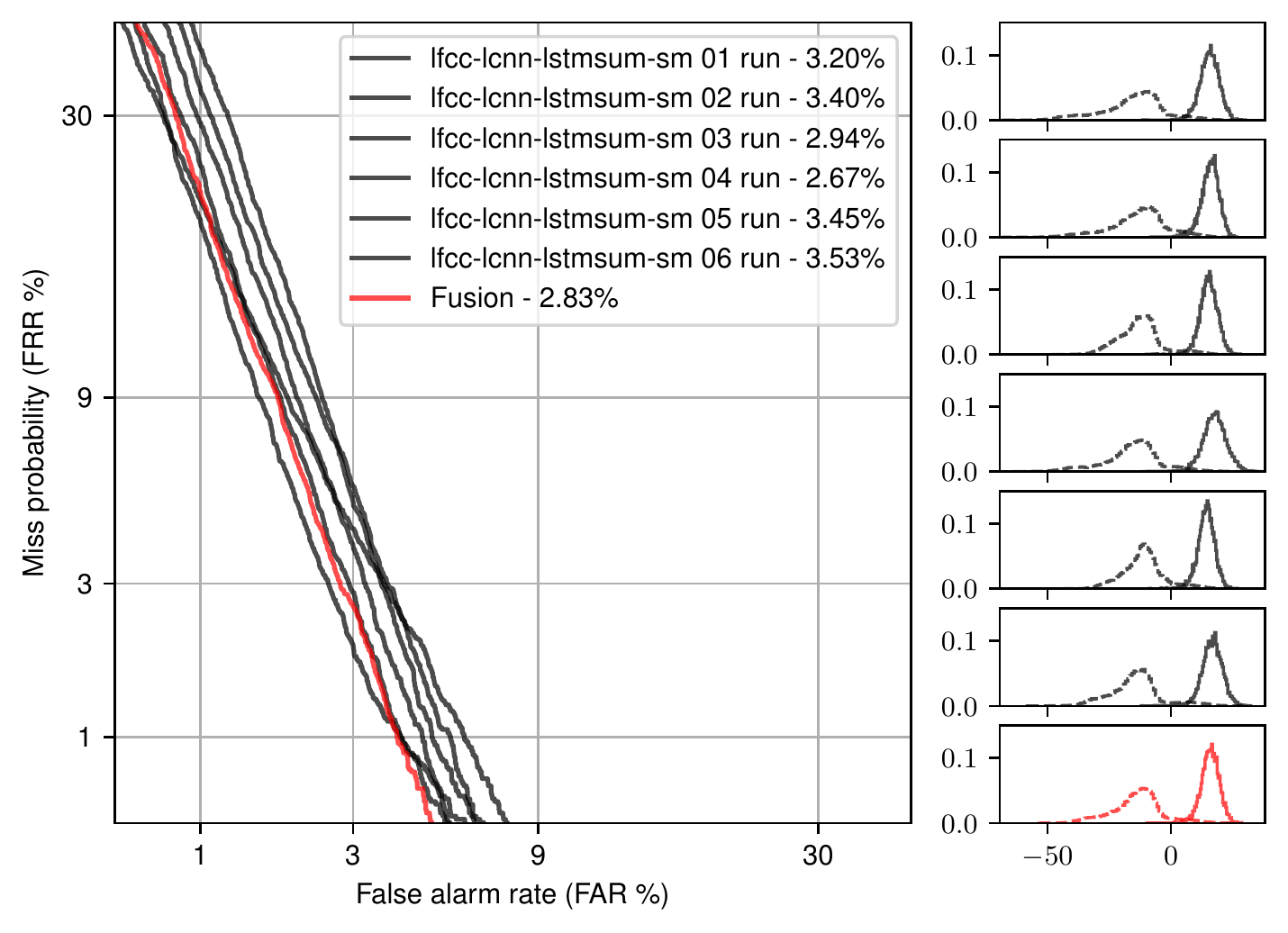}
\caption{DET curves (left) of \systemIII\ with LFCC and vanilla softmax (sm) in six training and evaluation rounds and their score distributions (right). Dashed and solid lines in the score distributions correspond to spoofed and bona fide trials, respectively.}
\label{fig:det_variance}
\end{figure}

\subsection{Results of Model Fusion}
Given the individual models discussed in the previous section, we explain in this section whether they can be fused into better models. We use the score average strategy, which requires no additional training process. 

\subsubsection{Fuse Multiple Runs of the Same Model}
We first attempted to fuse the individual models from the six rounds of training and evaluation. We chose the LFCC-\systemIII-sig because it achieved good performance in our experiment and is widely used in other studies. Fig.\ref{fig:det_variance} plots the DET curves of the six rounds and the fused result. Notice how the DET curves of different rounds are roughly parallel to each other. This indicates that the bona fide (and spoofed) trial score distributions are similar across the sub-models. In such a case, score averaging did not provide additional gains. 

Note that the models trained using different random initial states are different from the sub-models in a bagging-based model ensemble. In the latter, each sub-model is trained with a different sub-part of the training data. The sub-models in bagging-based ensemble are therefore more diverse.

\newpage
\clearpage
\begin{figure}[h]
      \begin{subfigure}[t]{\textwidth}
        \centering
        \includegraphics[width=1.0\textwidth]{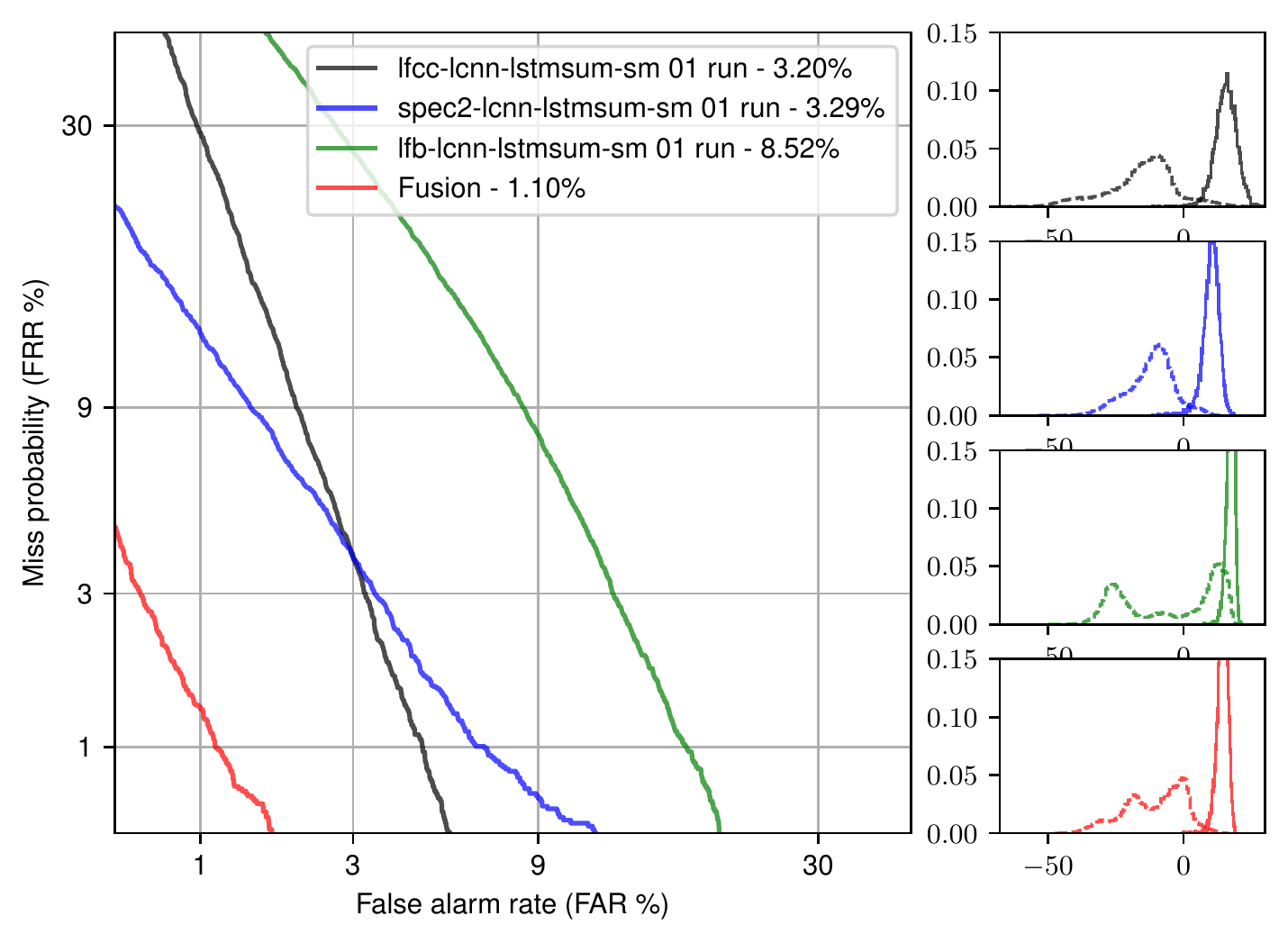}
        \caption{\systemIII\ with vanilla softmax (sm) and one of the three front ends.}
        \label{fig:det_fuse_front}
     \end{subfigure}
     \hfill
      \begin{subfigure}[t]{\textwidth}
        \centering
        \includegraphics[width=1.0\textwidth]{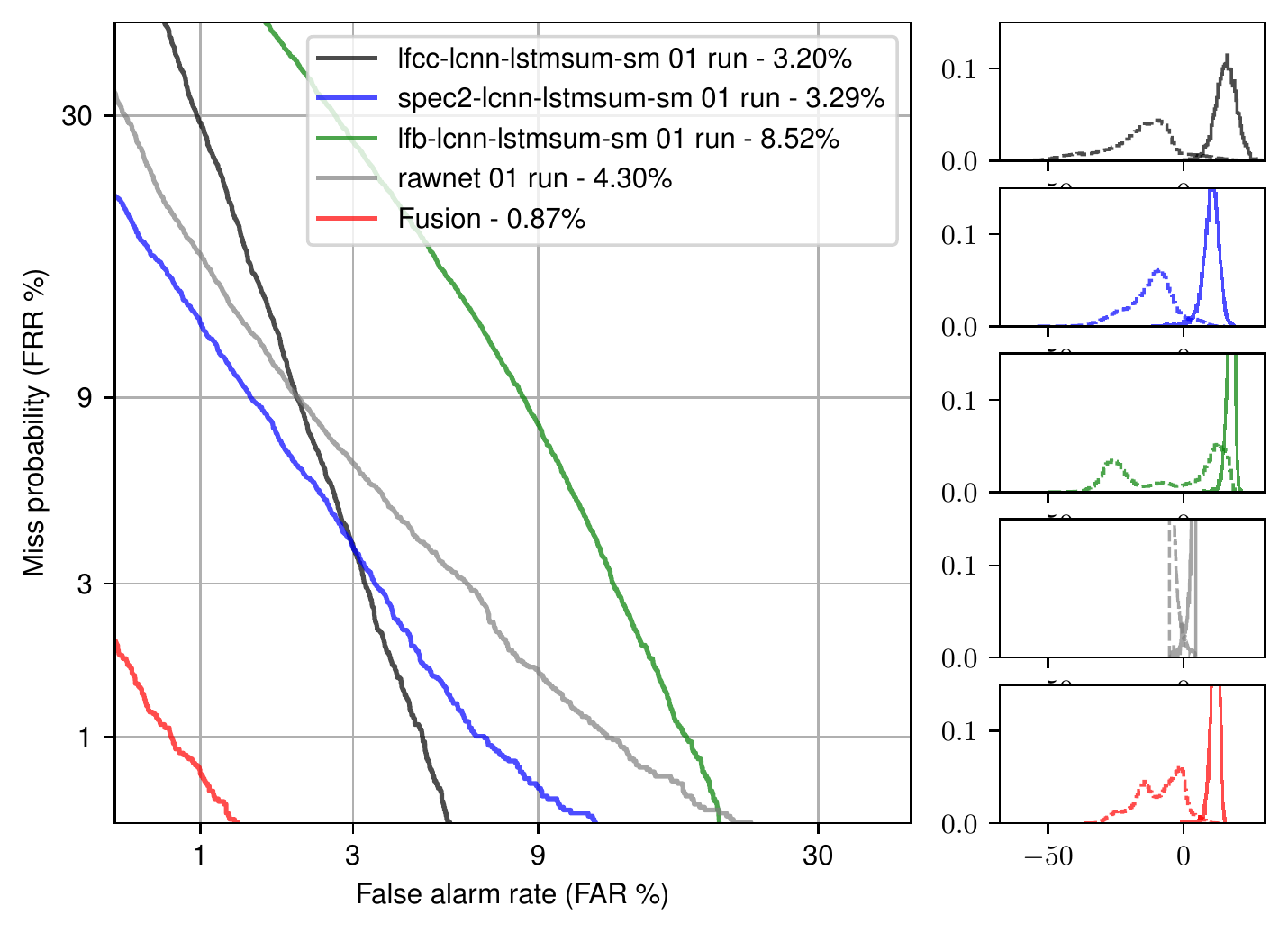}
        \caption{Same as Fig.\ref{fig:det_fuse_front} but with one RawNet model.}
        \label{fig:det_fuse_front_raw}
     \end{subfigure}
     \caption{DET curves (left) of single and fused models and score distributions (right). Only one training and evaluation round is included for each model. Dashed and solid lines in score distributions correspond to spoofed and bona fide trials, respectively.}
\end{figure}

\begin{figure}[h]
      \begin{subfigure}[t]{\textwidth}
        \centering
        \includegraphics[width=1.0\textwidth]{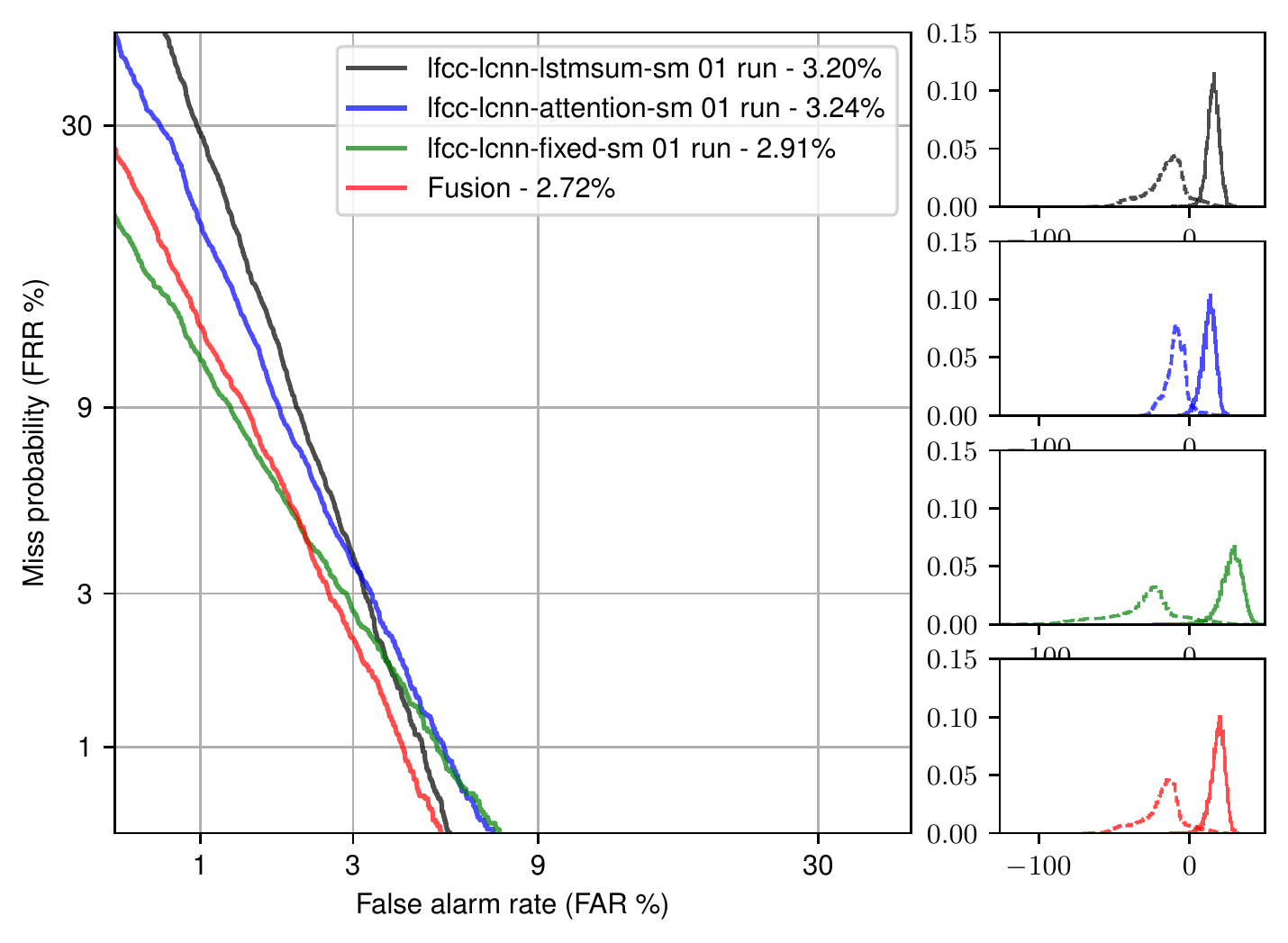}
        \caption{LFCC-LCNN with vanilla softmax (sm) and either attention, average pooling, or trim-and-pad strategy}
        \label{fig:det_fuse_pooling}
     \end{subfigure}
     \hfill
      \begin{subfigure}[t]{\textwidth}
        \centering
        \includegraphics[width=1.0\textwidth]{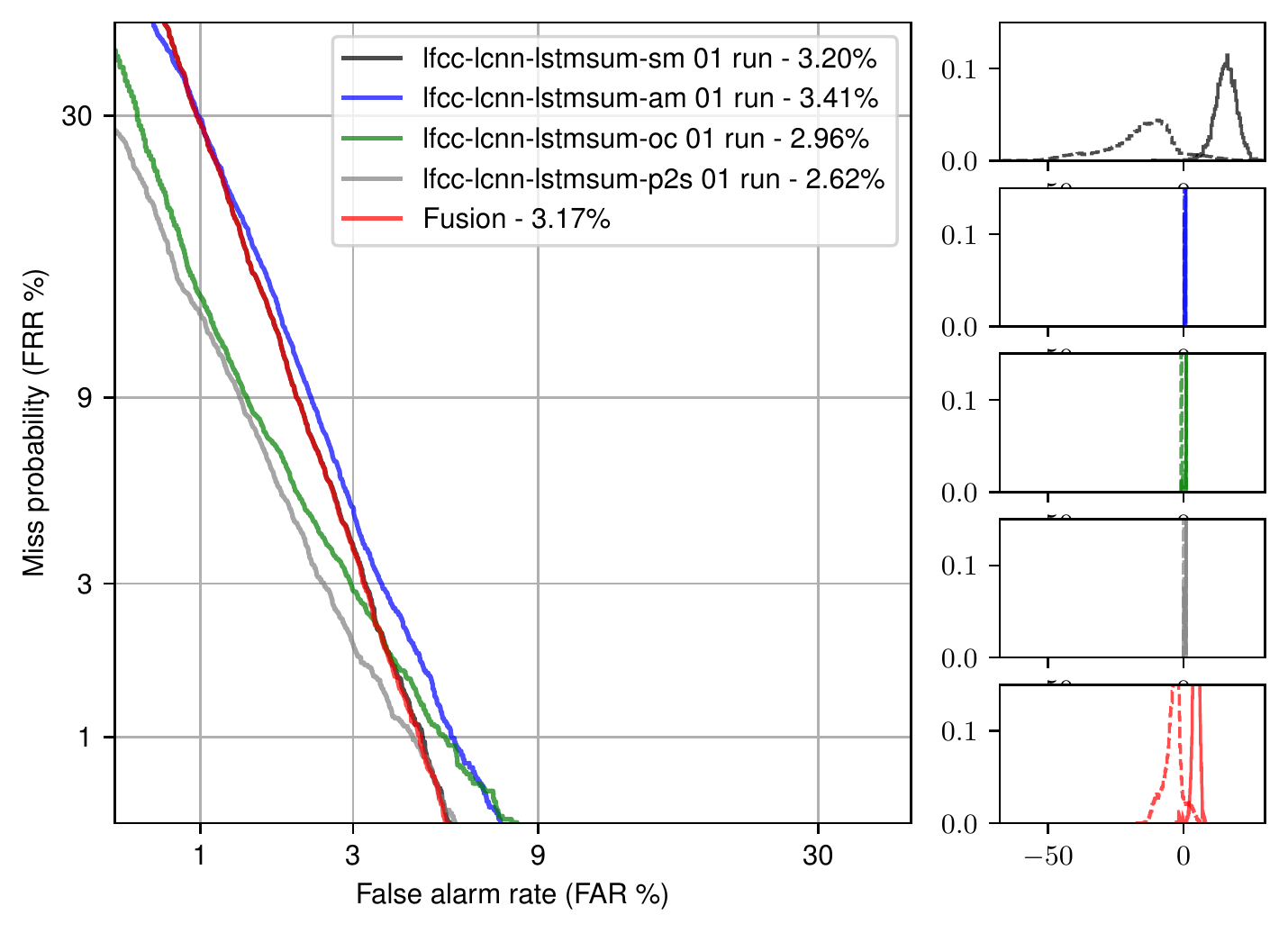}
        \caption{LFCC \systemIII\ and of the four training criteria.}
        \label{fig:det_fuse_criteria}
     \end{subfigure}
     \caption{DET curves (left) of single and fused models and score distributions (right). Only one training and evaluation round is included for each model. Dashed and solid lines in score distributions correspond to spoofed and bona fide trials, respectively.}
\end{figure}
\newpage
\clearpage

\subsubsection{Fuse Heterogeneous Models}
To create a better fused model, we need to fuse more heterogeneous models. On the front end, we fused only a single run of LFCC-\systemIII-sm, Spec-\systemIII-sm, and LFB-\systemIII-sm, which used the same random initial seed. The DET curves are illustrated in Fig.\ref{fig:det_fuse_front}. Interestingly, the fused model achieved an EER of 1.10\%, which not only outperformed the sub-models but also other single models in the experiment. The three sub-models have quite different DET curves, which is consistent with the decomposed EERs that have different patterns. Combining these heterogeneous sub-models is effective.

To further boost the performance, we added one RawNet2 model to the ensemble, and the resulting DET curve is illustrated in Fig.\ref{fig:det_fuse_front_raw}. The EER is further reduced to 0.87\%. Note that, although the fused model is based on score averaging, the EER is not a simple average of the sub-models' EERs. The score averaging changes the score distributions and affects the EER value. Accordingly, we cannot improve the performance by simply removing the worst sub-model. For example, if we remove the LFB-\systemIII-sm from the ensemble, the EER of the fused model decreased to 1.03\%. 

Using similar procedures, we explored the fusion of models with different back ends and training criteria. The results are plotted in Figs.\ref{fig:det_fuse_pooling} and \ref{fig:det_fuse_criteria}.
On the back end, the three sub-models are more or less similar to each other. Fusion over different training criteria is ineffective too. One possible reason is the difference of the score's dynamic range. The scores from margin-based softmax and P2Sgrad are cosine similarities between -1 and 1, while those based on vanilla softmax are not constrained. In this case, a simple score average is not the best strategy for model fusion. 

\subsection{Compared with Other PAD Models}
Given all the experimental results, we can compare them with other recent studies on the ASVspoof2019 LA task. Table \ref{tab:results_all} lists the results. Note that some of the studies used data augmentation or other methods to boost the amount of training data. This is marked in the table. 

The current best single model achieved an EER of 1.06\% \cite{tak2021end}, however, it is not reported how the result varied with different random initial seeds. This model introduced a new type of back end called a graph attention network. It directly works on the raw waveform. 

Among the fused models, the top entry in the ASVspoof2019 LA task is still the best. However, since that entry has not published a detailed system description, the implementation details remain unknown. The simple ensemble reported in Fig.\ref{fig:det_fuse_front_raw} worked quite well.

\begin{table}[t]
    \centering
    \caption{Comparison of recent PAD models on the ASVspoof 2019 LA task. There were 50 submissions to the task and many works after the challenge. Only a few works are listed, including the top single and primary models from the ASVspoof2019 challenge (2019LA top). Aug denotes data augmentation. The system in the last row used a slightly different evaluation set, and its EER is not directly comparable (marked with asterisk $^{*}$).}
    \setlength{\tabcolsep}{5pt}
    {\begin{tabular}{rrcccc}
    \multicolumn{6}{c}{Single model} \\
    \toprule
    \multirow{2}{*}{Ref.} & \multirow{2}{*}{Model} & \multirow{2}{*}{EER (\%)} & \multicolumn{2}{c}{min t-DCF} & \multirow{2}{*}{Aug.}\\
    \cline{4-5}
     & & & legacy & v2.0 & \\
    \midrule
    2019 LA top single & LFCC-LCNN (T45) & 5.06 & 0.1000 & 0.1562 & \\
    \cite{tak2020end} & RawNet2 (S1) & 5.64 & 0.1391 & - & \\
    \cite{Wu2020} & FG-LCNN  & 4.07 & 0.102 & - & \\
    \cite{das2021data} & CQT-LCNN (DASC) & 3.13 & 0.094 & - & $\checkmark$ \\ 
    \cite{zhang2020one} & LFCC-ResNet-OC  & 2.19 & 0.0560 & - & \\
    \cite{luo2021capsule} & LFCC-Capsule & 1.97 & 0.0538 & - & \\
    Table \ref{tab:eer} & LFCC-LCNN-LSTM-p2s & 1.92 & 0.0520 & 0.1119 & \\
    \cite{Chen2020Odyssey} & LFB-ResNet-AM   & 1.81 & 0.0520 & - & $\checkmark$ \\
    \cite{li2021channel} & CQT+MCG-Res2Net50 & 1.78 & 0.0520 & - &\\
    \cite{ge2021raw} & PC-DARTS Mel-F & 1.77 & 0.0517 & - & \\
    \cite{hua2021towards} & E2E Res-TSSDNet & 1.64 & - & -  & $\checkmark$ \\
    \cite{tak2021end} & RawGAT-ST (mul) & 1.06 & 0.0335 & - & $\checkmark$ \\
    \cite{chen2021spoofprint} & ResNet LDA cos-dis & 0.62$^{*}$ & - & - & $\checkmark$\\
    \bottomrule \\
    \multicolumn{6}{c}{Fused model} \\
    \toprule
    \multirow{2}{*}{Ref.} & \multirow{2}{*}{Model} & \multirow{2}{*}{EER (\%)} & \multicolumn{2}{c}{min t-DCF} & \multirow{2}{*}{Aug.}\\
    \cline{4-5}
     & & & legacy & v2.0 & \\
    \midrule
    \cite{tak2020end} &  GMM-RawNet2 (L+S1) & 1.12 & 0.0330  & - \\
    \cite{luo2021capsule} & LFCC-STFT Capsule & 1.07 & 0.0328 & - & \\
    Fig.\ref{fig:det_fuse_front_raw} & LCNNs \& RawNet & 0.87 & 0.0237 & 0.0849 &  \\
    2019 LA top primary & 7 sub-models (T05) & 0.22  & 0.0069 & 0.0692 & \\
    \bottomrule
    \end{tabular}}
    \label{tab:results_all}
\end{table}

\section{Remaining Issues and Challenges}
\label{sec:discussion}
This chapter is an introduction to the very basics of voice PADs. Although the models built in the experiments achieved reasonably good performances, they are by no means the best solutions. Many issues and challenges remain.

\subsubsection{Codec, transmission, and channel variability} The LA scenario introduced in this chapter assumes that both spoofed and bona fide trials are mounted behind the microphone on lossless transmission lines. This does not cover the scenarios when the speech data is transmitted across telephony or VoIP networks with various coding and transmission effects. If a PAD model is trained on data without the codec and transmission effects, this model may see performance degradation when it is deployed on a real telephony or VoIP network. This challenging scenario is partially explored in ASVspoof 2021 \cite{yamagishi21_asvspoof}, and many in-depth studies can be found from the ASVspoof 2021 workshop proceedings\footnote{\url{https://www.isca-speech.org/archive/asvspoof_2021/index.html}}.

\subsubsection{Generalizability} A few studies have demonstrated how the voice PAD models trained on one dataset performed badly on evaluation data from another database \cite{paul2017generalization,das2020assessing}. It is possible that the voice PAD models overfit the limited amount of data in the training set. When the evaluation trials from another data set are mounted, any mismatch could make the model fail to respond. This motivates a few studies using data augmentation, for example, augmentation based on waveform companding \cite{das2021data} and frequency mask \cite{Chen2020Odyssey}, to alleviate potential mismatches between training and unseen test data. Another potential direction is to use self-supervised front end trained on various speech data. It is shown that PAD models with a good pre-trained self-supervised front end achieved reasonable performance on ASVspoof 2015, 2019 LA, and 2021 LA test sets \cite{wang2021investigating}. However, further improvement is required to achieve the state-of-the-art EER on all the test sets.

\subsubsection{PAD paradigm} Currently, most of the studies including the ASVspoof challenges assume that the input trial is either bona fide or spoofed. However, attackers may not necessarily create a completely spoofed trial. They can use TTS and VC algorithms only to spoof or manipulate part of the trials. In other words, the input trial could be partially spoofed. There are a few recent studies that focus on this issue \cite{Kumar2021,yi2021half,zhang2021initial}. One strategy is to assign one score not only to the entire trial but also to each segment in the trial. It has been shown that conventional PAD models can be adapted to the partial spoof PAD task \cite{zhang2021initial}, but there are methods to improve the PAD for this new challening scenario \cite{zhang21_asvspoof}.

\section{Conclusion}
\label{sec:conclusion}
This chapter serves as an introduction to the voice PAD, with a focus on the logical access task. It covers the commonly used front ends, back ends, and training criteria, and the explanation is mainly from the machine learning point of view. Through experiments on some of the explained components, this chapter built a few voice PAD models that worked reasonably well on the ASVspoof 2019 LA task. It further explained the issue of intra-model variance and detailed statistical analysis on both intra- and inter-model differences. A model fusion experiment demonstrated the effectiveness of merging models with different front ends.
Source codes for all the experimental models have been published online, and we encourage the readers to review it.

For further introduction to the voice PAD task, readers are encouraged to read other papers. 
Studies in \cite{Wu2015,sahidullah2019introduction,evans2014speaker} provide an overview of voice PAD from different viewpoints. The former also includes a historic review of voice PAD in its early phase. 
Study in \cite{kamble_sailor_patil_li_2020} provides a detailed review on voice PAD, particularly on the front ends from a signal processing point of view. The overview paper of ASVspoof challenges \cite{Nautsch2021,wu2017asvspoof} are also good references. For more advanced studies, papers from ICASSP, INTERSPEECH, ISCA Odyssey, and related journals are highly recommended.

\subsubsection*{Acknowledgments} This study was supported by the Japanese-French VoicePersonae Project supported by JST CREST (JPMJCR18A6), JST CREST Grants (JPMJCR20D3), MEXT KAKENHI Grants (21K17775, 16H06302, 18H04112, 21H04906), Japan, and the Google AI for Japan program.

\bibliographystyle{splncs04}
\bibliography{library}

\begin{thebibliography}{100}
\providecommand{\url}[1]{\texttt{#1}}
\providecommand{\urlprefix}{URL }
\providecommand{\doi}[1]{https://doi.org/#1}

\bibitem{alegre2013one}
Alegre, F., Amehraye, A., Evans, N.: {A one-class classification approach to
  generalised speaker verification spoofing countermeasures using local binary
  patterns}. In: Proc. BTAS. pp.~1--8. IEEE (2013).
  \doi{10.1109/BTAS.2013.6712706}

\bibitem{Alluri2017}
Alluri, K.R., Achanta, S., Kadiri, S.R., Gangashetty, S.V., Vuppala, A.K.: {SFF
  Anti-Spoofer: IIIT-H Submission for Automatic Speaker Verification Spoofing
  and Countermeasures Challenge 2017}. In: Proc. Interspeech. pp. 107--111
  (2017). \doi{10.21437/Interspeech.2017-676}

\bibitem{baevski2019vq}
Baevski, A., Schneider, S., Auli, M.: {VQ-Wav2vec: Self-supervised learning of
  discrete speech representations}. Proc. ICLR  (2020)

\bibitem{NEURIPS2020_92d1e1eb}
Baevski, A., Zhou, Y., Mohamed, A., Auli, M.: {wav2vec 2.0: A Framework for
  Self-Supervised Learning of Speech Representations}. In: Proc. NIPS. vol.~33,
  pp. 12449--12460 (2020)

\bibitem{bengio2004statistical}
Bengio, S., Mari{\'{e}}thoz, J.: {A statistical significance test for person
  authentication}. In: Proc. Odyssey (2004)

\bibitem{bridle1990training}
Bridle, J.S.: {Training stochastic model recognition algorithms as networks can
  lead to maximum mutual information estimation of parameters}. In: Proc. NIPS.
  pp. 211--217 (1990)

\bibitem{bulusu2020anomalous}
Bulusu, S., Kailkhura, B., Li, B., Varshney, P.K., Song, D.: {Anomalous example
  detection in deep learning: A survey}. IEEE Access  \textbf{8},
  132330--132347 (2020). \doi{10.1109/ACCESS.2020.3010274}

\bibitem{Cai2017}
Cai, W., Cai, D., Liu, W., Li, G., Li, M.: {Countermeasures for Automatic
  Speaker Verification Replay Spoofing Attack : On Data Augmentation, Feature
  Representation, Classification and Fusion}. In: Proc. Interspeech. pp. 17--21
  (2017). \doi{10.21437/Interspeech.2017-906}

\bibitem{Cai2019}
Cai, W., Wu, H., Cai, D., Li, M.: {The DKU Replay Detection System for the
  ASVspoof 2019 Challenge: On Data Augmentation, Feature Representation,
  Classification, and Fusion}. In: Proc. Interspeech. pp. 1023--1027 (2019).
  \doi{10.21437/Interspeech.2019-1230}

\bibitem{chakroborty2007improved}
Chakroborty, S., Roy, A., Saha, G.: {Improved closed set text-independent
  speaker identification by combining MFCC with evidence from flipped filter
  banks}. International Journal of Signal Processing  \textbf{4}(2),  114--122
  (2007)

\bibitem{chen2015robust}
Chen, N., Qian, Y., Dinkel, H., Chen, B., Yu, K.: {Robust deep feature for
  spoofing detection—The SJTU system for ASVspoof 2015 challenge}. In: Proc.
  Interspeech. pp. 2097--2101 (2015). \doi{10.21437/Interspeech.2015-474}

\bibitem{chen2021spoofprint}
Chen, T., Khoury, E.: {Spoofprint: A New Paradigm for Spoofing Attacks
  Detection}. In: Proc. SLT. pp. 538--543. IEEE (2021).
  \doi{10.1109/SLT48900.2021.9383572}

\bibitem{Chen2020Odyssey}
Chen, T., Kumar, A., Nagarsheth, P., Sivaraman, G., Khoury, E.: {Generalization
  of Audio Deepfake Detection}. In: Proc. Odyssey. pp. 132--137 (2020).
  \doi{10.21437/Odyssey.2020-19}

\bibitem{Chen2017}
Chen, Z., Xie, Z., Zhang, W., Xu, X.: {ResNet and Model Fusion for Automatic
  Spoofing Detection}. In: Proc. Interspeech. pp. 102--106 (2017).
  \doi{10.21437/Interspeech.2017-1085}

\bibitem{cheng2019replay}
Cheng, X., Xu, M., Zheng, T.F.: {Replay detection using CQT-based modified
  group delay feature and ResNeWt network in ASVspoof 2019}. In: Proc. APSIPA
  ASC. pp. 540--545 (2019). \doi{10.1109/APSIPAASC47483.2019.9023158}

\bibitem{chettri2020dataset}
Chettri, B., Benetos, E., Sturm, B.L.T.: {Dataset artefacts in anti-spoofing
  systems: a case study on the ASVspoof 2017 benchmark}. IEEE/ACM Transactions
  on Audio, Speech, and Language Processing  \textbf{28},  3018--3028 (2020).
  \doi{10.1109/TASLP.2020.3036777}

\bibitem{Chettri2019}
Chettri, B., Stoller, D., Morfi, V., Ram{\'{i}}rez, M.A.M., Benetos, E., Sturm,
  B.L.: {Ensemble Models for Spoofing Detection in Automatic Speaker
  Verification}. In: Proc. Interspeech. pp. 1018--1022 (2019).
  \doi{10.21437/Interspeech.2019-2505}

\bibitem{rahman2018attention}
rahman Chowdhury, F.A.R., Wang, Q., Moreno, I.L., Wan, L.: {Attention-based
  models for text-dependent speaker verification}. In: Proc. ICASSP. pp.
  5359--5363. IEEE (2018). \doi{10.1109/ICASSP.2018.8461587}

\bibitem{chung2021similarity}
Chung, Y.A., Belinkov, Y., Glass, J.: {Similarity Analysis of Self-Supervised
  Speech Representations}. In: Proc. ICASSP. pp. 3040--3044. IEEE (2021).
  \doi{10.1109/ICASSP39728.2021.9414321}

\bibitem{das2020assessing}
Das, R.K., Yang, J., Li, H.: {Assessing the scope of generalized
  countermeasures for anti-spoofing}. In: Proc. ICASSP. pp. 6589--6593. IEEE
  (2020). \doi{10.1109/ICASSP40776.2020.9053086}

\bibitem{das2021data}
Das, R.K., Yang, J., Li, H.: {Data Augmentation with Signal Companding for
  Detection of Logical Access Attacks}. Proc. ICASSP pp. 6349--6353 (2021).
  \doi{10.1109/ICASSP39728.2021.9413501},
  \url{https://ieeexplore.ieee.org/document/9413501/}

\bibitem{davis1980comparison}
Davis, S., Mermelstein, P.: {Comparison of parametric representations for
  monosyllabic word recognition in continuously spoken sentences}. IEEE
  transactions on acoustics, speech, and signal processing  \textbf{28}(4),
  357--366 (1980). \doi{10.1109/TASSP.1980.1163420}

\bibitem{de2012evaluation}
{De Leon}, P.L., Pucher, M., Yamagishi, J., Hernaez, I., Saratxaga, I.:
  {Evaluation of speaker verification security and detection of HMM-based
  synthetic speech}. IEEE Transactions on Audio, Speech, and Language
  Processing  \textbf{20}(8),  2280--2290 (2012).
  \doi{10.1109/TASL.2012.2201472}

\bibitem{dempster1977maximum}
Dempster, A.P., Laird, N.M., Rubin, D.B.: {Maximum likelihood from incomplete
  data via the EM algorithm}. Journal of the Royal Statistical Society: Series
  B (Methodological)  \textbf{39}(1),  1--22 (1977)

\bibitem{deng2019arcface}
Deng, J., Guo, J., Xue, N., Zafeiriou, S.: {Arcface: Additive angular margin
  loss for deep face recognition}. In: Proc. CVPR. pp. 4690--4699 (2019).
  \doi{10.1109/CVPR.2019.00482}

\bibitem{dinkel2017end}
Dinkel, H., Chen, N., Qian, Y., Yu, K.: {End-to-end spoofing detection with raw
  waveform CLDNNS}. In: Proc. ICASSP. pp. 4860--4864. IEEE (2017).
  \doi{10.1109/ICASSP.2017.7953080}

\bibitem{evans2013spoofing}
Evans, N., Kinnunen, T., Yamagishi, J.: {Spoofing and countermeasures for
  automatic speaker verification}. In: Proc. Interspeech. pp. 925--929 (2013).
  \doi{10.21437/Interspeech.2013-288}

\bibitem{evans2014speaker}
Evans, N., Kinnunen, T., Yamagishi, J., Wu, Z., Alegre, F., {De Leon}, P.:
  {Speaker recognition anti-spoofing}. In: Handbook of biometric anti-spoofing,
  pp. 125--146. Springer (2014)

\bibitem{frank2021wavefake}
Frank, J., Sch{\"{o}}nherr, L.: {WaveFake: A Data Set to Facilitate Audio
  DeepFake Detection}. Proc. NeurIPS Datasets and Benchmarks 2021 p. accepted
  (2021)

\bibitem{ge2021raw}
Ge, W., Patino, J., Todisco, M., Evans, N.: {Raw Differentiable Architecture
  Search for Speech Deepfake and Spoofing Detection}. In: Proc. ASVspoof
  Challenge workshop. pp. 22--28 (2021). \doi{10.21437/ASVSPOOF.2021-4}

\bibitem{gomez2019gated}
Gomez-Alanis, A., Peinado, A.M., Gonzalez, J.A., Gomez, A.M.: {A gated
  recurrent convolutional neural network for robust spoofing detection}.
  IEEE/ACM Transactions on Audio, Speech, and Language Processing
  \textbf{27}(12),  1985--1999 (2019). \doi{10.1109/TASLP.2019.2937413}

\bibitem{hansen2015speaker}
Hansen, J.H.L., Hasan, T.: {Speaker recognition by machines and humans: A
  tutorial review}. IEEE Signal processing magazine  \textbf{32}(6),  74--99
  (2015). \doi{10.1109/MSP.2015.2462851}

\bibitem{he2016deep}
He, K., Zhang, X., Ren, S., Sun, J.: {Deep residual learning for image
  recognition}. In: Proc. CVPR. pp. 770--778 (2016). \doi{10.1109/CVPR.2016.90}

\bibitem{heckert2002handbook}
Heckert, N.A., Filliben, J.J., Croarkin, C.M., Hembree, B., Guthrie, W.F.,
  Tobias, P., Prinz, J., Others: {Handbook 151: NIST/SEMATECH e-Handbook of
  Statistical Methods}  (2002). \doi{https://doi.org/10.18434/M32189}

\bibitem{holm1979simple}
Holm, S.: {A simple sequentially rejective multiple test procedure}.
  Scandinavian journal of statistics pp. 65--70 (1979)

\bibitem{hsu2021hubert}
Hsu, W.N., Bolte, B., Tsai, Y.H.H., Lakhotia, K., Salakhutdinov, R., Mohamed,
  A.: {HuBERT: Self-Supervised Speech Representation Learning by Masked
  Prediction of Hidden Units}. IEEE/ACM Transactions on Audio, Speech, and
  Language Processing  \textbf{29},  3451--3460 (2021).
  \doi{10.1109/TASLP.2021.3122291}

\bibitem{hua2021towards}
Hua, G., Bengjinteoh, A., Zhang, H.: {Towards End-to-End Synthetic Speech
  Detection}. IEEE Signal Processing Letters  (2021).
  \doi{10.1109/LSP.2021.3089437}

\bibitem{hubert2010minimum}
Hubert, M., Debruyne, M.: {Minimum covariance determinant}. Wiley
  interdisciplinary reviews: Computational statistics  \textbf{2}(1),  36--43
  (2010)

\bibitem{ISO-IEC-30107-1-PAD-Framework-160115}
{ISO/IEC JTC1 SC37 Biometrics}: {ISO/IEC 30107-1. Information Technology -
  Biometric presentation attack detection - Part 1: Framework} (2016)

\bibitem{ISOIEC30107}
{ISO/IEC JTC1 SC37 Biometrics}: {ISO/IEC 30107: Information technology —
  Biometric presentation attack detection} (2016)

\bibitem{jia2018transfer}
Jia, Y., Zhang, Y., Weiss, R., Wang, Q., Shen, J., Ren, F., Nguyen, P., Pang,
  R., Moreno, I.L., Wu, Y., Others: {Transfer learning from speaker
  verification to multispeaker text-to-speech synthesis}. In: Proc. NIPS. pp.
  4480--4490 (2018)

\bibitem{kamble_sailor_patil_li_2020}
Kamble, M.R., Sailor, H.B., Patil, H.A., Li, H.: {Advances in anti-spoofing:
  from the perspective of ASVspoof challenges}. APSIPA Transactions on Signal
  and Information Processing  \textbf{9}, ~e2 (2020).
  \doi{10.1017/ATSIP.2019.21}

\bibitem{khoury2014introducing}
Khoury, E., Kinnunen, T., Sizov, A., Wu, Z., Marcel, S.: {Introducing i-vectors
  for joint anti-spoofing and speaker verification}. In: Proc. Interspeech. pp.
  61--65 (2014)

\bibitem{kingma2014adam}
Kingma, D.P., Ba, J.: {Adam: A method for stochastic optimization}. In: Proc.
  ICLR (2014)

\bibitem{kinnunen2020tandem}
Kinnunen, T., Delgado, H., Evans, N., Lee, K.A., Vestman, V., Nautsch, A.,
  Todisco, M., Wang, X., Sahidullah, M., Yamagishi, J., Reynolds, D.A.: {Tandem
  Assessment of Spoofing Countermeasures and Automatic Speaker Verification:
  Fundamentals}. IEEE/ACM Transactions on Audio, Speech, and Language
  Processing  \textbf{28},  2195--2210 (2020). \doi{10.1109/TASLP.2020.3009494}

\bibitem{kinnunen2018t}
Kinnunen, T., Lee, K.A., Delgado, H., Evans, N., Todisco, M., Sahidullah, M.,
  Yamagishi, J., Reynolds, D.A.: {t-DCF: a detection cost function for the
  tandem assessment of spoofing countermeasures and automatic speaker
  verification}. In: Proc. Odyssey. pp. 312--319 (2018).
  \doi{10.21437/Odyssey.2018-44}

\bibitem{kinnunen2010overview}
Kinnunen, T., Li, H.: {An overview of text-independent speaker recognition:
  From features to supervectors}. Speech communication  \textbf{52}(1),  12--40
  (2010). \doi{10.1016/j.specom.2009.08.009}

\bibitem{Kinnunen2017}
Kinnunen, T., Sahidullah, M., Delgado, H., Todisco, M., Evans, N., Yamagishi,
  J., Lee, K.A.: {The ASVspoof 2017 Challenge: Assessing the Limits of Replay
  Spoofing Attack Detection}. In: Proc. Interspeech. pp.~2--6 (2017).
  \doi{10.21437/Interspeech.2017-1111}

\bibitem{korshunov2018use}
Korshunov, P., Gon{\c{c}}alves, A.R., Violato, R.P.V., Sim{\~{o}}es, F.O.,
  Marcel, S.: {On the use of convolutional neural networks for speech
  presentation attack detection}. In: Proc. ISBA. pp.~1--8. IEEE (2018).
  \doi{10.1109/ISBA.2018.8311474}

\bibitem{korshunov2016overview}
Korshunov, P., Marcel, S., Muckenhirn, H., Gon{\c{c}}alves, A.R., Mello,
  A.G.S., Violato, R.P.V., Simoes, F.O., Neto, M.U., {de Assis Angeloni}, M.,
  Stuchi, J.A., Others: {Overview of BTAS 2016 speaker anti-spoofing
  competition}. In: Proc. BTAS. pp.~1--6. IEEE (2016).
  \doi{10.1109/BTAS.2016.7791200}

\bibitem{Kumar2021}
Kumar, A.K., Paul, D., Pal, M., Sahidullah, M., Saha, G.: {Speech frame
  selection for spoofing detection with an application to partially spoofed
  audio-data}. International Journal of Speech Technology  \textbf{24}(1),
  193--203 (mar 2021)

\bibitem{lai2019attentive}
Lai, C.I., Abad, A., Richmond, K., Yamagishi, J., Dehak, N., King, S.:
  {Attentive filtering networks for audio replay attack detection}. In: Proc.
  ICASSP. pp. 6316--6320. IEEE (2019). \doi{10.1109/ICASSP.2019.8682640}

\bibitem{Lai2019}
Lai, C.I., Chen, N., Villalba, J., Dehak, N.: {ASSERT: Anti-Spoofing with
  Squeeze-Excitation and Residual Networks}. In: Proc. Interspeech. pp.
  1013--1017 (2019). \doi{10.21437/Interspeech.2019-1794}

\bibitem{Lavrentyeva2017}
Lavrentyeva, G., Novoselov, S., Malykh, E., Kozlov, A., Kudashev, O.,
  Shchemelinin, V.: {Audio Replay Attack Detection with Deep Learning
  Frameworks}. In: Proc. Interspeech. pp. 82--86 (2017).
  \doi{10.21437/Interspeech.2017-360}

\bibitem{lavrentyeva2019stc}
Lavrentyeva, G., Novoselov, S., Tseren, A., Volkova, M., Gorlanov, A., Kozlov,
  A.: {STC Antispoofing Systems for the ASVspoof2019 Challenge}. In: Proc.
  Interspeech. pp. 1033--1037 (2019). \doi{10.21437/Interspeech.2019-1768}

\bibitem{Lee2018mdistance}
Lee, K., Lee, K., Lee, H., Shin, J.: {A Simple Unified Framework for Detecting
  Out-of-Distribution Samples and Adversarial Attacks}. In: Proc. NIPS. pp.
  7167--7177 (2018)

\bibitem{li2021channel}
Li, X., Wu, X., Lu, H., Liu, X., Meng, H.: {Channel-Wise Gated Res2Net: Towards
  Robust Detection of Synthetic Speech Attacks}. Proc. Interspeech pp.
  4314--4318 (aug 2021). \doi{10.21437/Interspeech.2021-2125}

\bibitem{lin2020wav2spk}
Lin, W.W., Mak, M.W.: {Wav2Spk: A Simple DNN Architecture for Learning Speaker
  Embeddings from Waveforms.} In: Proc. Interspeech. pp. 3211--3215 (2020).
  \doi{10.21437/Interspeech.2020-1287}

\bibitem{ling21_interspeech}
Ling, H., Huang, L., Huang, J., Zhang, B., Li, P.: {Attention-Based
  Convolutional Neural Network for ASV Spoofing Detection}. In: Proc.
  Interspeech. pp. 4289--4293 (2021). \doi{10.21437/Interspeech.2021-1404}

\bibitem{liu2017sphereface}
Liu, W., Wen, Y., Yu, Z., Li, M., Raj, B., Song, L.: {Sphereface: Deep
  hypersphere embedding for face recognition}. In: Proc. CVPR. pp. 212--220
  (2017). \doi{10.1109/CVPR.2017.713}

\bibitem{Lorenzo-Trueba2018}
Lorenzo-Trueba, J., Yamagishi, J., Toda, T., Saito, D., Villavicencio, F.,
  Kinnunen, T., Ling, Z.: {The Voice Conversion Challenge 2018: Promoting
  Development of Parallel and Nonparallel Methods}. In: Proc. Odyssey. pp.
  195--202 (2018). \doi{10.21437/Odyssey.2018-28}

\bibitem{luo2021capsule}
Luo, A., Li, E., Liu, Y., Kang, X., Wang, Z.J.: {A Capsule Network Based
  Approach for Detection of Audio Spoofing Attacks}. In: Proc. ICASSP. pp.
  6359--6363. IEEE (2021). \doi{10.1109/ICASSP39728.2021.9414670}

\bibitem{ma2021improved}
Ma, X., Liang, T., Zhang, S., Huang, S., He, L.: {Improved LightCNN with
  Attention Modules for ASV Spoofing Detection}. In: Proc. ICME. pp.~1--6. IEEE
  (2021). \doi{10.1109/ICME51207.2021.9428313}

\bibitem{madhyastha2019model}
Madhyastha, P.S., Jain, R.: {On Model Stability as a Function of Random Seed}.
  In: Proc. CoNLL. pp. 929--939 (2019)

\bibitem{martin1997det}
Martin, A., Doddington, G., Kamm, T., Ordowski, M., Przybocki, M.: {The DET
  Curve in Assessment of Detection Task Performance}. In: Proc. Eurospeech. pp.
  1895--1898 (1997)

\bibitem{muckenhirn2017end}
Muckenhirn, H., Magimai-Doss, M., Marcel, S.: {End-to-end convolutional neural
  network-based voice presentation attack detection}. In: 2017 IEEE
  international joint conference on biometrics (IJCB). pp. 335--341. IEEE
  (2017). \doi{10.1109/BTAS.2017.8272715}

\bibitem{Nautsch2021}
Nautsch, A., Wang, X., Evans, N., Kinnunen, T.H., Vestman, V., Todisco, M.,
  Delgado, H., Sahidullah, M., Yamagishi, J., Lee, K.A.: {ASVspoof 2019:
  Spoofing Countermeasures for the Detection of Synthesized, Converted and
  Replayed Speech}. IEEE Transactions on Biometrics, Behavior, and Identity
  Science  \textbf{3}(2),  252--265 (apr 2021).
  \doi{10.1109/TBIOM.2021.3059479}

\bibitem{wavenet}
van~den Oord, A., Dieleman, S., Zen, H., Simonyan, K., Vinyals, O., Graves, A.,
  Kalchbrenner, N., Senior, A.W., Kavukcuoglu, K.: {WaveNet: A Generative Model
  for Raw Audio}. CoRR  \textbf{abs/1609.0} (2016)

\bibitem{paul2017generalization}
Paul, D., Sahidullah, M., Saha, G.: {Generalization of spoofing
  countermeasures: A case study with ASVspoof 2015 and BTAS 2016 corpora}. In:
  Proc. ICASSP. pp. 2047--2051. IEEE (2017). \doi{10.1109/ICASSP.2017.7952516}

\bibitem{prabhu2021uncertainty}
{Prabhu Teja}, S., Fleuret, F.: {Uncertainty reduction for model adaptation in
  semantic segmentation}. In: Proc.CVPR. pp. 9613--9623 (2021).
  \doi{10.1109/CVPR46437.2021.00949}

\bibitem{ravanelli2018speaker}
Ravanelli, M., Bengio, Y.: {Speaker Recognition from raw waveform with
  SincNet}. In: Proc. SLT. pp. 1021--1028. IEEE (2018).
  \doi{10.1109/SLT.2018.8639585}

\bibitem{reynolds2000speaker}
Reynolds, D.A., Quatieri, T.F., Dunn, R.B.: {Speaker verification using adapted
  Gaussian mixture models}. Digital signal processing  \textbf{10}(1-3),
  19--41 (2000)

\bibitem{rousseeuw1999fast}
Rousseeuw, P.J., Driessen, K.V.: {A fast algorithm for the minimum covariance
  determinant estimator}. Technometrics  \textbf{41}(3),  212--223 (1999)

\bibitem{sahidullah2019introduction}
Sahidullah, M., Delgado, H., Todisco, M., Kinnunen, T., Evans, N., Yamagishi,
  J., Lee, K.A.: {Introduction to voice presentation attack detection and
  recent advances}. In: Handbook of Biometric Anti-Spoofing, pp. 321--361.
  Springer (2019)

\bibitem{sahidullah2017robust}
Sahidullah, M., Thomsen, D.A.L., Hautam{\"{a}}ki, R.G., Kinnunen, T., Tan,
  Z.H., Parts, R., Pitk{\"{a}}nen, M.: {Robust voice liveness detection and
  speaker verification using throat microphones}. IEEE/ACM Transactions on
  Audio, Speech, and Language Processing  \textbf{26}(1),  44--56 (2017).
  \doi{10.1109/TASLP.2017.2760243}

\bibitem{Sailor2017}
Sailor, H.B., Kamble, M.R., Patil, H.A.: {Unsupervised Representation Learning
  Using Convolutional Restricted Boltzmann Machine for Spoof Speech Detection}.
  In: Proc. Interspeech. pp. 2601--2605 (2017).
  \doi{10.21437/Interspeech.2017-1393}

\bibitem{Schneider2019a}
Schneider, S., Baevski, A., Collobert, R., Auli, M.: {wav2vec: Unsupervised
  Pre-Training for Speech Recognition}. In: Proc. Interspeech. pp. 3465--3469.
  ISCA, ISCA (sep 2019). \doi{10.21437/Interspeech.2019-1873}

\bibitem{shen2018natural}
Shen, J., Pang, R., Weiss, R.J., Schuster, M., Jaitly, N., Yang, Z., Chen, Z.,
  Zhang, Y., Wang, Y., Skerrv-Ryan, R., Others: {Natural TTS synthesis by
  conditioning WaveNet on Mel spectrogram predictions}. In: Proc. ICASSP. pp.
  4779--4783 (2018). \doi{10.1109/ICASSP.2018.8461368}

\bibitem{shiota2015voice}
Shiota, S., Villavicencio, F., Yamagishi, J., Ono, N., Echizen, I., Matsui, T.:
  {Voice liveness detection algorithms based on pop noise caused by human
  breath for automatic speaker verification}. In: Proc. Interspeech. pp.
  239--243 (2015). \doi{10.21437/Interspeech.2015-92}

\bibitem{sizov2015joint}
Sizov, A., Khoury, E., Kinnunen, T., Wu, Z., Marcel, S.: {Joint speaker
  verification and antispoofing in the i-vector space}. IEEE Transactions on
  Information Forensics and Security  \textbf{10}(4),  821--832 (2015).
  \doi{10.1109/TIFS.2015.2407362}

\bibitem{snyder2018x}
Snyder, D., Garcia-Romero, D., Sell, G., Povey, D., Khudanpur, S.: {X-vectors:
  Robust dnn embeddings for speaker recognition}. In: Proc. ICASSP. pp.
  5329--5333. IEEE (2018). \doi{10.1109/ICASSP.2018.8461375}

\bibitem{srivastava2014dropout}
Srivastava, N., Hinton, G., Krizhevsky, A., Sutskever, I., Salakhutdinov, R.:
  {Dropout: A simple way to prevent neural networks from overfitting}. The
  Journal of Machine Learning Research  \textbf{15}(1),  1929--1958 (2014)

\bibitem{tak2021end}
Tak, H., Jung, J.w., Patino, J., Kamble, M., Todisco, M., Evans, N.:
  {End-to-end spectro-temporal graph attention networks for speaker
  verification anti-spoofing and speech deepfake detection}. Proc. ASVspoof
  Challenge workshop pp.~1--8 (sep 2021). \doi{10.21437/ASVSPOOF.2021-1},
  \url{https://www.isca-speech.org/archive/asvspoof{\_}2021/tak21{\_}asvspoof.html}

\bibitem{tak2020end}
Tak, H., Patino, J., Todisco, M., Nautsch, A., Evans, N., Larcher, A.:
  {End-to-end anti-spoofing with RawNet2}. Proc. ICASSP pp. 6369----6373
  (2020). \doi{0.1109/ICASSP39728.2021.9414234}

\bibitem{Tamamori2017}
Tamamori, A., Hayashi, T., Kobayashi, K., Takeda, K., Toda, T.:
  {Speaker-dependent WaveNet vocoder}. In: Proc. Interspeech. pp. 1118--1122
  (2017). \doi{10.21437/Interspeech.2017-314}

\bibitem{tan2021survey}
Tan, X., Qin, T., Soong, F., Liu, T.Y.: {A survey on neural speech synthesis}.
  arXiv preprint arXiv:2106.15561  (2021)

\bibitem{tian2016spoofing}
Tian, X., Xiao, X., Chng, E.S., Li, H.: {Spoofing speech detection using
  temporal convolutional neural network}. In: Proc. APSIPA. pp.~1--6. IEEE
  (2016). \doi{10.1109/APSIPA.2016.7820738}

\bibitem{Todisco2017}
Todisco, M., Delgado, H., Evans, N.: {Constant Q cepstral coefficients: A
  spoofing countermeasure for automatic speaker verification}. Computer Speech
  {\&} Language  \textbf{45},  516--535 (2017). \doi{10.1016/j.csl.2017.01.001}

\bibitem{todisco2016new}
Todisco, M., Delgado, H., Evans, N.W.D.: {A New Feature for Automatic Speaker
  Verification Anti-Spoofing: Constant Q Cepstral Coefficients.} In: Odyssey.
  vol.~2016, pp. 283--290 (2016). \doi{10.21437/Odyssey.2016-41}

\bibitem{Todisco2019}
Todisco, M., Wang, X., Vestman, V., Sahidullah, M., Delgado, H., Nautsch, A.,
  Yamagishi, J., Evans, N., Kinnunen, T.H., Lee, K.A.: {ASVspoof 2019: future
  horizons in spoofed and fake audio detection}. In: Proc. Interspeech. pp.
  1008--1012 (2019). \doi{10.21437/Interspeech.2019-2249}

\bibitem{Tom2018}
Tom, F., Jain, M., Dey, P.: {End-To-End Audio Replay Attack Detection Using
  Deep Convolutional Networks with Attention}. In: Proc. Interspeech. pp.
  681--685 (2018). \doi{10.21437/Interspeech.2018-2279}

\bibitem{tomilov21_asvspoof}
Tomilov, A., Svishchev, A., Volkova, M., Chirkovskiy, A., Kondratev, A.,
  Lavrentyeva, G.: {STC Antispoofing Systems for the ASVspoof2021 Challenge}.
  In: Proc. ASVspoof Challenge workshop. pp. 61--67 (2021).
  \doi{10.21437/ASVSPOOF.2021-10}

\bibitem{variani2014deep}
Variani, E., Lei, X., McDermott, E., Moreno, I.L., Gonzalez-Dominguez, J.:
  {Deep neural networks for small footprint text-dependent speaker
  verification}. In: Proc. ICASSP. pp. 4052--4056. IEEE (2014).
  \doi{10.1109/ICASSP.2014.6854363}

\bibitem{waibel1989phoneme}
Waibel, A., Hanazawa, T., Hinton, G., Shikano, K., Lang, K.J.: {Phoneme
  recognition using time-delay neural networks}. IEEE Transactions on
  Acoustics, Speech, and Signal Processing  \textbf{37}(3),  328--339 (1989).
  \doi{10.1109/29.21701}

\bibitem{wang2018additive}
Wang, F., Cheng, J., Liu, W., Liu, H.: {Additive margin softmax for face
  verification}. IEEE Signal Processing Letters  \textbf{25}(7),  926--930
  (2018). \doi{10.1109/LSP.2018.2822810}

\bibitem{wang2018cosface}
Wang, H., Wang, Y., Zhou, Z., Ji, X., Gong, D., Zhou, J., Li, Z., Liu, W.:
  {Cosface: Large margin cosine loss for deep face recognition}. In: Proc.
  CVPR. pp. 5265--5274 (2018). \doi{10.1109/CVPR.2018.00552}

\bibitem{wang2021comparative}
Wang, X., Yamagishi, J.: {A comparative study on recent neural spoofing
  countermeasures for synthetic speech detection}. Proc. Interspeech pp.
  4259--4263 (2021). \doi{10.21437/Interspeech.2021-702}

\bibitem{wang2021investigating}
Wang, X., Yamagishi, J.: {Investigating self-supervised front ends for speech
  spoofing countermeasures}. arXiv preprint arXiv:2111.07725  (2021)

\bibitem{WANG2020101114}
Wang, X., Yamagishi, J., Todisco, M., Delgado, H., Nautsch, A., Evans, N.,
  Sahidullah, M., Vestman, V., Kinnunen, T., Lee, K.A., Juvela, L., Alku, P.,
  Peng, Y.H., Hwang, H.T., Tsao, Y., Wang, H.M., Maguer, S.L., Becker, M.,
  Henderson, F., Clark, R., Zhang, Y., Wang, Q., Jia, Y., Onuma, K., Mushika,
  K., Kaneda, T., Jiang, Y., Liu, L.J., Wu, Y.C., Huang, W.C., Toda, T.,
  Tanaka, K., Kameoka, H., Steiner, I., Matrouf, D., Bonastre, J.F., Govender,
  A., Ronanki, S., Zhang, J.X., Ling, Z.H.: {ASVspoof 2019: A large-scale
  public database of synthesized, converted and replayed speech}. Computer
  Speech {\&} Language  \textbf{64},  101114 (nov 2020).
  \doi{10.1016/j.csl.2020.101114}

\bibitem{wu2018light}
Wu, X., He, R., Sun, Z., Tan, T.: {A light CNN for deep face representation
  with noisy labels}. IEEE Transactions on Information Forensics and Security
  \textbf{13}(11),  2884--2896 (2018). \doi{10.1109/TIFS.2018.2833032}

\bibitem{Wu2020}
Wu, Z., Das, R.K., Yang, J., Li, H.: {Light Convolutional Neural Network with
  Feature Genuinization for Detection of Synthetic Speech Attacks}. In: Proc.
  Interspeech. pp. 1101--1105 (2020). \doi{10.21437/Interspeech.2020-1810}

\bibitem{Wu2015}
Wu, Z., Evans, N., Kinnunen, T., Yamagishi, J., Alegre, F., Li, H.: {Spoofing
  and countermeasures for speaker verification: A survey}. Speech Communication
   \textbf{66},  130--153 (feb 2015). \doi{10.1016/j.specom.2014.10.005}

\bibitem{wu2014study}
Wu, Z., Gao, S., Cling, E.S., Li, H.: {A study on replay attack and
  anti-spoofing for text-dependent speaker verification}. In: Proc. APSIPA.
  pp.~1--5. IEEE (2014). \doi{10.1109/APSIPA.2014.7041636}

\bibitem{li2017study}
Wu, Z., Gao, S., Cling, E.S., Li, H.: {A study on replay attack and
  anti-spoofing for text-dependent speaker verification}. Proc.APSIPA pp.~1--5
  (dec 2014). \doi{10.1109/APSIPA.2014.7041636},
  \url{http://ieeexplore.ieee.org/document/7041636/}

\bibitem{wu2015sas}
Wu, Z., Khodabakhsh, A., Demiroglu, C., Yamagishi, J., Saito, D., Toda, T.,
  King, S.: {SAS: A speaker verification spoofing database containing diverse
  attacks}. In: Proc. ICASSP. pp. 4440--4444. IEEE (2015).
  \doi{10.1109/ICASSP.2015.7178810}

\bibitem{wu2015asvspoof}
Wu, Z., Kinnunen, T., Evans, N., Yamagishi, J., Hanil{\c{c}}i, C., Sahidullah,
  M., Sizov, A.: {ASVspoof 2015: the first automatic speaker verification
  spoofing and countermeasures challenge}. In: Proc. Interspeech. pp.
  2037--2041 (2015). \doi{10.21437/Interspeech.2015-462}

\bibitem{wu2017asvspoof}
Wu, Z., Yamagishi, J., Kinnunen, T., Hanil{\c{c}}i, C., Sahidullah, M., Sizov,
  A., Evans, N., Todisco, M., Delgado, H.: {ASVspoof: the automatic speaker
  verification spoofing and countermeasures challenge}. IEEE Journal of
  Selected Topics in Signal Processing  \textbf{11}(4),  588--604 (2017).
  \doi{10.1109/JSTSP.2017.2671435}

\bibitem{xie21_interspeech}
Xie, Y., Zhang, Z., Yang, Y.: {Siamese network with wav2vec feature for
  spoofing speech detection}. In: Proc. Interspeech. pp. 4269--4273 (2021).
  \doi{10.21437/Interspeech.2021-847}

\bibitem{yamagishi21_asvspoof}
Yamagishi, J., Wang, X., Todisco, M., Sahidullah, M., Patino, J., Nautsch, A.,
  Liu, X., Lee, K.A., Kinnunen, T., Evans, N., Delgado, H.: {ASVspoof 2021:
  accelerating progress in spoofed and deepfake speech detection}. In: Proc.
  ASVspoof Challenge workshop. pp. 47--54 (2021).
  \doi{10.21437/ASVSPOOF.2021-8}

\bibitem{Yang2019}
Yang, Y., Wang, H., Dinkel, H., Chen, Z., Wang, S., Qian, Y., Yu, K.: {The SJTU
  Robust Anti-Spoofing System for the ASVspoof 2019 Challenge}. In: Proc.
  Interspeech 2019. pp. 1038--1042 (2019). \doi{10.21437/Interspeech.2019-2170}

\bibitem{yi2021half}
Yi, J., Bai, Y., Tao, J., Ma, H., Tian, Z., Wang, C., Wang, T., Fu, R.:
  {Half-Truth: A Partially Fake Audio Detection Dataset}. Proc. Interspeech pp.
  1654--1658 (aug 2021). \doi{10.21437/Interspeech.2021-930}

\bibitem{Yi2020}
Yi, Z., Huang, W.C., Tian, X., Yamagishi, J., Das, R.K., Kinnunen, T., Ling,
  Z.H., Toda, T.: {Voice Conversion Challenge 2020 –- Intra-lingual
  semi-parallel and cross-lingual voice conversion –-}. In: Proc. Joint
  Workshop for the Blizzard Challenge and Voice Conversion Challenge 2020. pp.
  80--98 (2020). \doi{10.21437/VCCBC.2020-14}

\bibitem{yu2017dnn}
Yu, H., Tan, Z.H., Zhang, Y., Ma, Z., Guo, J.: {DNN Filter Bank Cepstral
  Coefficients for Spoofing Detection}. IEEE Access  \textbf{5},  4779--4787
  (2017). \doi{10.1109/ACCESS.2017.2687041}

\bibitem{zeghidour2021leaf}
Zeghidour, N., Teboul, O., Quitry, F.d.C., Tagliasacchi, M.: {Leaf: A learnable
  frontend for audio classification}. Proc. ICLR  (2021)

\bibitem{Zeinali2019}
Zeinali, H., Stafylakis, T., Athanasopoulou, G., Rohdin, J., Gkinis, I.,
  Burget, L., {\v{C}}ernock{\'{y}}, J.: {Detecting Spoofing Attacks Using VGG
  and SincNet: BUT-Omilia Submission to ASVspoof 2019 Challenge}. In: Proc.
  Interspeech 2019. pp. 1073--1077 (2019). \doi{10.21437/Interspeech.2019-2892}

\bibitem{zhang2017investigation}
Zhang, C., Yu, C., Hansen, J.H.L.: {An investigation of deep-learning
  frameworks for speaker verification antispoofing}. IEEE Journal of Selected
  Topics in Signal Processing  \textbf{11}(4),  684--694 (2017).
  \doi{10.1109/JSTSP.2016.2647199}

\bibitem{zhang2017mixup}
Zhang, H., Cisse, M., Dauphin, Y.N., Lopez-Paz, D.: {mixup: Beyond empirical
  risk minimization}. Proc. ICLR  (2018)

\bibitem{zhang21_asvspoof}
Zhang, L., Wang, X., Cooper, E., Yamagishi, J.: {Multi-task Learning in
  Utterance-level and Segmental-level Spoof Detection}. In: Proc. ASVspoof
  Challenge workshop. pp. 9--15 (2021). \doi{10.21437/ASVSPOOF.2021-2}

\bibitem{zhang2021initial}
Zhang, L., Wang, X., Cooper, E., Yamagishi, J., Patino, J., Evans, N.: {An
  Initial Investigation for Detecting Partially Spoofed Audio}. Proc.
  Interspeech pp. 4264--4268 (aug 2021). \doi{10.21437/Interspeech.2021-738}

\bibitem{zhang2020viblive}
Zhang, L., Tan, S., Wang, Z., Ren, Y., Wang, Z., Yang, J.: {VibLive: A
  Continuous Liveness Detection for Secure Voice User Interface in IoT
  Environment}. In: Annual Computer Security Applications Conference. pp.
  884--896 (2020)

\bibitem{Zhang2017}
Zhang, L., Tan, S., Yang, J.: {Hearing Your Voice is Not Enough}. In:
  Proceedings of the 2017 ACM SIGSAC Conference on Computer and Communications
  Security. pp. 57--71. ACM, New York, NY, USA (oct 2017).
  \doi{10.1145/3133956.3133962}

\bibitem{zhang2016voicelive}
Zhang, L., Tan, S., Yang, J., Chen, Y.: {Voicelive: A phoneme localization
  based liveness detection for voice authentication on smartphones}. In:
  Proceedings of the 2016 ACM SIGSAC Conference on Computer and Communications
  Security. pp. 1080--1091 (2016). \doi{10.1145/2976749.2978296}

\bibitem{zhang2019p2sgrad}
Zhang, X., Zhao, R., Yan, J., Gao, M., Qiao, Y., Wang, X., Li, H.: {P2sgrad:
  Refined gradients for optimizing deep face models}. In: Proc. CVPR. pp.
  9906--9914 (2019). \doi{10.1109/CVPR.2019.01014}

\bibitem{zhang2020one}
Zhang, Y., Jiang, F., Duan, Z.: {One-Class Learning Towards Synthetic Voice
  Spoofing Detection}. IEEE Signal Processing Letters  \textbf{28},  937--941
  (2021). \doi{10.1109/LSP.2021.3076358}

\bibitem{zhang2021fmfcc}
Zhang, Z., Gu, Y., Yi, X., Zhao, X.: {FMFCC-A: A Challenging Mandarin Dataset
  for Synthetic Speech Detection}. arXiv preprint arXiv:2110.09441  (2021)

\bibitem{Zhu2018}
Zhu, Y., Ko, T., Snyder, D., Mak, B., Povey, D.: {Self-Attentive Speaker
  Embeddings for Text-Independent Speaker Verification}. In: Proc. Interspeech.
  pp. 3573--3577 (2018). \doi{10.21437/Interspeech.2018-1158}

\end{thebibliography}

\section*{Appendix}

\subsection*{Open-sourced Code}
\begin{itemize}
    \item Source code for experiments in Section~\ref{sec:exp} is available at \url{https://github.com/nii-yamagishilab/project-NN-Pytorch-scripts}. Please check the project called \emph{Speech anti-spoofing for ASVspoof 2019 LA task};
    \item Official source code for ASVspoof2021 is available at \url{https://github.com/asvspoof-challenge/2021};
    \item Official source code for EER and min-tDCF evaluation is available at \url{https://www.asvspoof.org/resources/tDCF_python_v2.zip}. A derived version for ASVspoof2021 is also available in the 2nd link above.
\end{itemize}

\subsection*{ASVspoof challenges}
Official website for ASVspoof challenges is \url{https://www.asvspoof.org}.

\subsection*{Database}
\begin{itemize}
    \item ASVspoof 2015: \url{https://doi.org/10.7488/ds/298} 
    \item ASVspoof 2017: \url{https://doi.org/10.7488/ds/2332} 
    \item ASVspoof 2019: \url{https://doi.org/10.7488/ds/2555} 
    \item ASVspoof 2021: 
      \begin{itemize}
      \item LA scenario \url{https://doi.org/10.5281/zenodo.4837263}
      \item PA scenario \url{https://doi.org/10.5281/zenodo.4834716}
      \item DF scenario \url{https://doi.org/10.5281/zenodo.4835108}
      \end{itemize}
    \item AVSpoof \url{https://www.idiap.ch/en/dataset/avspoof}
    \item VoicePA \url{https://www.idiap.ch/en/dataset/voicepa}
    \item SAS \url{https://datashare.ed.ac.uk/handle/10283/771}
    \item WaveFake \url{https://doi.org/10.5281/zenodo.5642694}
    \item FMFCC-A \url{https://github.com/Amforever/FMFCC-A}
\end{itemize}


\end{document}